\documentclass[aps,prd,reprint,superscriptaddress,nofootinbib, longbibliography]{revtex4-1}

\usepackage[utf8]{inputenc}

\usepackage{epsf}  
\usepackage{graphicx}
\usepackage{amsmath}
\usepackage{amsfonts,color}
\usepackage{amssymb,float}
\usepackage{ulem}
\usepackage{booktabs}
\usepackage{mathrsfs}
\usepackage[caption=false]{subfig}
\usepackage{tikz}
\usetikzlibrary{shapes.misc, positioning}
\usepackage{microtype}
\usepackage[english]{babel}
\usepackage{blindtext}
\usepackage{bm}
\usepackage[breaklinks,colorlinks,citecolor=blue]{hyperref}

\renewcommand{\vec}[1]{{\bm #1}}

\def\apjs{{\ ApJS}}
\def\mnras{{\ MNRAS}}
\def\aap{{\ A\&A}}

\def\araa{{\ ARA\&A}}

\def\apjl{Astroph.J.Lett.}
\def\apjs{Astroph.J.S.}
\def\mnras{MNRAS}
\def\araa{ARA\&A}

\begin{document}
\title{Extreme Mass-Ratio Inspirals in Active Galactic Nucleus Disks: The Role of Circumsingle Disks}

\author{Ya-Ping Li}
\email{liyp@shao.ac.cn}
\affiliation{Shanghai Astronomical Observatory, Chinese Academy of Sciences, Shanghai 200030, China}

\author{Huan Yang}
\email{hyangdoa@tsinghua.edu.cn}
\affiliation{Department of Astronomy, Tsinghua University, Beijing 100084, China}

\author{Zhen Pan}
\email{zhpan@sjtu.edu.cn}
\affiliation{Tsung-Dao Lee Institute, Shanghai Jiao-Tong University, Shanghai, 520 Shengrong Road, 201210,  China}
\affiliation{School of Physics \& Astronomy, Shanghai Jiao-Tong University, Shanghai, 800 Dongchuan Road, 200240,  China}

\begin{abstract}
In this work, we numerically explore the dynamics of a point mass (e.g., a stellar-mass black hole) moving within a thin accretion disk of a massive object (i.e., a supermassive black hole) with three dimensional hydrodynamical simulations using \texttt{Athena++}. We are particularly interested in the regime that the Hill radius of the point mass is greater than the disk thickness, but the point mass is not sufficiently massive to open a ``gap" in a high viscosity disk. This parameter regime may be  seen for stellar-mass objects moving within thin accretion disks of massive black holes, but less studied for the planet migration scenario in protoplanetary disks. We find that the disk migration may be significantly slower than the type I migration, depending on the surface density gradient of the disk. Furthermore, the circumsingle disk around the point mass plays an important role in damping the orbital eccentricity. If the gravitational interaction between the point mass and circumsingle disk material is turned off, the orbital eccentricity may be pumped to $\mathcal{O}(10^{-2})$ level. 
Because space-borne gravitational wave detectors such as LISA (Laser Interferometer Space Antenna) is able to measure the eccentricity of extreme mass-ratio inspirals to the level of $\mathcal{O}(10^{-5})$, this finding highlights the importance of understanding stellar-mass black hole feedback mechanisms which will modify the structures of the circumsingle disk, as they will impact LISA observables such as the eccentricity.  
\end{abstract}
\maketitle

\section{Introduction}

With the ongoing enhancement of ground-based detector sensitivity and the upcoming launch of space-borne detectors in the next decade \cite{Evans:2021gyd,Punturo:2010zz,Baker:2019nia,Hu:2017mde,luo2016tianqin}, gravitational wave (GW) astronomy is set to enter a new era of precision science, akin to the advancements brought by the Planck mission in the study of the Cosmic Microwave Background \cite{Planck:2018nkj}. One promising avenue for these future detections is the exploration of the astrophysical environments surrounding gravitational wave sources, achieved through direct measurement of waveform modulation and/or indirect measurement of source parameters like mass, spin, eccentricity, and more.

\begin{figure*}
\centering
\includegraphics[width=0.47\textwidth,clip=true]{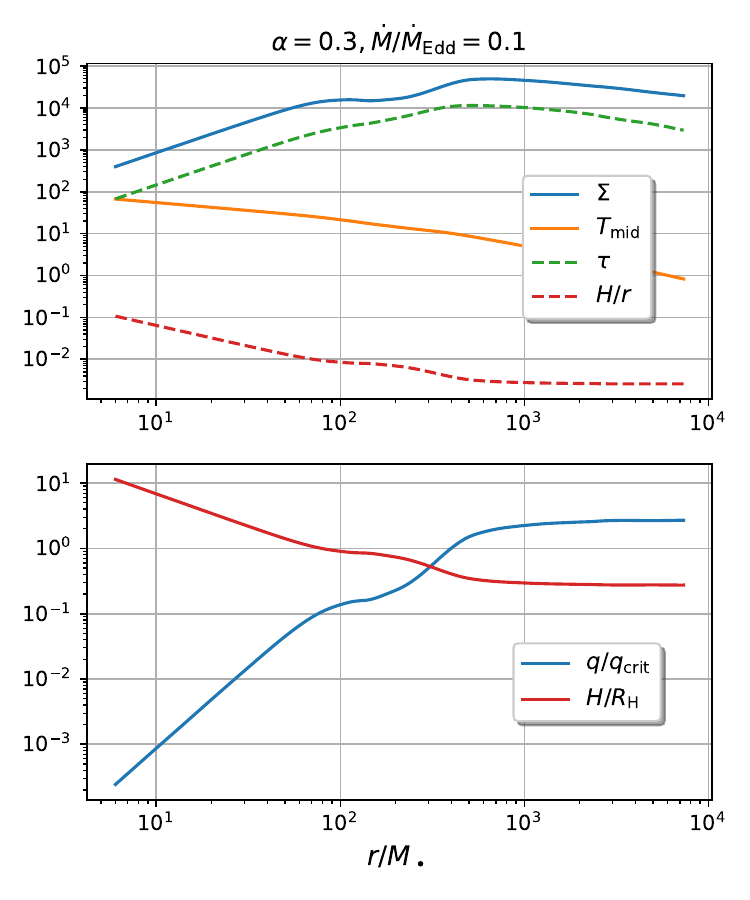}
\includegraphics[width=0.47\textwidth,clip=true]{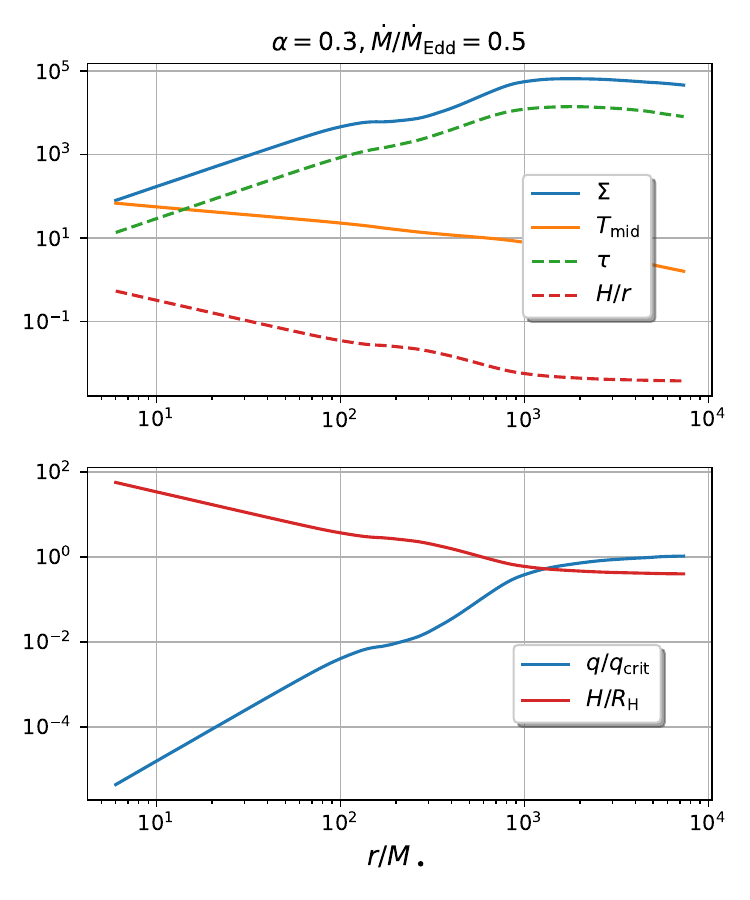}
\caption{Upper panels: Structure of two $\alpha$-disks with 
the central SMBH mass is $M_\bullet=4\times10^6 M_\odot$, the stellar-mass black hole mass is $M_{\rm p}=10 M_\odot$, the dimensionless viscosity coefficient $\alpha=0.3$ and 
accretion rates $\dot M/\dot M_{\rm Edd} = 0.1$ and $0.5$
respectively. Here  $\Sigma\ [{\rm g/cm}^2]$, $T_{\rm mid}\ [{\rm eV}]$, $H\ [M_\bullet]$ and $\tau$ 
are the disk surface density, the mid-plane temperature,  the disk scale height from the mid-plane and the disk optical depth, respectively.
Lower panels: We compare the disk height $H$ with the Hill radius of the sBH $R_{\rm H}=(q/3)^{1/3} r_{\rm p}= (M_{\rm p}/3M_\bullet)^{1/3} r_{\rm p}$, and check the gap opening condition $q\gtrsim q_{\rm crit}$, where $q_{\rm crit} = \sqrt{25 \alpha h^5}$, according to the gap-opening condition $K:= q^2/a h^5 \gtrsim 25$ \citep{Kanagawa2015} (see also \citep{Lin1986,Duffell2013}), with the disk aspect ratio $h=H/r$, and $q=M_{\rm p}/M_{\bullet}$.}
 \label{fig:no_gap}
\end{figure*}

Among various gravitational wave sources it is known that Extreme Mass-Ratio Inspiral (EMRI), which usually comprises a stellar-mass object orbiting around a massive black hole, is a particularly sensitive probe to weak, astrophysical environmental forces \cite{LISAConsortiumWaveformWorkingGroup:2023arg,LISA:2022yao,Barausse:2020rsu}. 
The gravitational radiation timescale of such systems is generally much longer than the orbital timescale, so that it generally stays in-band for many orbital cycles during the observation period.   Therefore, small effects may be able to accumulate for many cycles to generate detectable signals. This level of sensitivity is also reflected in the measurement precision of intrinsic parameters of the system. For a typical EMRI with detection signal-to-noise ratio $\sim 20$, the relative measurement uncertainties for the mass, spin of the massive black hole and EMRI's orbital eccentricity are $\sim \mathcal{O}(10^{-5})$ \cite{Babak:2017tow,Fan:2020zhy}.

It has been suggested some (wet) EMRIs may form within Active Galactic Nuclei (AGNs) \cite{Levin:2003ej}, with the rate comparable or even higher than those formed through scattering in the nuclear cluster \cite{Pan:2021ksp,Pan:2021oob,Pan:2021lyw,Wang2023}. Recent analyses of X-ray quasi-periodic eruptions provide observational evidence of the common existence of low-eccentricity stellar-mass orbiters around supermassive black holes (SMBHs) that are consistent with wet EMRIs \cite{Zhou:2024bjt,Zhou:2024nuu,Zhou:2024vwj}.
Disk-driven migration is the key factor for the accelerated EMRI formation within AGN \cite{Pan:2021ksp}. It is in some aspects similar to planet migrations extensively studied in protoplanetary disks, such as physical processes associated with density wave excitation/dissipation, gap-opening phenomena, etc, although the physical condition between AGN disks and protoplanetary disks are quite different \citep[e.g.,][]{GT1980,Lin1986,KleyNelson2012,Chen2020,Paardekooper2023,Li2024}. However, stellar-mass black hole (sBH) motion in  AGN disks also bears some important differences from planet migration in planetary disks.  For example, because of the low level of turbulence, effective viscosity parameter of protoplanetary disks $\alpha$ is constrained to be on the order of $\le \mathcal{O}(10^{-3})$ in most studies \cite{Pinte2016,Flaherty2017}. On the other hand, the $\alpha$ viscosity of fully ionized disks (e.g., inner parts of AGN disks) can reach the level of $\mathcal{O}(0.1\sim 1)$ as inferred from various accretion systems \cite{Martin:2019qfg}. Second, the disk aspect ratio $H/r$ for protoplanetary disks is usually on the order of $\sim0.05$ with a flaring geometry \citep{Chiang1997}, which is much thicker than that of AGN disk as shown in Figure~\ref{fig:no_gap}. Note that here we do not consider the slim disks in the super-Eddington regime \citep{Abramowicz1988} or the advection-dominated accretion flows for the low-luminosity AGN regime \citep{Narayan1994,Yuan2014}, for which the disk aspect ratio is usually much larger as shown in recent simulations \citep{Guo2024,Hopkins2024}. The relative scale length of the disk aspect ratio and the Hill radius may determine whether the sBH can carve deep gaps in the global disk.
Third, sBHs immersed in AGN disks do not simply swallow disk materials within the Bondi/Hill radius, instead through the supercritical accretion process most gravitationally captured may actually join the outflow \cite{Pan:2021xhv}, which may dramatically modify the local disk structure around the sBHs. From the study of this work, we shall argue that the feedback process of sBHs may modify the eccentricity  of wet EMRIs, which is potentially detectable by space-borne gravitational wave detectors.

In this work, we focus on EMRIs with orbital radius of $\mathcal{O}(10^2)-\mathcal{O}(10^3)$ gravitational radii of the massive black hole - the evolution stage of EMRI migration right before gravitational wave decay takes over \cite{Pan:2021ksp}. We are particularly interested in the regime that the disk thickness is smaller than the Hill radius of the sBH, but yet  the sBH is not sufficiently massive to open a gap (see Figure~\ref{fig:no_gap}, where we show two examples of AGN disks \citep{Sirko:2002ex}).  This possibility is difficult to reach for planetary systems as protoplanetary disks generally have much smaller effective viscosity.

We compare the dynamics of EMRIs for two sets of disk parameters, with the Hill radius set to be $0.015$ in code unit and disk aspect ratio $h$ chosen to be $0.05$ or $0.01$, respectively. In the $h=0.05$ case, the three-dimensional joint hydrodynamical evolution of the disk and the point mass approximately reproduces the expectation of the type-I migration theory with a nearly negligible eccentricity, which is consistent with linear theory and planet-disk interaction simulations \citep{Papaloizou2000,Goldreich2003,Li2019,Ida2020}. In the $h=0.01$ scenario, however, we find that the point mass migration speed is significantly slower than the type-I migrate rate, especially where the disk surface density is more homogeneous. In addition, the density wave emission seems to prefer to drive the orbit out of circular profiles, which has not been discussed in previous literature (to the best of our knowledge). Note that since there is no deep gap around the sBH, the eccentricity excitation could be different from the scenario associated with the tidally induced gap shown in \citep{Papaloizou2001,Goldreich2003,DAngelo2006,Duffell2015}, but the first-order external Lindblad resonances likely still play a role in exciting the eccentricity \citep{Goldreich2003,Duffell2015}. Meanwhile the fluid within the circumsingle disk (CSD) tends to damp  the orbital eccentricity, bringing back the orbit to quasi-circles. This dynamical effect of CSDs, which is non-perturbative in nature as fluids within are circling around the point mass, has also not been identified before. It appears to be quite efficiently in damping out the eccentricity as simulations with CSDs removed may have $\mathcal{O}(10^{-2})$ eccentricity, as compared to $\mathcal{O}(10^{-5})$ detection limit by LISA. Since we expect the gas structure of CSDs should be intimately related to the feedback process of the less massive object, it may play a vital role in determining the eccentricity of wet EMRIs measured by LISA.

This paper is organized as follows. In Sec.~\ref{sec:method} we show the system setup for our hydrodynamical numerical simulations and analytical method for eccentricity evolution. In Sec.~\ref{sec:results} we present the numerical results from our simulations. The discussion and conclusion are shown in Sec.~\ref{sec:con}.

\section{Methods}\label{sec:method}

In this section, we first discuss the setup of the disk and orbital profile of the system, and then diagnose the dynamical evolution of the point mass for the study of eccentricity evolution in Sec.~\ref{sec:ecc_ana}.

\subsection{System Setup}

We simulate globally  the object embedded in a 3D thin and non-self-gravitating disk with \texttt{Athena++}
\citep{Stone2020}. 
Our simulations are mostly targeting at the scenario of a stellar-mass object embedded in AGN disks, but can also be applied to stars in AGN disks or low-mass planets in some special regions of protoplanetary disks. From now on, we will use the term sBH to refer to the embedded object.

We numerically solve the continuity equation, and Navier–Stokes equation of the fluid. 
The origin of our coordinate system is located at the position of the central SMBH with 
mass $M_{\bullet }$. 

We adopt a simple model in which the disk temperature $T_{\rm disk}$ is independent of the distance above the mid-plane $z$, with $h_{0}\equiv H/r$ at $r_{0}$, where $r_{0}$ is a typical scale length of the disk. Two disk scale heights are considered: with $h_{0}=0.05$ and $h_{0}=0.01$.
For these 3D models,  we use spherical coordinates $(r, \theta, \phi)$ for the simulations,
and initialize density structure in the cylindrical radial ($R= r \sin \theta$) and vertical ($z = r \cos \theta$) direction as \citep{Nelson2013}. 

\begin{equation}
\rho = \rho_{0}\left(\frac{R}{r_{0}}\right)^{p}\exp\left[\frac{GM_{\bullet }}{c_{\rm s}^2} \left(\frac{1}{\sqrt{R^2+z^2}}-\frac{1}{R}\right) \right], 
\end{equation}
and the rotational velocity is

\begin{equation}
v_{\phi}(R,z) = v_{\rm K}\left[ p\left(\frac{c_{\rm s}}{v_{\rm K}}\right)^{2}+1\right]^{1/2}, 
\end{equation}
where $p$ is the radial power-law index for gas density, and $v_{\rm K}=\sqrt{GM_{\bullet }/R}$. In all cases, a globally isothermal equation-of-state is further adopted such that the disk temperature is also independent of radius $r$.
Density profile with different power-law index $p$ has been considered to explore its effect on the dynamics of the embedded object. $\rho_{0}=4\times10^{-3}M_{\bullet}/r_{0}^{3}$ is the gas density at $r_{0}$, which is about one order of magnitude lower than the disk density as implied from Figure~\ref{fig:no_gap} with $M_{\bullet}\sim\mathcal{O}(10^{6})M_{\odot}$ and $r_{0}\sim\mathcal{O}(10^{2})\ M_{\bullet}$.
Such a low gas density slows the active migration of the sBH for the dynamically evolving (``live") orbit simulations, allowing us to select a narrow radial region for the disk simulations and justify the omission of the disk's self-gravity.
The disk viscosity is adopted from the Shakura-Sunyaev prescription $\nu_{\rm g}=\alpha c_{\rm s}H$ with
a constant $\alpha$ across the whole disk \citep{Shakura1973}. 

For the gravitational potential of the point mass (e.g., the sBH) at $\vec{r}_{\rm p}$, we use a smoothed potential 
of the form \citep[e.g.,][]{GT1980}

\begin{equation}
\Phi_{\rm p}=-\frac{G M_{_{\rm p}}}{(\left|\vec{r}_{\rm p}-\vec{r}\right|^2+\epsilon^2)^{1/2}}+q \Omega_{\rm p}^{2} \vec r_{\rm p} \cdot \vec r,
\label{eq:potential}
\end{equation}
where $\vec{r}_{\rm p}$ indicates the location of the embedded object, $\epsilon$ is the softening length and $\epsilon=0.1\ R_{\rm H}$ is adopted unless otherwise stated, $q=M_{\rm p}/M_{\bullet}$ is the mass ratio between the sBH and the central massive object. A mass ratio of $q=10^{-5}$, which leads to a Hill radius for the sBH of $R_{\rm H}=0.015\ r_{\rm p}$, is adopted for the regime where an sBH orbiting around an SMBH in the galaxy center.

In some of our simulations, the sBH feels the disk force so that it actively migrates within the disk and its dynamics within the disk deviates from the free Keplerian motion. The equations of motion of the sBH are solved by the adaptive Runge-Kutta-Verner method, which is good enough for the evolution of $N$-body systems in $\mathcal{O}(10^{3})$ orbits and in which multi-objects do not suffer close encounters \citep[e.g.,][]{Li2009,Benitez-Llambay2016}. 
The inclination the sBH is fixed at zero, i.e., fixed at the mid-plane of the disk. 
To further diagnosis the dynamics evolution under the influence of the disk force in the very thin disk limit, we also run several controlled simulations by fixing the orbit of the sBH, i.e., the sBH does not feels the disk force so that it orbits around the SMBH with a fixed Keperian orbit described by a time-independent semi-major axis $a_{\rm p}$ and eccentricity $e_{\rm p}$. 

To explore the effect of CSD on the dynamics of the embedded object, we may choose to exclude the force within a fraction of the Hill radius $r_{\rm c}$ (e.g., $0.2\ R_{\rm H}$ and $0.5\ R_{\rm H}$) around the the embedded object, which is comparable to the size of CSD \citep{Machida2008,Fung2019,Li2023}. The accretion and feedback effect for the embedded object is neglected. Their effect on the CSD structure is modelled by the CSD cut-off parameter $r_{\rm c}$, which parameterize the region within which the sBH feedback can significantly change the CSD gravitational torque on the sBH. This is similar to the case that a different thermodynamics prescription for the CSD region compared with the global disk can modify the CSD structures and in turn modify the gravitational torque on the embedded objects \citep{Li2022b}.

An adaptive mesh refinement is adopted to resolve the region around the sBH. We use a base grid with 128 radial grids spaced  uniform between $r_{\rm min}=0.5\ r_{0}$, $r_{\rm max}=2.5\ r_{0}$, and 384 uniform grids in azimuth. The vertical direction extends to 10 disk scale height for $h_{0}=0.01$ and 4 disk scale height for $h_{0}=0.05$. The base grid in the vertical direction ensures the same resolution as the radial and azimuth directions. A convergence test with a higher resolution by a factor of 2 has been carried out to confirm that the dynamics of the embedded object is not significantly modified.
Only half disk above the mid-plane is simulated to save the computation expense considering the symmetry. Four levels of mesh refinement are implemented, with the grid resolution increased by a factor of 16 within the region $\delta r < R_{\rm H}$. This treatment significantly reduces computation cost and make 3D global simulations possible.
With such kind of mesh refinement, $1\ R_{\rm H}$ can be resolved by $\sim15$ grids in each dimension.

A fixed boundary condition is applied at both the inner and outer radial edges, a periodic boundary is implemented in the azimuthal direction, and a reflecting boundary is used in the $\theta$ direction.
To prevent wave reflections at the radial boundaries, we implement wave-killing regions at both the inner and outer edges to minimize the effects from these boundaries \citep{deValborro2006}. The wave-killing region for the outer boundary is $[2.2, 2.5]r_0$, while for the inner boundary, it is $[0.5, 0.56]r_0$. The damping rate is set to $5 \Omega$, where $\Omega$ represents the local orbital frequency, resulting in a wave-damping timescale of one-fifth of an orbital period \citep[e.g.,][]{Li2022a,Li2024}.

\subsection{Eccentricity Evolution}\label{sec:ecc_ana}

In this subsection, we discuss the mathematical formalism that will be used to analyze eccentricity evolution of the point mass motion within the AGN disk. 

\subsubsection{Orbital Perturbation Based on the Equation of Motion} \label{sec:ecc}

Consider an sBH perturbed by the  gravitational force from the disk. The equation of motion for the point mass ($r_{\rm p}$, $\phi_{\rm p}$) is
\begin{align}
&\ddot{r}_{\rm p} - r_{\rm p} \dot{\phi}^2_{\rm p} +\frac{GM_{\bullet}}{r_{\rm p}^2} = f_r\,, \nonumber \\
& \ddot{\phi}_{\rm p}r_{\rm p}^2+2 \dot{\phi}_{\rm p} \dot{r}_{\rm p} r_{\rm p} =t_\phi\,,
\end{align}
with $f_r$ being the force along the radial direction and $t_\phi$ being the torque along the angular direction. We assume that sBH is weakly perturbed from a circular orbit:

\begin{align}
r_{\rm p}=r_0 +\delta r_{\rm p},\quad \phi_{\rm p} = \omega_0 t+ \delta \phi_{\rm p},
\end{align}
with $\omega^2_0 = GM_{\bullet}/r^3_0$. The equation of motion can be simplified as
\begin{align}
&\ddot{r}_{\rm p} -3\omega_0^2 \delta r_{\rm p} -2 r_0 \omega_0 \delta \dot{\phi}_{\rm p} =f_r,\nonumber \\
& \delta \ddot{\phi}_{\rm p}r^2_0+2 \omega_0 r_0 \delta \dot{r}_{\rm p} =t_\phi\,.
\end{align}

In the frequency domain the derivative operators can be replaced as $\partial_t \rightarrow - i \omega, \partial^2_t \rightarrow -\omega^2$, so the equation of motion becomes
\begin{align}\label{eq:eom1}
&-\omega^2 \delta r_{\rm p}-3 \omega^2_0 \delta r_{\rm p}+2 i r_0 \omega_0 \omega \delta \phi_{\rm p} = f_r(\omega),\nonumber \\
& -\omega^2 \delta \phi_{\rm p} r^2_0-2 i \omega_0\omega r_0 \delta r_{\rm p} =t_\phi(\omega)\,.
\end{align}

The solution for the radial motion is given by
\begin{align}\label{eq:dr}
\delta r_{\rm p}  = -\frac{\omega r_0 f_r + 2 i \omega_0 t_\phi}{\omega r_0(\omega^2-\omega^2_0)},
\end{align}
so that we may view $f_{\rm eff}(\omega) \equiv f_r+2 i \omega_0 t_\phi/(\omega r_0)$ in the frequency domain as the effective radial force. In particular, in the time domain, we should have an energy balance law in the radial direction:
\begin{align}\label{eq:edot_r}
\dot{E}_{\rm r} = f_{\rm eff} \delta \dot{r}_{\rm p}
\end{align}
where
\begin{align}
E_{\rm r} =\frac{\omega^2_0 \delta r^2_{\rm p}}{2} +\frac{\delta \dot{r}^2_{\rm p}}{2}
\end{align}
is the radial energy. The balance law above can be used to analyze the growing or decay trend of radial oscillations with instantaneous disk forces for our `live' orbit simulations.

\subsubsection{Calculating $\dot{a}_{\rm p}$ and $\dot{e}_{\rm p}$} \label{sec:eccdot}

We can also analyze the orbital evolution due to the disk force by calculating the torque and power delivered onto the sBH. 

Consider the specific angular momentum $\ell_{\rm p}=\sqrt{GM_{\bullet}a_{\rm p}(1-e_{\rm p}^{2})}$ and specific energy $\varepsilon_{\rm p}=-GM_{\bullet}/2a_{\rm p}$ for the sBH-SMBH system with $q\ll1$, the rates of change for $\ell_{\rm p}$ and $\varepsilon_{\rm p}$ due to the disk force acting on the sBH are $\dot{\ell}_{\rm p}=\tau_{\phi}=r_{\rm p}\times f_{\phi}$ and $\dot{\varepsilon}_{\rm p}=\bm{r}_{\rm p}\cdot \bm{f}$, where $\bm{f}$ includes the contribution from the radial and azimuthal disk forces $f_{r}$, $f_{\phi}$, respectively. When the accretion onto the sBH-SMBH system is neglected, the migration rate and the eccentricity change rate are then

\begin{equation}\label{eq:adot}
\frac{\dot{a}_{\rm p}}{a_{\rm p}}=-\frac{\dot{\varepsilon}_{\rm p}}{\varepsilon_{\rm p}},
\end{equation}

\begin{equation}\label{eq:eccdot}
\frac{\dot{e}_{\rm p}}{e_{\rm p}}=-\left(2\frac{\dot{\ell}_{\rm p}}{\ell_{\rm p}}+\frac{\dot{\varepsilon}_{\rm p}}{\varepsilon_{\rm p}}\right)\frac{1-e_{\rm p}^{2}}{2e_{\rm p}^{2}}.
\end{equation}

We can apply the above method to evaluate the orbital evolution of sBH driven by disk force.
This relies on the accurate orbital eccentricity estimation for the sBH. In the case of a ``live'' orbit undergoing gradual radial inspiral, the orbit is no longer Keplerian because of the disk interaction, so that eccentricity estimates derived from closed Keplerian models may fail to account for the perturbations from the mean motion, particularly for sBH-SMBH systems with low eccentricity. This $a_{\rm p}, e_{\rm p}$ description is more reliable for our `fixed' orbit simulations where the eccentricity and semi-major axis are time-independent.


\section{Numerical Evolution}\label{sec:results}

In this section we discuss numerical evolution of the sBH in two separate regimes. In the first scenario aspect ratio is set to be $h_{0}=0.05$, so that the Hill radius $R_{\rm H}$ is smaller than the disk thickness. The sBH motion is consistent with the description of type-I migration. In the second scenario, the aspect ratio is set to be $h_{0}=0.05$ and the Hill radius is $R_{\rm H}\gtrsim H$. We find distinctive dynamical features of the sBH with an important role played by the CSD, which we detail later.

\subsection{Spherical Hill region }\label{sec:s}

\begin{figure}
\centering
\includegraphics[width=0.45\textwidth,clip=true]{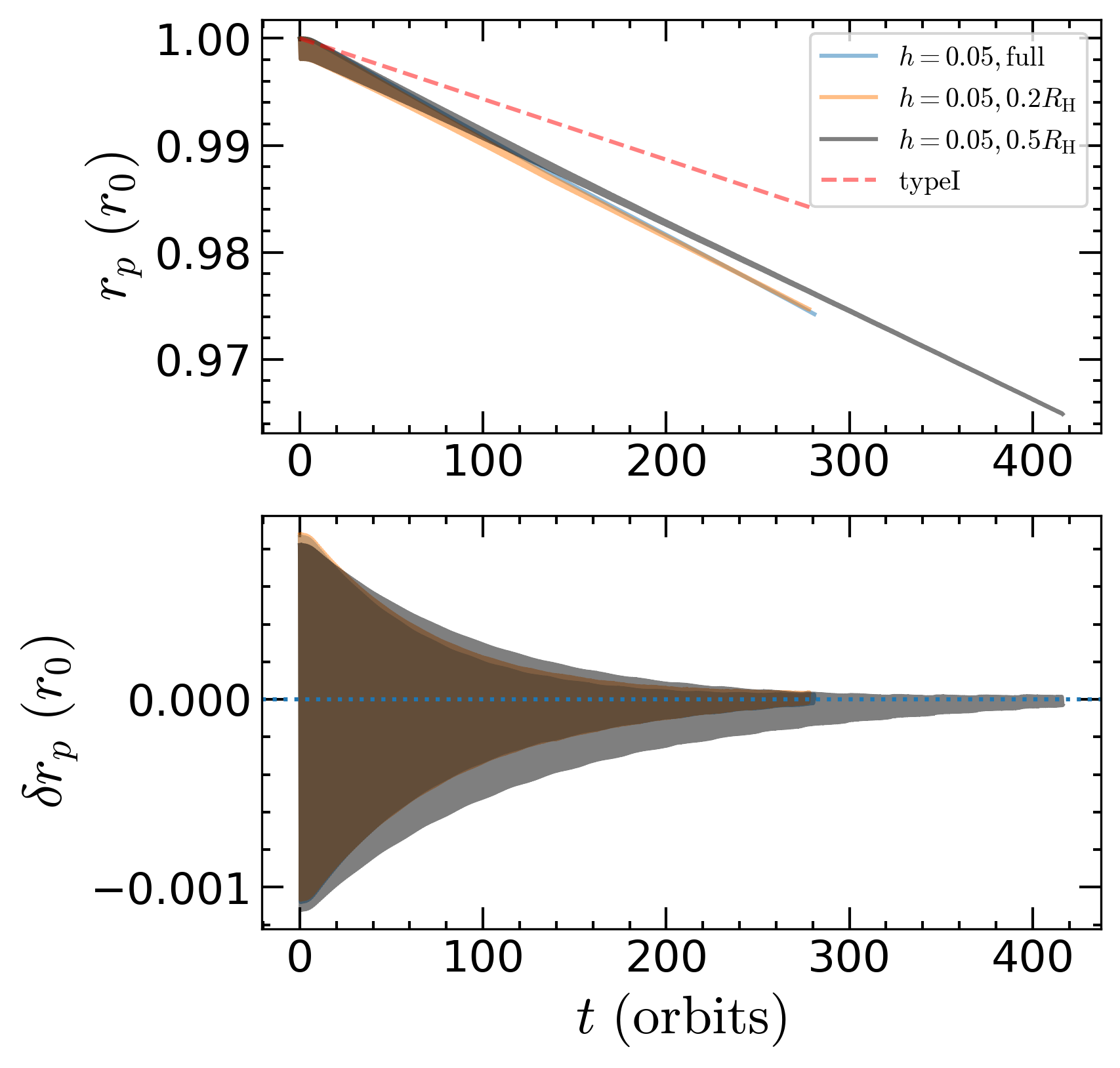}
\caption{Semi-major axis (upper panel) and eccentricity (lower panel) of the embedded object in the disk with a disk aspect ratio of $h_{0}=0.05$. Different colors correspond to models with different torque cut-off. The blue lines include all the torque in the disk, while the orange (black) exclude the torque within $0.2\ R_{\rm H}$ ($0.5\ R_{\rm H}$) from the embedded object. The dashed line in the upper panel corresponds to the classical type I migration rate. }
 \label{fig:ecc_h05}
\end{figure}

We first consider the classical case with $h_{0}=0.05$ such that $R_{\rm H}<H$. In this case, the sBH and its self-gravity dominated region are fully embedded within the disk, therefore, it is termed as the spherical Hill region. This is similar to Earth-like planet interacting gravitationally with the protoplanetary disk. The dynamical evolution of the embedded object is shown as the blue lines in Figure~\ref{fig:ecc_h05}. The upper panel shows the evolution profile of the separation between the sBH and the central SMBH, and the low panel shows the eccentricity evolution. As expected, the embedded object migrates inward quickly. This migration speed ($\sim-1.6\times10^{-5}\Omega_{0}$) roughly matches the classical type I migration rate of $\dot{a}_{\rm p}/a_{\rm p}\sim-1.0\times10^{-5}\Omega_{0}$ with a weak dependence on the surface density and temperature slope at the embedded object location \citep{Tanaka2002}. 

Estimating the eccentricity for this type of quasi-stationary inspriral can be subtle. On the one hand, a direct calculation, assuming Keplerian motion and using the instantaneous orbital radius and velocity, leads to $e_{\rm p} \sim0.001$, which is consistent with a negligible eccentricity due to the strong eccentricity damping in the type-I linear regime \citep{Papaloizou2000,Goldreich2003,Li2019,Ida2020}. On the other hand, as we subtract out the mean orbital motion, the relative radial motion quickly damps out to be under $\mathcal{O}(10^{-4})$, as shown in the bottom panel of Figure~\ref{fig:ecc_h05}. The difference comes from the fact that the orbit is {\it not} strictly Keplerian, because  orbit decays in time and  the sBH-disk interaction also matters. For the gravitational wave emission, the higher-order modes due to radial motion should directly depends on $\delta r$ in Figure~\ref{fig:ecc_h05}, instead of the ``canonical" eccentricity obtained assuming instantaneous Keplerian motion. Since we are considering the dynamics of EMRIs with orbital radii $\sim \mathcal{O}(10^2) M_{\bullet}$ (c.f. Figure~\ref{fig:no_gap}), by the time they enter the LISA band at $\mathcal{O}(10) M_{\bullet}$, the relative orbital oscillation $\delta r/r$ is likely below $\mathcal{O}(10^{-5})$, which is not observable by LISA.

In Sec.~\ref{sec:ph} we will discuss the role played by materials within CSD to the dynamics of the sBH. Here it is instructive to perform a similar analysis for later comparisons. 
To this end, we remove some fraction of the torque within the Hill sphere (as the size of CSD is a fraction of the Hill radius in Sec.~\ref{sec:ph}) with a torque cut-off radius $r_{\rm c}=0.2\ R_{\rm H}$ and $r_{\rm c}=0.5\ R_{\rm H}$. Specifically, the torque within a distance of $r_{\rm c}$ from the sBH are excised for the dynamical evolution of the embedded object, while the gravitational force from the other part of the disk remains the same. The two cases with $r_{\rm c}=0.2\ R_{\rm H}$ and $r_{\rm c}=0.5\ R_{\rm H}$ are shown in Figure~\ref{fig:ecc_h05}. We can see that their evolution, both the semi-major axis migration and the evolution for radial oscillation, is quite similar to the full evolution case. The contribution from materials within the ``CSD regime" under this circumstance is almost negligible. In fact, for such a fully embedded object with $R_{\rm H}<H$, we find that the flow around the embedded object intrinsically behaves like a 3D envelope, and there is essentially no  rotationally-supported disk around the sBH \citep[e.g.,][]{Szulagyi2016,Li2023}.

\subsection{Planar Hill region}\label{sec:ph}

As shown in Figure~\ref{fig:no_gap}, for an EMRI object at $\mathcal{O}(10^2-10^3)$ gravitational radii, it is possible to have a parameter regime that $R_{\rm H}\gtrsim H$, while there is no gap opening. This parameter regime is interesting as the Hill region no longer has spherical-type geometry, which we refer as ``planar Hill region". We realize this case by setting a typical disk parameter with $h_{0}=0.01$ and a high viscosity parameter of $\alpha=0.1$ \citep{Duffell2013,Kanagawa2015,Li2023}. A higher viscosity parameter with $\alpha=0.3$ will make the gap even more shallow.  In this scenario, we will show distinctive dynamical features for sBHs in this scenario.

\subsubsection{Flow Structure}

\begin{figure*}
\centering
\includegraphics[width=0.95\textwidth,clip=true]{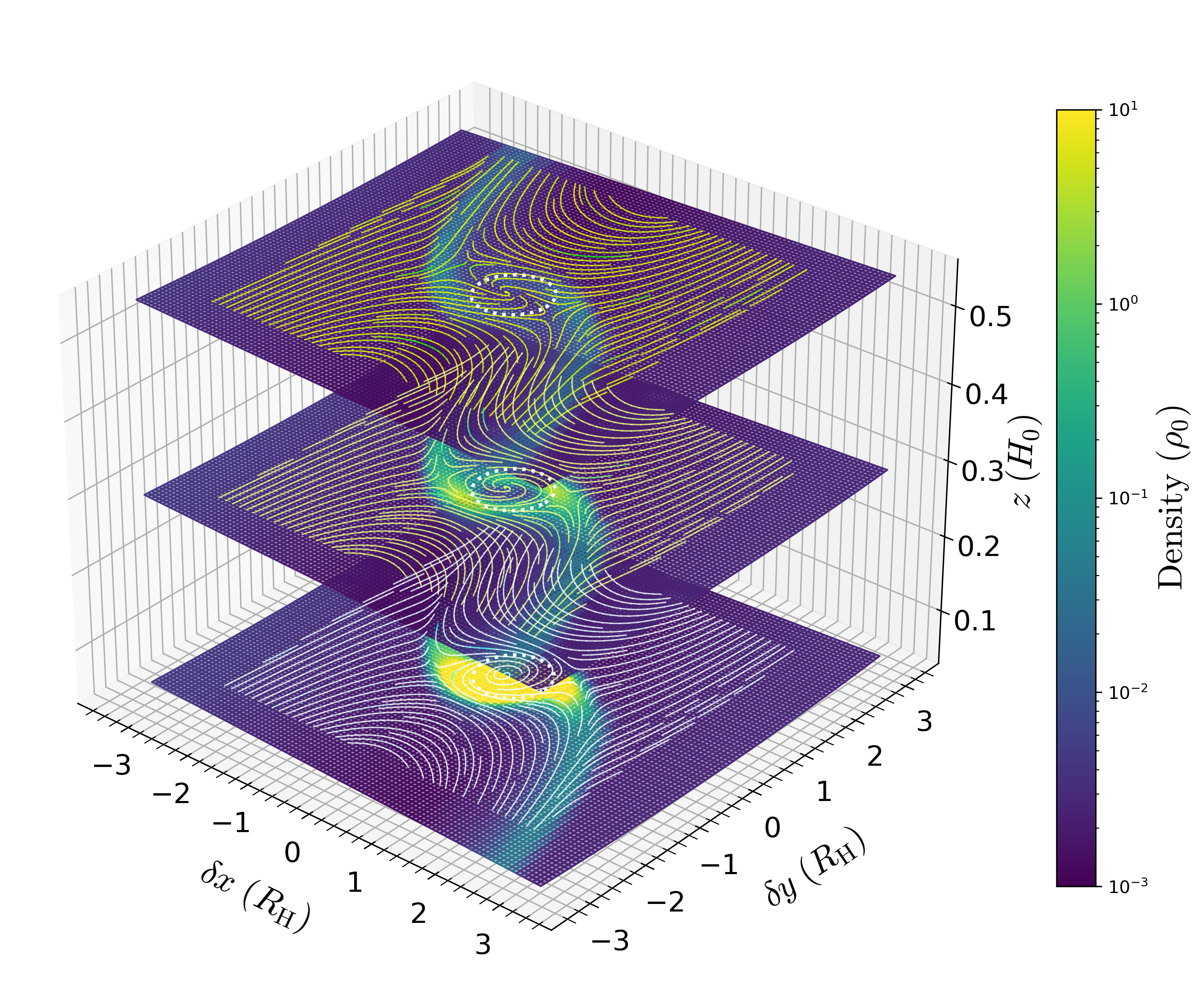}
\caption{
Density (contour) and flow pattern relative to the sBH (solid lines) around the sBH for the model with $h_0=0.01$ and without CSD cut-off at 500 orbits. Three contours and streamlines correspond to the slices at $z=0$ (midplane), $z=0.25\ H_0$, and $z=0.5\ H_0$, respectively. The three white circles indicates the region of $\delta r=0.5\ R_{\rm H}$ from the embedded object. The greenish streamlines in the vicinity of the sBH at different slices indicates the region with the vertical inflow onto the sBH.}
 \label{fig:flow}
\end{figure*}

The flow structures around the sBH are shown in Figure~\ref{fig:flow}. We show the gas density at different heights $z$ overlaid with the arrows to represent the velocity relative to the sBH. We can find that there is a partially rotationally supported disk around the sBH within the Hill sphere, which is the region of CSD. The two dotted circulars show $0.5\ R_{\rm H}$ from the sBH. The CSD is dominated by rotation up to $0.2-0.3\ R_{\rm H}$, but can still be partially rotationally supported up to a fraction of the Hill radius, as shown from the streamline \citep[see also e.g.,][]{Li2023}. The outer disk material feeds the CSD from the upper horseshoe and is deflected into the inner disk by the sBH. Only a small fraction of the inflow with an appropriate pitch angle will enter the CSD region, which eventually joins the outflow from the two spiral arms.

By further examining the velocity structures in the vertical direction,  we find that the most of the inflow flux comes from the high latitude as indicated by the greenish streamlines at different slices in the vicinity of the sBH, while the mid-plane is the outflow region.  
In the realistic scenario,  some fraction of the material around the CSD are likely bound to the embedded object, or even disrupted due to the feedback process associated with the radiation/outflow from the sBH. Therefore, their contribution to the dynamics may be significantly modified.


\subsubsection{Dynamical Evolution}

\begin{figure}
\centering
\includegraphics[width=0.45\textwidth,clip=true]{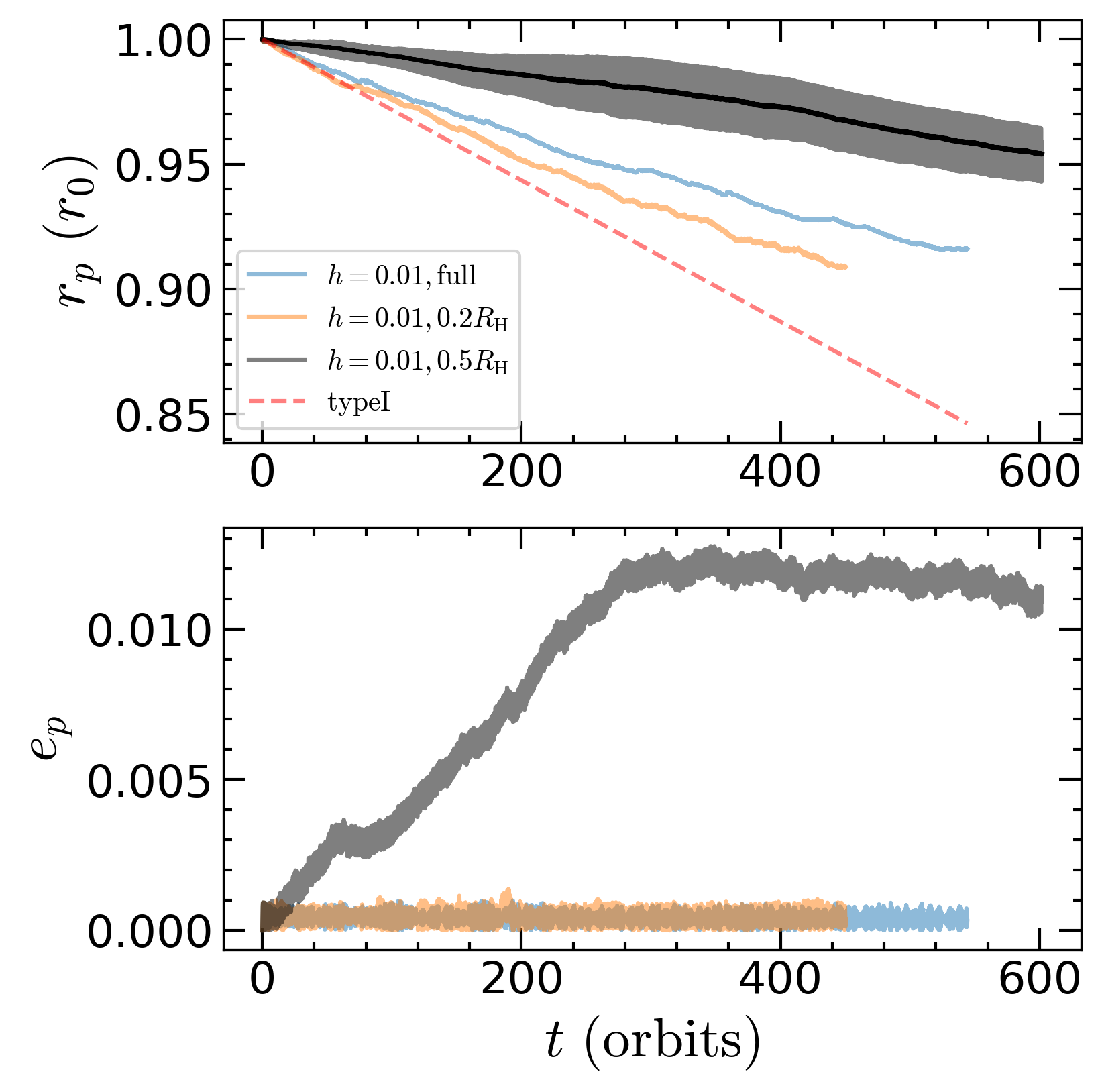}
\caption{Similar to Figure~\ref{fig:ecc_h05} but with a disk aspect ratio of $h_{0}=0.01$. The shaded region in the upper panel shows the instantaneous separation from the central object. Note that the lower panel shows the orbital eccentricity instead of the oscillation amplitude as in Figure~\ref{fig:ecc_h05}. These two are the same for the fixed eccentric Keplerian orbit.} 
 \label{fig:ecc_h01}
\end{figure}

The evolution of sBH-SMBH separation and eccentricity is shown in Figure~\ref{fig:ecc_h01}. We first present  the case where the full disk torque is included for the evolution of the sBH, which is shown as the blue lines in Figure~\ref{fig:ecc_h01}. The inward migration rate is a factor of $2-3$ faster than that of the $h_{0}$ case, which is, however, lower by a factor of $\sim2$ slower than the theoretical expectation of the type I migration rate, based on the scaling of $h_{0}^{-1}$ with a constant mid-plane density at $r_{0}$ \citep{Tanaka2002}.

The eccentricity evolution is shown as the blue line in the lower panel of Figure~\ref{fig:ecc_h01}. The equilibrium eccentricity is around $\sim 10^{-3}$, with the level of radial oscillation being $\delta r/r \sim5\times10^{-4}$. Once the EMRI enters the LISA band, the residual eccentricity may still be detectable ($\mathcal{O}(10^{-5})$).

Motivated by the uncertainty of CSD contribution to the sBH dynamics as discussed above, we apply two cut-off schemes in the CSD region. Two cut-off radii are chosen, $r_{\rm c}=0.2\ R_{\rm H}$ and $r_{\rm c}=0.5\ R_{\rm H}$, based on the CSD structures shown in Figure~\ref{fig:flow}. The results are shown in Figure~\ref{fig:ecc_h01} with different colors. With $r_{\rm c}=0.2\ R_{\rm H}$, the migration is slightly faster than that without any CSD cut-off, but still slower than that of the theoretical expectation of classical type I migration. The eccentricity evolution is similar to the case of no CSD cut-off with a similar level of $\delta r/r$.

\begin{figure}
\centering
\includegraphics[width=0.45\textwidth,clip=true]{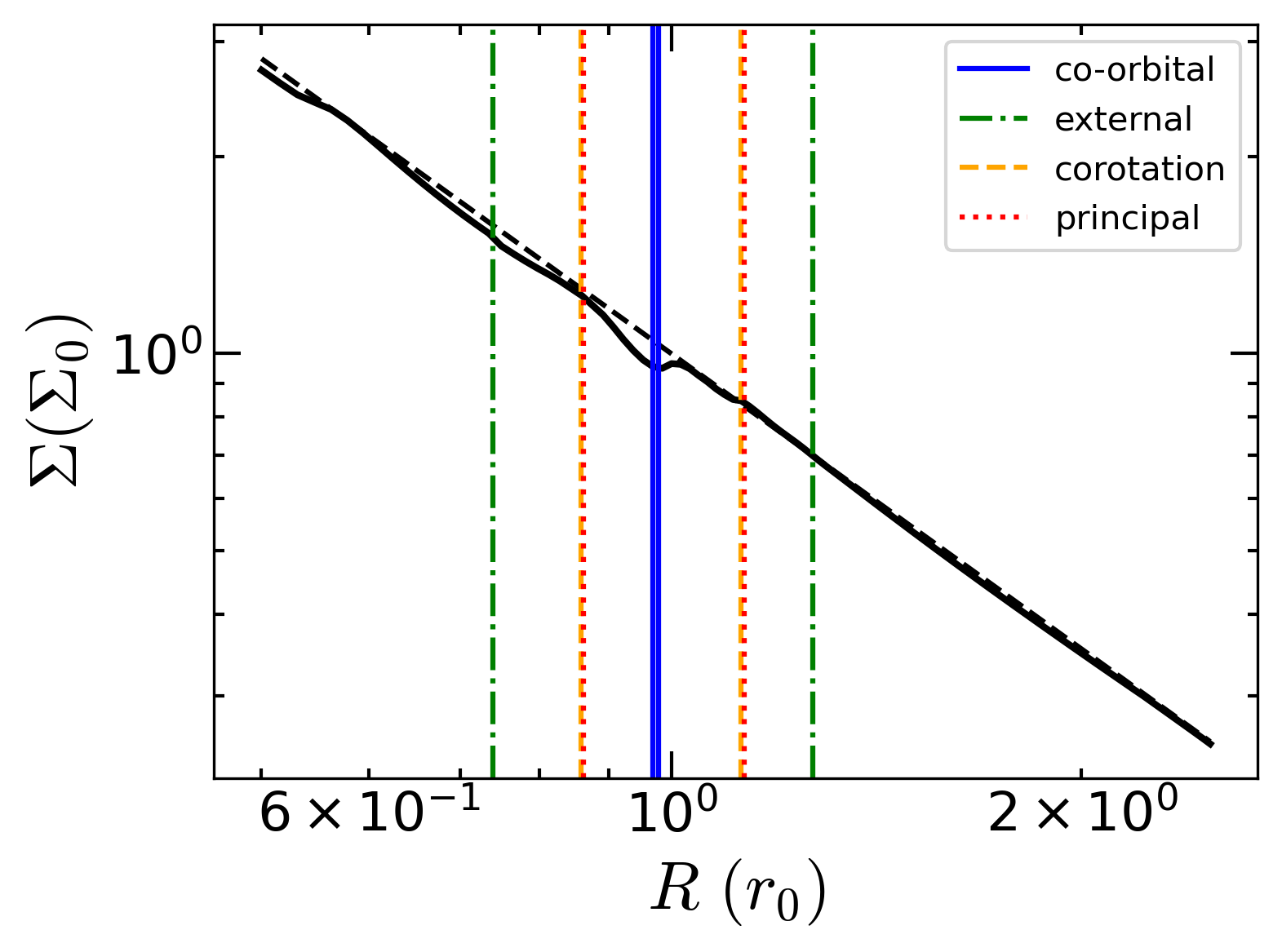}
\caption{Azimuthally-averaged surface density of the global disk for the disk model with $h_{0}=0.01$ and $r_{\rm c}=0.5\ R_{\rm H}$ at $500$ orbits. The black dashed line shows the initial surface density profile. The embedded object is located at the vertical solid line, which indicates the co-orbital region. We have also plot first-order external resonances (dash-dotted lines), first-order corotation resonances (dashed lines), and the principal inner and outer Lindblad resonances (dotted lines) for $m=5$, with their locations defined in Appendix~\ref{sec:edot_ana}. We plot some overlapping resonances (corotation and principal resonances) with small arbitrary offsets for visual clarity only.}
 \label{fig:sigma1d}
\end{figure}

Furthermore, we apply a larger CSD cut-off with $r_{\rm c}=0.5\ R_{\rm H}$. The dynamical evolution is shown as black line in Figure~\ref{fig:ecc_h01}. The shaded region in the upper panel corresponds to the instantaneous separation between sBH-SMBH, while the black line is the semi-major axis. The migration rate is further suppressed compared to the no cut-off case.
This means that the CSD material is promoting the inward migration in this setup.

\begin{figure}
\centering
\includegraphics[width=0.45\textwidth,clip=true]{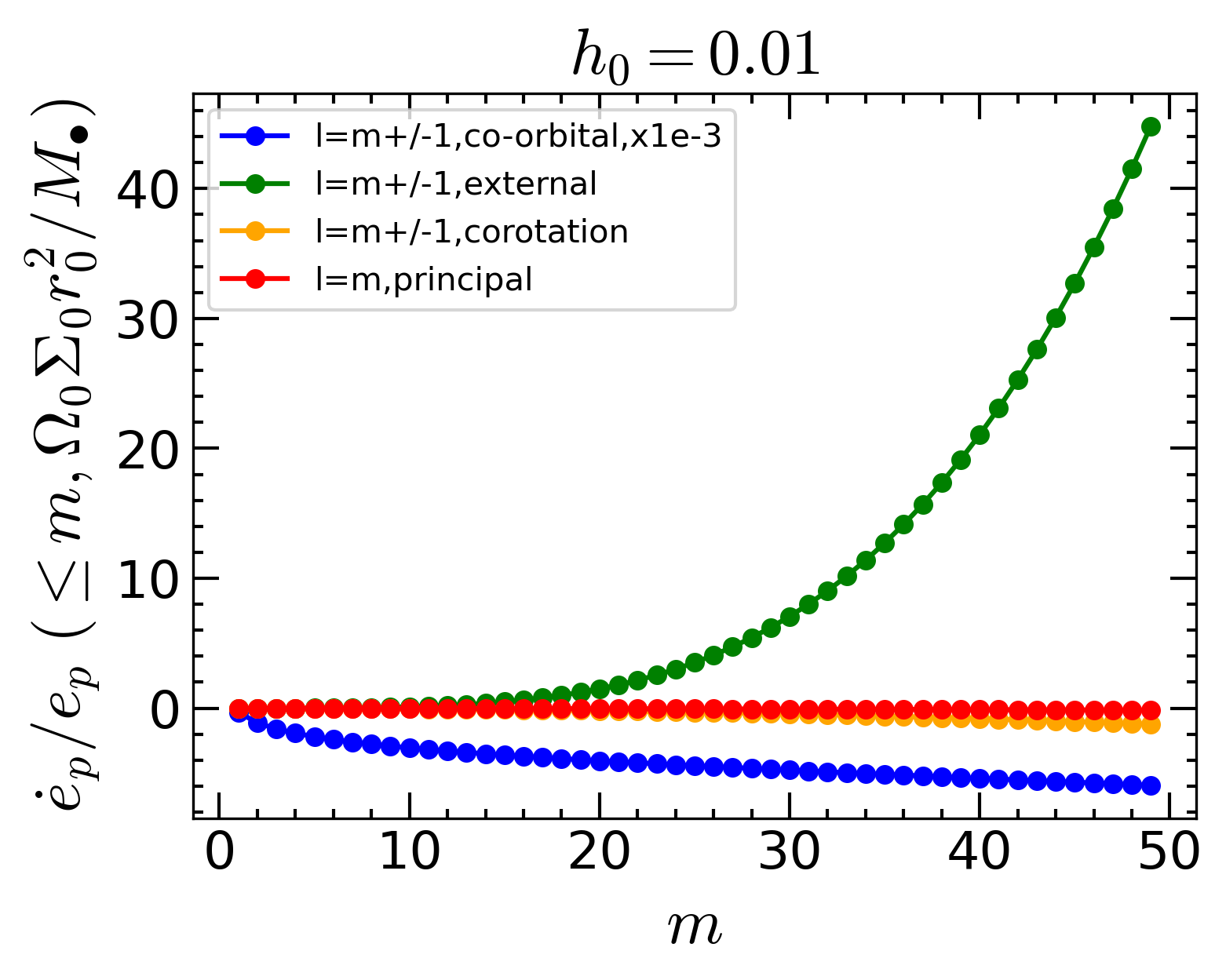}
\includegraphics[width=0.45\textwidth,clip=true]{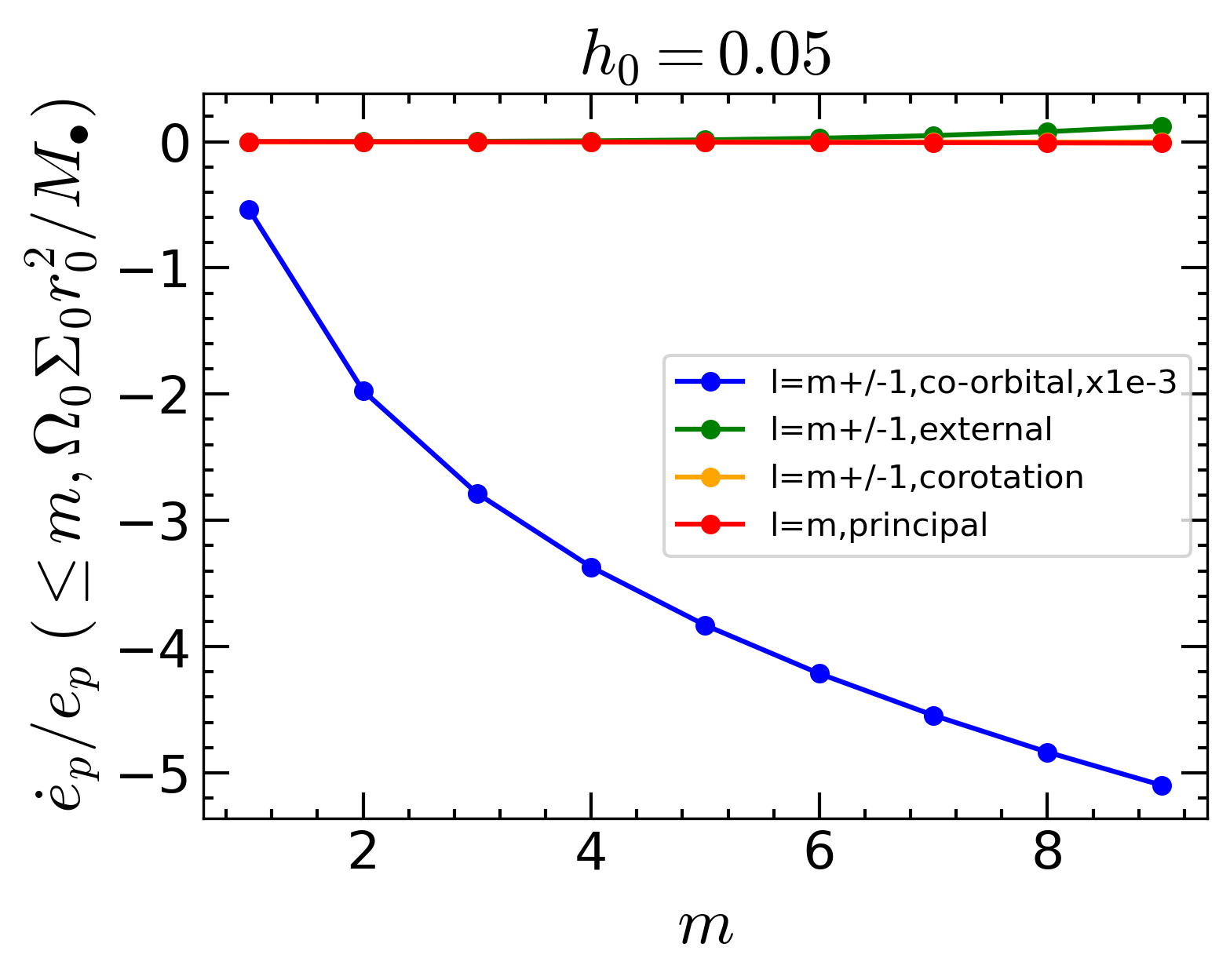}
\caption{Upper panel: the $\dot{e}_{\rm p}/e_{\rm p}$ as a function of different azimuthal wavenumber  $m$ for our disk model with $h_{0}=0.01$ and $e_{\rm p}=0.01$. Different lines correspond to first-order co-orbital (blue line), first-order external (green line), first-order corotation (orange line),  principal inner and outer (ILR+ORL; red line) resonances. The first-order external co-orbital resonance contribution are multiplied by a factor of $10^{-3}$ to compare with the other components.
Lower panel: similar to the upper panel but for $h_{0}=0.05$ and $e_{\rm p}=0.001$. Note that the orange and red lines are overlapped with each other.
}
 \label{fig:edot_ana}
\end{figure}

Accompanied with the suppression of inward migration, there exists significant radial oscillation of the sBH orbit, which indicate the strong excitation of the sBH's orbital eccentricity, as confirmed in the lower panel. We can see that the eccentricity of sBH can be as large as $0.01$, much larger than the other two cases. To the best of our knowledge, this effect has not been reported in the past literature. In the traditional type-I regime, as discussed in Sec.~\ref{sec:s}, removing material within similar distance to the sBH has negligible impact on the sBH dynamics. This is consistent with the fact that no CSD structure is observed in the spherical Hill region case, but a clear CSD structure exists in the planar Hill region case. 

On the other hand, the excitation of eccentricity at least reveals two interesting observations. 
First, the dynamic interaction between the sBH and disk materials outside the CSD regime seems to give rise to eccentricity excitation. Unlike the traditional type-I migration, circular orbit is not a stable fixed point of the orbital evolution. The eccentricity excitation mechanism should be different from the scenario of the direct resonant wave excitation at 3:1 Lindblad resonance (i.e., at $2.08\ r_{0}$) where the gap width needs to wide enough \citep{Papaloizou2001}.
The exact eccentricity mechanism is still largely uncertain but it could be driven by the first-order external resonance \citep{Goldreich2003,Duffell2015}, although the sBH does not open a deep gap in the disk for our high viscosity disk as shown in Figure~\ref{fig:sigma1d}. To better understand this mechanism, we plot all the contribution of $\dot{e}_{\rm p}/e_{\rm p}$ at different azimuthal wavenumbers from the first-order Lindblad and corotation resonances in Figure~\ref{fig:edot_ana} based on the linear theory \citep{Goldreich2003,Duffell2015}. We group them into four catalogs, i.e., the first-order co-orbital resonances, first-order external resonances, first-order corotation resonances, and principal inner/outer Lindblad resonances (ILR+OLR) as in \citep{Goldreich2003,Duffell2015}. The resonant locations, as defined in Appendix~\ref{sec:edot_ana}, for a typical wavenumber $m=5$ are labeled in Figure~\ref{fig:sigma1d}. The main equations for the $\dot{e}_{\rm p}$ are summarized in Appendix~\ref{sec:edot_ana}. Note that the linear theory does not include the torque in the vicinity of the embedded object.
We can see that for both $h_{0}=0.01$ and $h_{0}=0.05$ disks, the first-order co-orbital resonances damp the eccentricity while the first-order external resonances excite the eccentricity although in a lower magnitude. Other components are much smaller, although with a dependence on the gravitational softening and other disk parameters, e.g., the local disk density profile. The first-order external resonances are much stronger for the thinner disk with $h_{0}=0.01$ compared with that of $h_{0}=0.05$. With the CSD-cutoff which removes some fractions of the torque associated with the first-order co-orbital resonances, the first-order external resonances in the $h_{0}=0.01$ case could dominate over the eccentricity evolution. It is still changeling for the first-order external resonances to dominate over the eccentricity evolution in the $h_{0}=0.05$ case even when CSD-cutoff takes away some faction of material in the co-orbital region.

Second, the fact that the $\mathcal{O}(10^{-2})$ is only observed if most of the CSD materials are removed, suggest that the CSD actually plays a role damping out the eccentricity driven by density wave emissions. Notice that the effect of CSD is fully {\it non-perturbative}, so it is difficult to describe in a linear theory. Nonetheless, we may further analyze its dynamical effects using the tool developed in Sec.~\ref{sec:ecc}.

\begin{figure*}
\centering
\includegraphics[width=0.45\textwidth,clip=true]{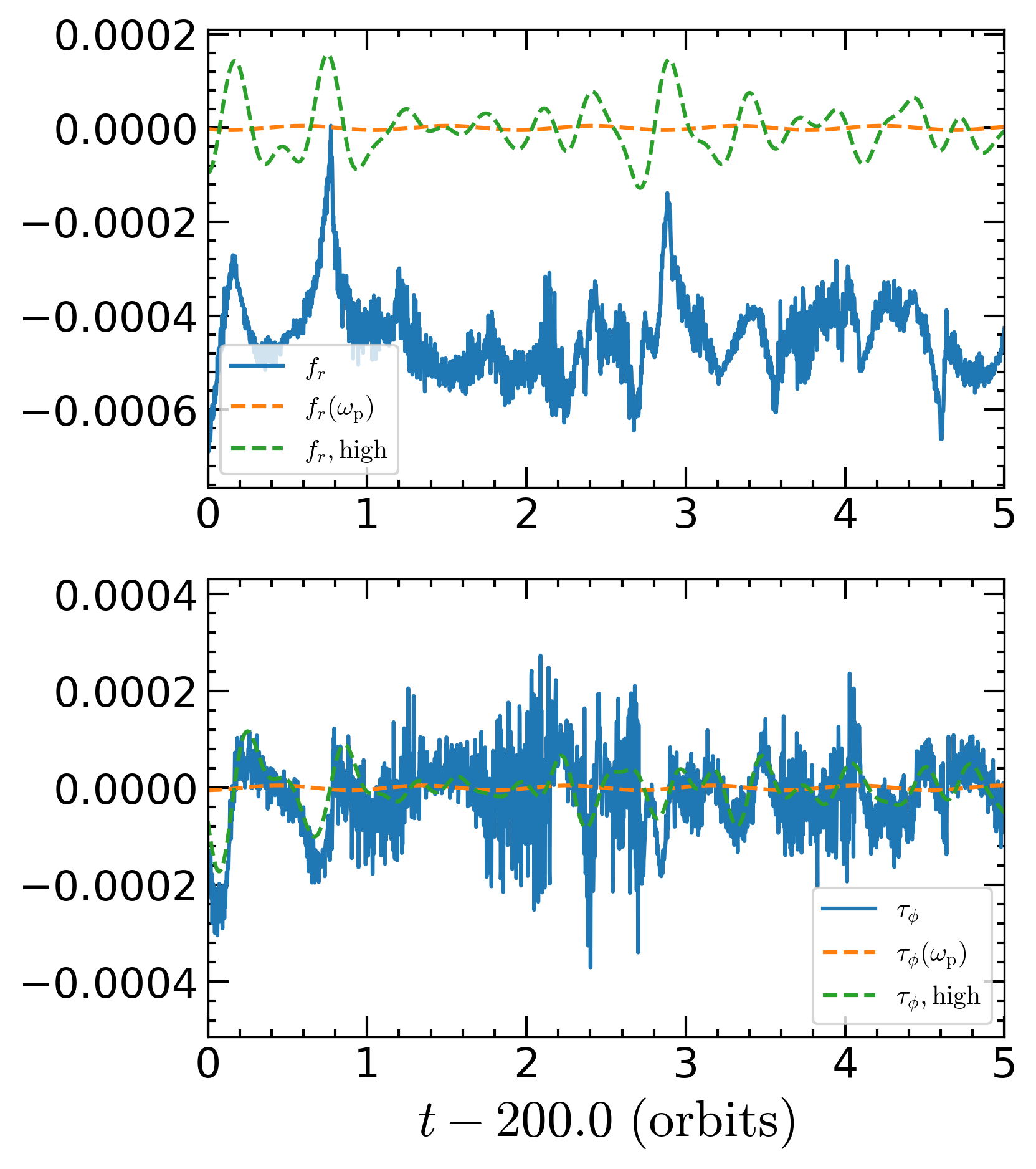}
\includegraphics[width=0.45\textwidth,clip=true]{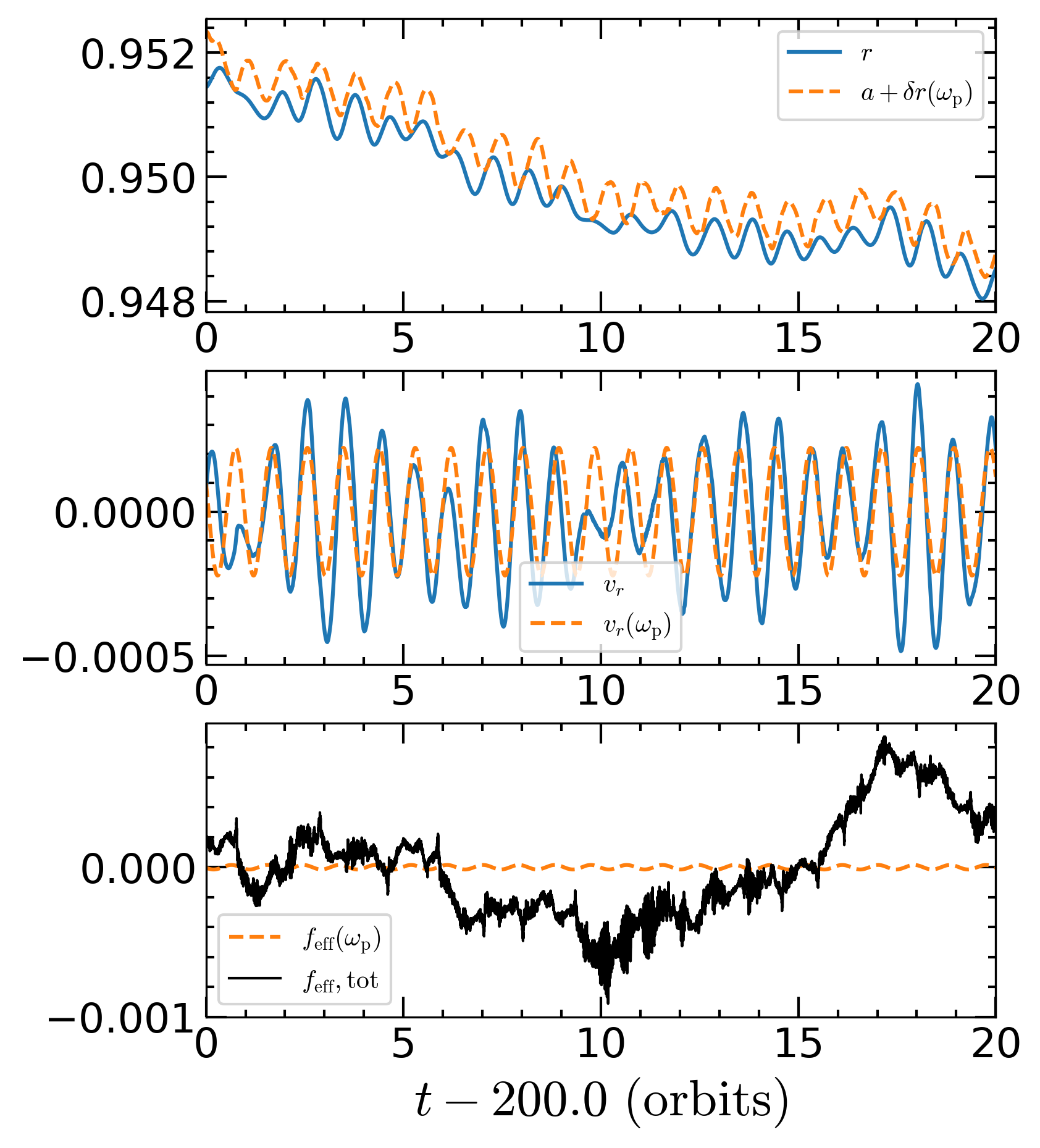}
\caption{Left panels: Radial force (upper left) and torque (lower left) profiles for the model of $h/r=0.01$ with a torque cut-off of $r_{\rm c}=0.2R_{\rm H}$ in code unit. Right panels: Radial distance from the central object (upper right panel), radial velocity (middle right panel), and effective force $f_{\rm eff}$ on the embedded object (lower right panel) for the model of $h/r=0.01$ with a torque cut-off of $r_{\rm c}=0.2R_{\rm H}$. The profile of $f_{\rm eff}$ is also shown in the middle right panel to check the possible phase offset relative to $v_{r}$. The predicated separation between the sBH and the SMBH is shown as the dashed line in the upper panel, where $a$ is the semi-major axis of the sBH. 
}
 \label{fig:torq_rc02}
\end{figure*}

\subsubsection{Diagnostic of Eccentricity Evolution}

\begin{figure}
\centering
\includegraphics[width=0.45\textwidth,clip=true]{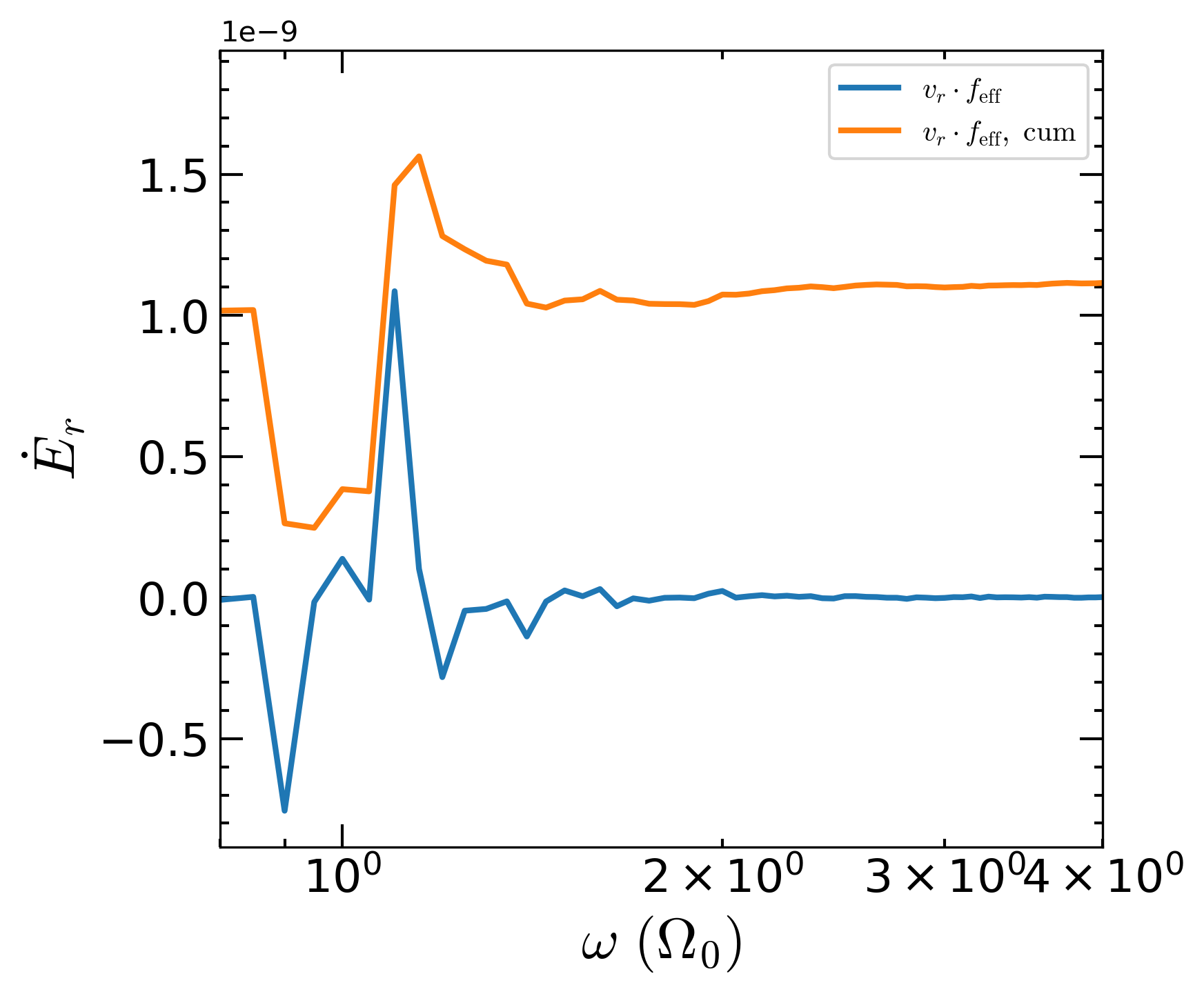}
\caption{$\dot{E}_{r}$ spectrum (blue line) in code unit due to the effective force acting on the embedded object for the model of $h/r=0.01$ with a torque cut-off of $r_{\rm c}=0.2R_{\rm H}$. The orange line shows the cumulative contribution for the frequency domain $<\omega$.}
 \label{fig:power_rc02}
\end{figure}

We track the radial force and torque evolution at different times for different CSD cut-off. For the case with $r_{\rm c}=0.2\ R_{\rm H}$,
we show the time profiles of $f_{r}$ and $\tau_{\phi}$ in the left panel of Figure~\ref{fig:torq_rc02}. Fourier transform is carried out to decompose different frequency components. 
As a sanitary check, we confirm that the summation of all frequency component can reproduce the time profiles of $f_{r}$ and $\tau_{\phi}$. 
$f_{r}(\omega_{\rm p})$ and $\tau_{\phi}(\omega_{\rm p})$ correspond to the component at the frequency $\omega_{\rm p}$, where $\omega_{\rm p}$ is the peak frequency of the power spectrum of the sBH's radial velocity $v_{r}$. Because of the disk-sBH interaction, $\omega_{\rm p}$ slightly deviates from $\omega_0$, which is the Keplerian frequency assuming only the gravitational interaction with the central object.  
The summation of the component from $1.1\omega_{\rm p}$  to $4.0\omega_{\rm p}$ shown as the lines (with the label  ``\texttt{high}") is to visualize the high frequency contribution  as compared to the total force. 
With $r_{\rm c}=0.2\ R_{\rm H}$, we can see that both the high frequency component of $f_{r}$ (i.e., $f_{r,\rm high}$) and $\tau_{\phi}$ (i.e., $\tau_{\phi, \rm high}$) is much more prominent than that of $\omega_{\rm p}$.
Intuitively the circulation frequency of disk material around the sBH at a distance of $1.4\ R_{\rm H}$ from the sBH is about $\omega_{\rm p}$, and materials within that distance should have higher circulation frequency. Therefore, the component of $f_{r},\tau_{\phi}$ with  $\omega \ge \omega_{\rm p}$ should be dominated by  disk materials closer to the sBH, which suggests the important contribution of CSD materials around the sBH.

\begin{figure*}
\centering
\includegraphics[width=0.45\textwidth,clip=true]{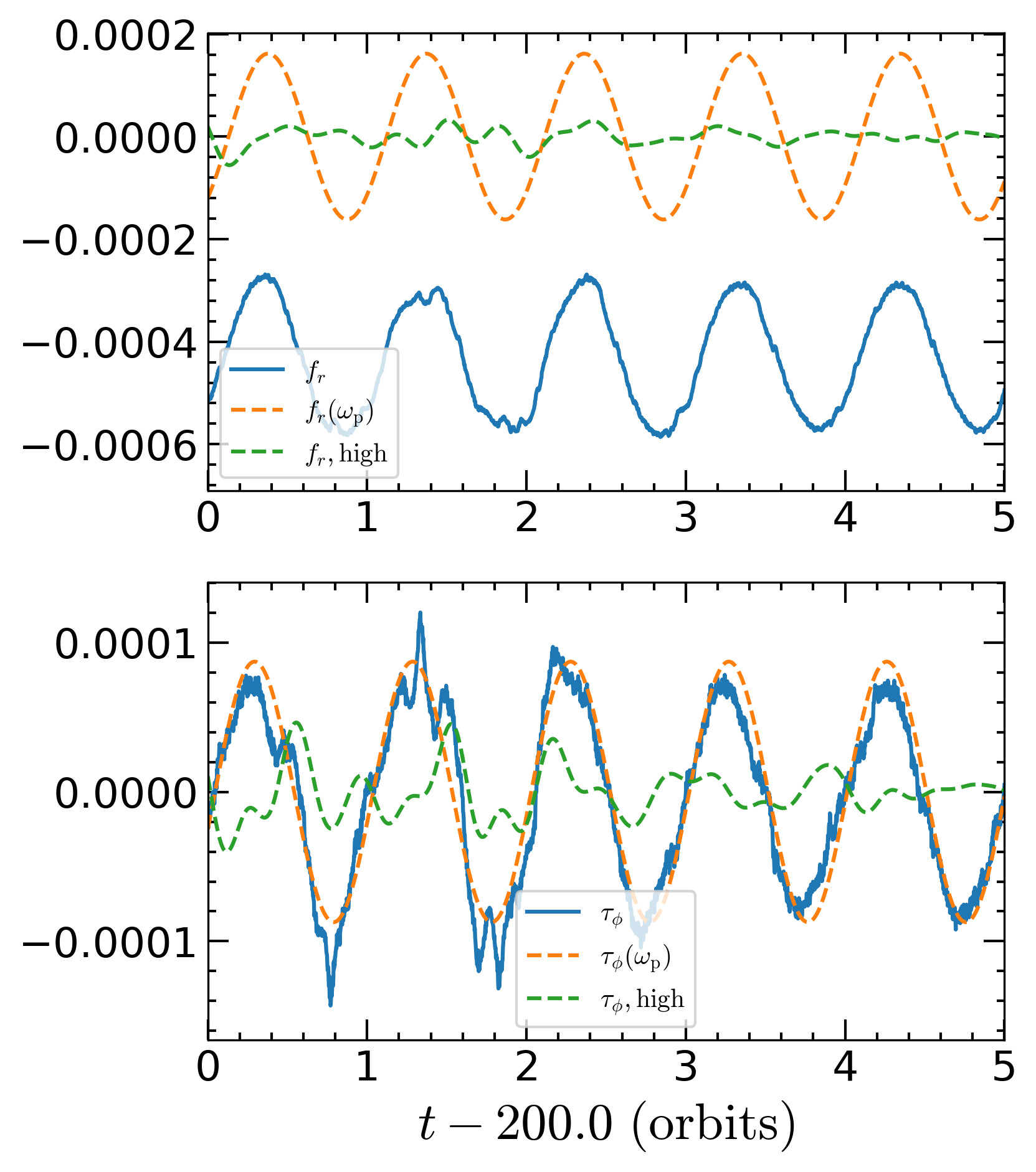}
\includegraphics[width=0.45\textwidth,clip=true]{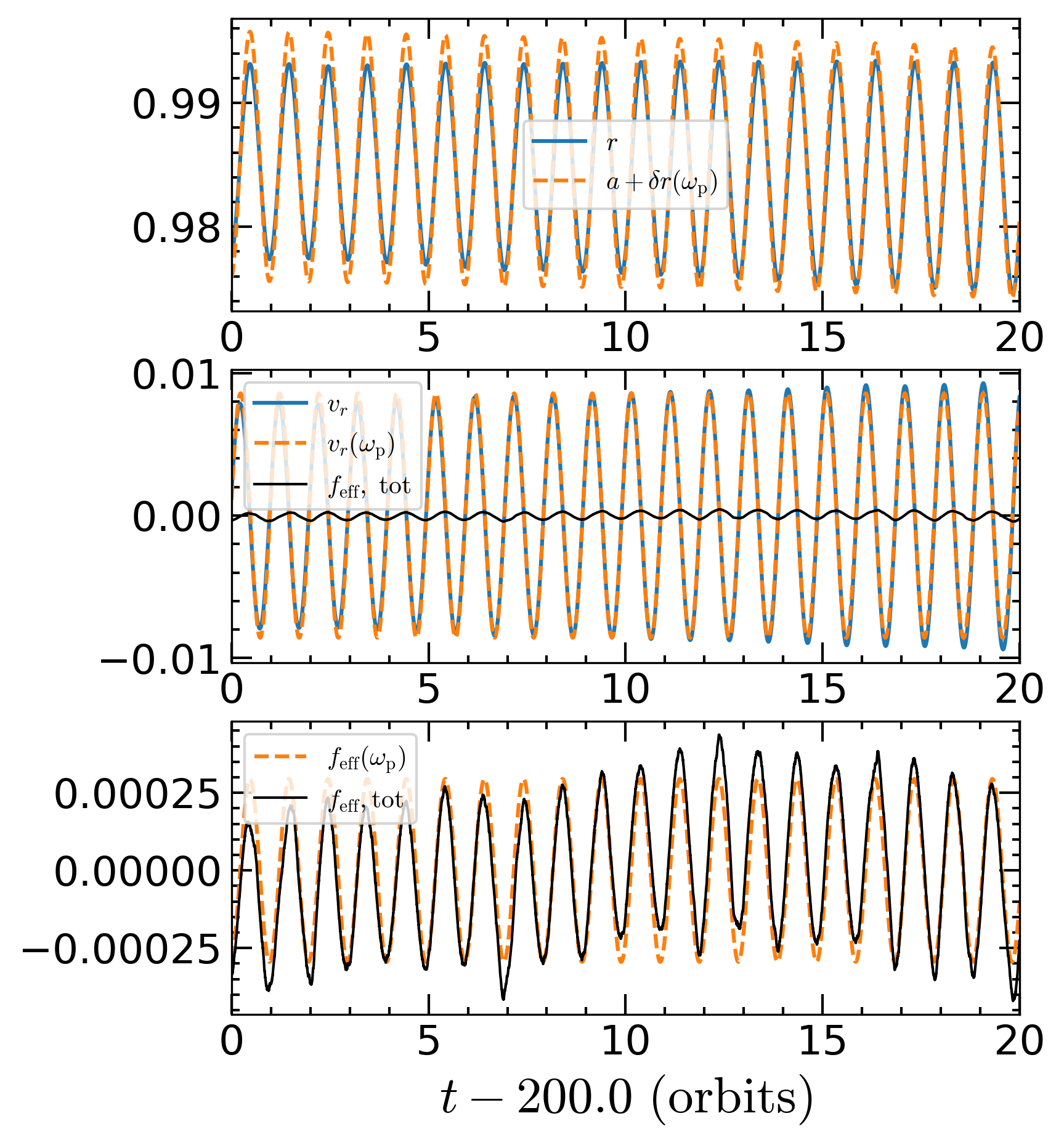}
\caption{Similar to Figure~\ref{fig:torq_rc02} but for the model with a torque cut-off of $r_{\rm c}=0.5R_{\rm H}$.}
 \label{fig:torq_rc05}
\end{figure*}

With the power spectrum of $f_{r}$ and $\tau_{\phi}$ in hand (not shown in the paper), 
we show the effective radial force $f_{\rm eff}$ integrated over all frequency and at the peak frequency $\omega_{\rm p}$ in the right panels of Figure~\ref{fig:torq_rc02}. 
The predicated $\delta r$ at a given frequency $\omega_{\rm p}$ is overlaid in the upper right panel of Figure~\ref{fig:torq_rc02}.
We can see that the discrepancy from the actual evolution is large, which suggests that there are important contributions from other frequency components as for the profiles of $f_{r}$ and $\tau_{\phi}$.

The eccentricity evolution is governed by the power $\dot{E}_{r}$ defined by Equation~(\ref{eq:edot_r}) in the time domain. The power spectrum for different frequency component of $\dot{E}_{r}$ is shown in Figure~\ref{fig:power_rc02} with the frequency in unit of $\Omega_{0}$, where $\Omega_{0}$ is the Keplerican frequency at the \textit{initial} position $r=1.0\ r_{0}$
Both positive and negative contribution are seen for different frequency bands, which indicates a in-phase (phase offset $<\pi/2$)  and out-phase (phase offset $>\pi/2$) correlation between $f_{\rm eff}$ and $v_{r}$.  The cumulative contribution by the integration over $\omega$ ($<\omega$), which is represented as the orange line, has a slightly positive value, which should reflect the instantaneous ``heating rate" of the radial motion around $t=0$.

\begin{figure}
\centering
\includegraphics[width=0.45\textwidth,clip=true]{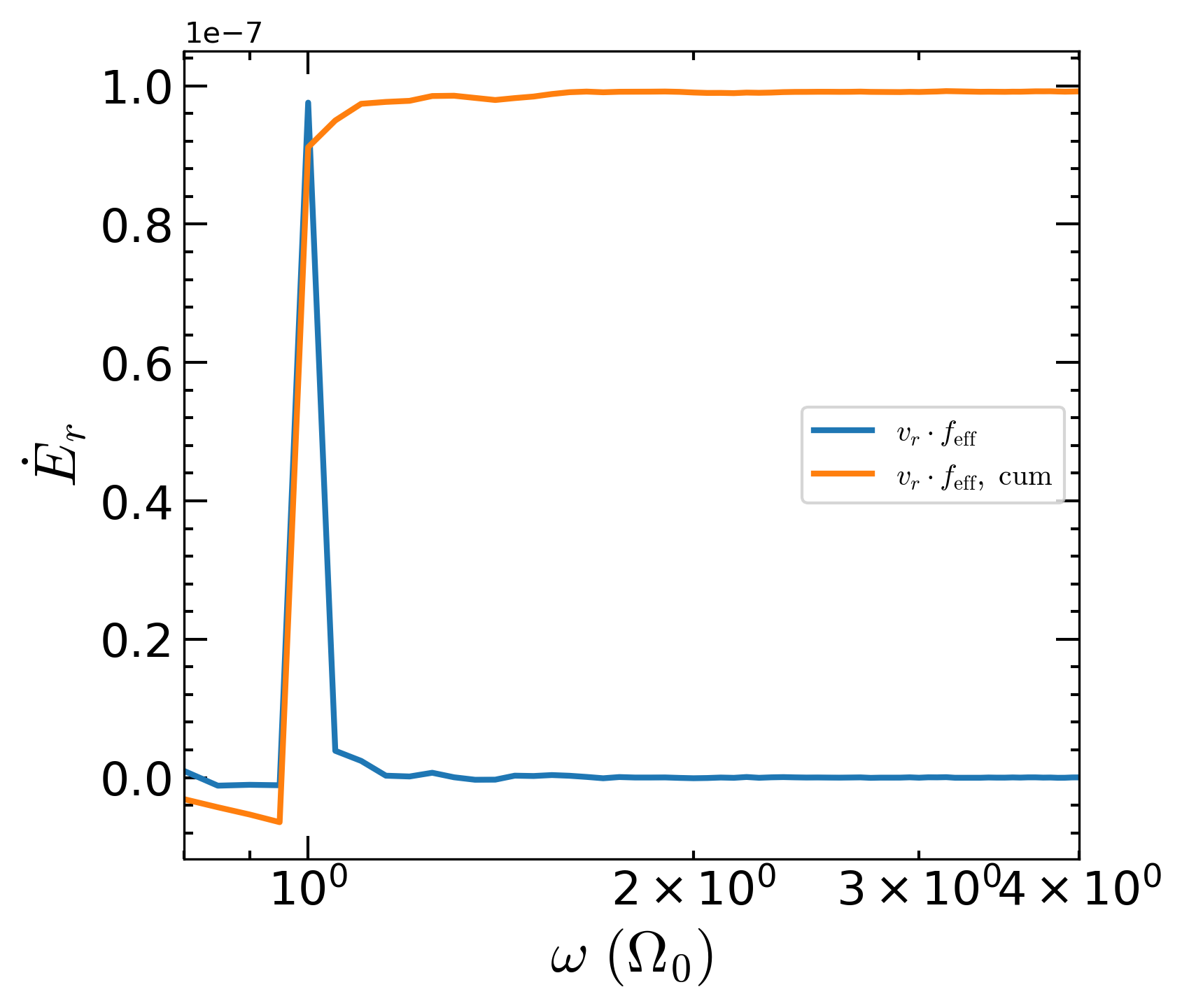}
\caption{Similar to Figure~\ref{fig:power_rc02} but with a torque cut-off of $r_{\rm c}=0.5\ R_{\rm H}$. }
 \label{fig:power_rc05}
\end{figure}

Similarly, we plot the time evolution of the radial force $f_{r}$ and torque $\tau_{\phi}$ on the sBH for the case with $r_{\rm c}=0.5\ R_{\rm H}$ around the 200th orbit, which is shown in Figure~\ref{fig:torq_rc05}. Compared with the smaller CSD cut-off case, both $f_{r}$ and $\tau_{\phi}$ are now mainly contributed by the $\omega_{\rm p}$ component. 
The total effective force $f_{\rm eff}$ is shown in the lower right panel of Figure~\ref{fig:torq_rc05}, which is, as expected, dominated by the $\omega_{\rm p}$ component. There is almost a fixed phase offset $\delta\phi<\pi/2$ between $f_{\rm eff}$ and $v_{r}$ in the time domain, as shown in the right middle panel.

\begin{figure*}
\centering
\includegraphics[width=0.45\textwidth,clip=true]{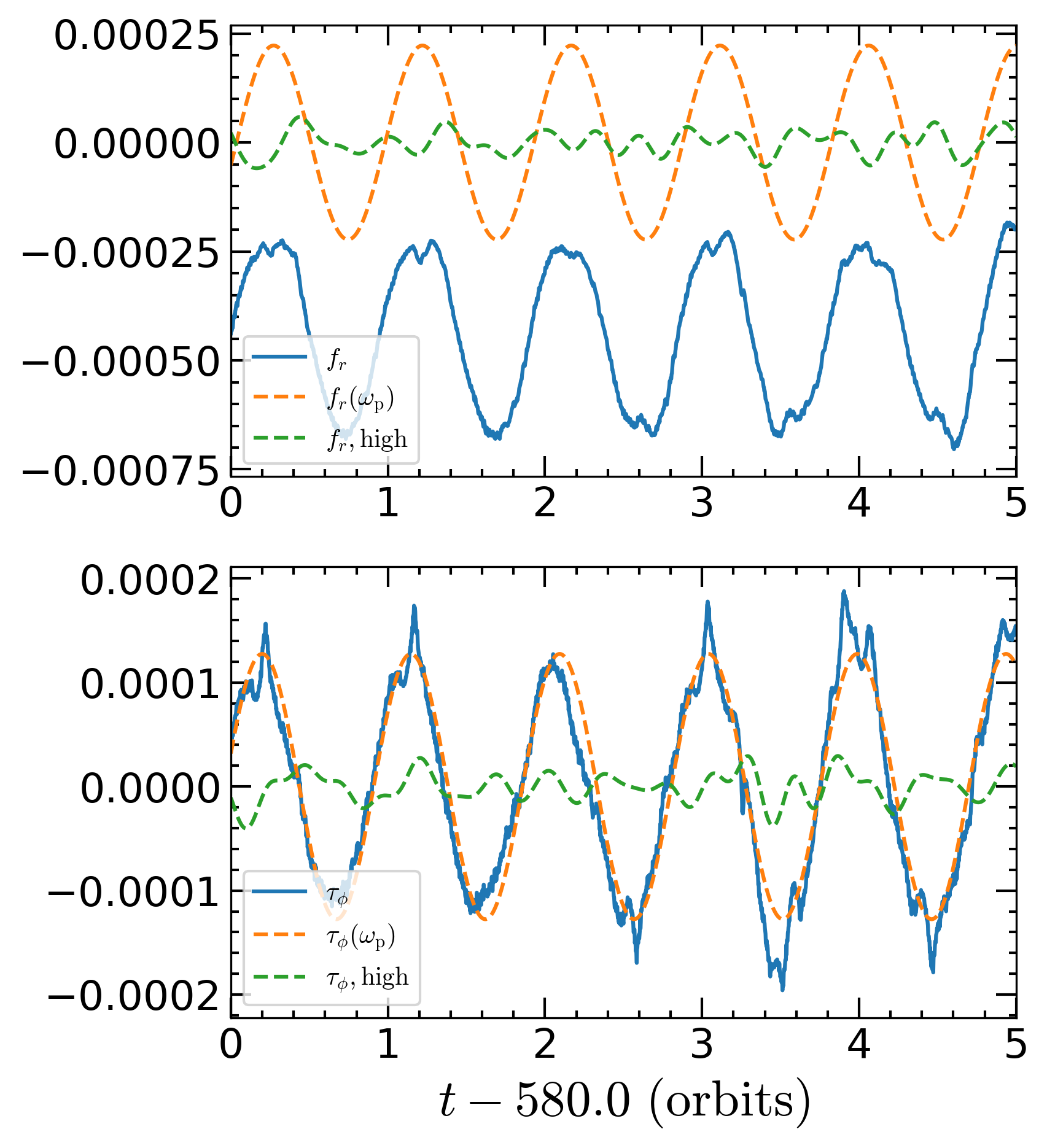}
\includegraphics[width=0.45\textwidth,clip=true]{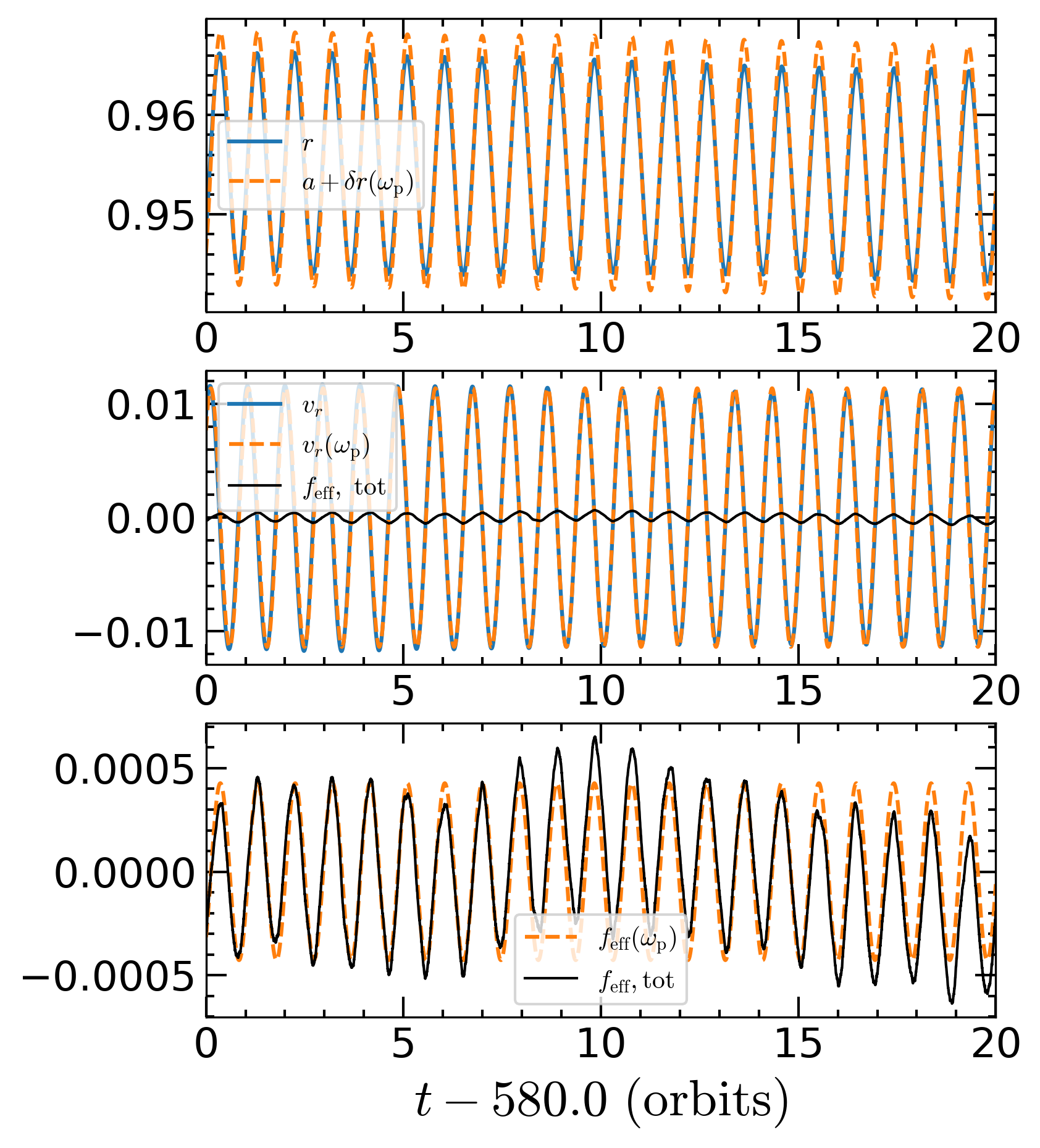}
\caption{Similar to Figure~\ref{fig:torq_rc05} but at 580-600 orbits.}
 \label{fig:torq_rc05_t600}
\end{figure*}

The predicted $\delta r$ (based on Equation~\ref{eq:dr}) at the peak frequency $\omega_{\rm p}$ of $v_{r}$ power spectrum)  is shown in the upper right panel of Figure~\ref{fig:torq_rc05}. Note that $\omega_{\rm p}$ is slightly different from $\omega_{\rm 0}$ since the sBH also feels the disk forces, i.e. no divergence when Equation~\ref{eq:dr} is applied here.
A better match with the full time-domain result is found than what we observe in Figure~\ref{fig:torq_rc02},  although there is small discrepancy for the overall oscillation magnitude.

\begin{figure}
\centering
\includegraphics[width=0.45\textwidth,clip=true]{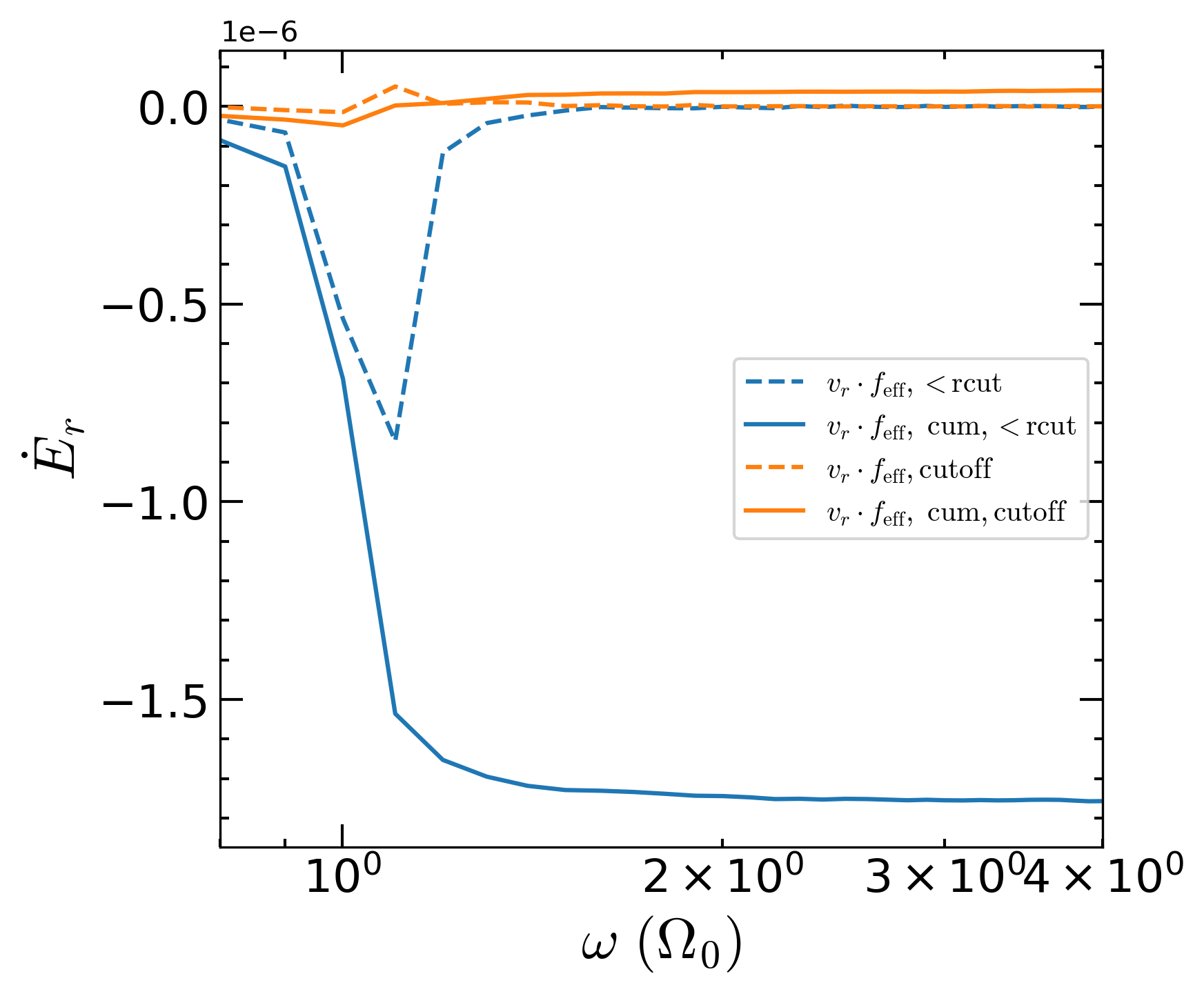}
\caption{Similar to Figure~\ref{fig:power_rc05} but around 600 orbits. The orange lines show the $\dot{E}_{r}$ from the CSD cut-off force, while the blue lines correspond to the force within the cut-off region although the actual evolution of the sBH is still governed by the cut-off torque. The summation of these two indicates the total force from the disk without any CSD cut-off, which damps the eccentricity with the negative $\dot{E}_{r}$.}
 \label{fig:power_rc05_t600}
\end{figure}

A coherent phase offset $\delta\phi<\pi/2$ indicates a positive $\dot{E}_{r}$, which further suggests an eccentricity pumping force from the disk. The spectrum of $\dot{E}_{r}$ and its cumulative distribution is shown in Figure~\ref{fig:power_rc05}. The $\dot{E}_{r}$ is mostly peaked around $\omega_{\rm p}$ as $f_{\rm eff}$, and thus dominated by the $\omega_{\rm p}$ component. The cumulative $\dot{E}_{r}$ is indeed positive and significantly larger than that of $r_{\rm c}=0.2\ R_{\rm H}$ shown in Figure~\ref{fig:power_rc02}, which suggests a strong eccentricity excitation around the 200th orbit with $r_{\rm c}=0.5\ R_{\rm H}$. This is indeed what we observe in Figure~\ref{fig:ecc_h01}, where around the 200th orbit the eccentricity corresponding to the grey line does quickly increase with time. 

After evolving 300 orbits, the eccentricity evolution almost reaches an equilibrium state. 
In Figure~\ref{fig:torq_rc05_t600}, we show the similar time profiles for different quantities as Figure~\ref{fig:torq_rc05} at 580-600 orbits. The quantities $f_{r}$, $\tau_{\phi}$, and $f_{\rm eff}$ are all dominated by the $\omega_{\rm p}$ component, and furthermore, the predicted $\delta r$ (based on Equation~\ref{eq:dr}) at frequency $\omega_{\rm p}$ is consistent with the time-domain simulation results.
The cumulative $\dot{E}_{r}$ spectrum shown in Figure~\ref{fig:power_rc05_t600} is now much smaller compared to that at the 200th orbit, as the orbit eccentricity shows much less variation in time.

In order to understand the role of the CSD for the eccentricity evolution, for the large CSD cut-off case with $r_{\rm c}=0.5\ R_{\rm H}$, we also calculate the $f_{r}$ and $\tau_{\phi}$ contributed by the cut-off region around 600 orbits, although these forces are not applied to the actual dynamical evolution of the sBH. The spectrum of $f_{r}$ and $\tau_{\phi}$ are now not centered on $\omega_{\rm p}$ but instead show quite broad distribution. There is no coherent phase offset between $f_{\rm eff}$ and $v_{r}$ in the time domain.
The high frequency component of $f_{\rm eff}$  becomes more important than the low frequency counterpart, and the spectrum of $\dot{E}_{r}$ is more negative with a very high magnitude, shown as the blues line in Figure~\ref{fig:power_rc05_t600}. This confirms the strong eccentricity damping effect from the CSD region. This suggests that the eccentricity of the sBH should be damped more efficiently if this CSD torque is applied to the sBH's active evolution, as we have observed in the full evolution with no CSD cutoff.

In summary, for this particular disk density profile assumed, the existence of CSD has modest impact on the migration rate of the sBH. It also shows a strong eccentricity-damping effect, with the underlying mechanism not presently understood. On the other hand, the emission of density waves seems to drive the orbit out of the circular profile, with another eccentricity pumping mechanism. The balance of two mechanisms give rise to $\mathcal{O}(10^{-3})$ eccentricity in the full evolution, and $\mathcal{O}(10^{-2})$ eccentricity if the CSD is fully removed. Both eccentricities may leave detectable imprint on the gravitational waveform as the sBH enters the detection band of space-borne detectors.  It’s important to note that in a realistic scenario, the accretion and feedback from the sBH are likely to significantly alter the gas profile in the CSD region, which may also affect the associated CSD dynamical effects. Nevertheless, the analysis provided here underscores the significance of understanding the CSD of sBHs for gravitational wave observations of EMRIs.

\begin{figure}
\centering
\includegraphics[width=0.45\textwidth,clip=true]{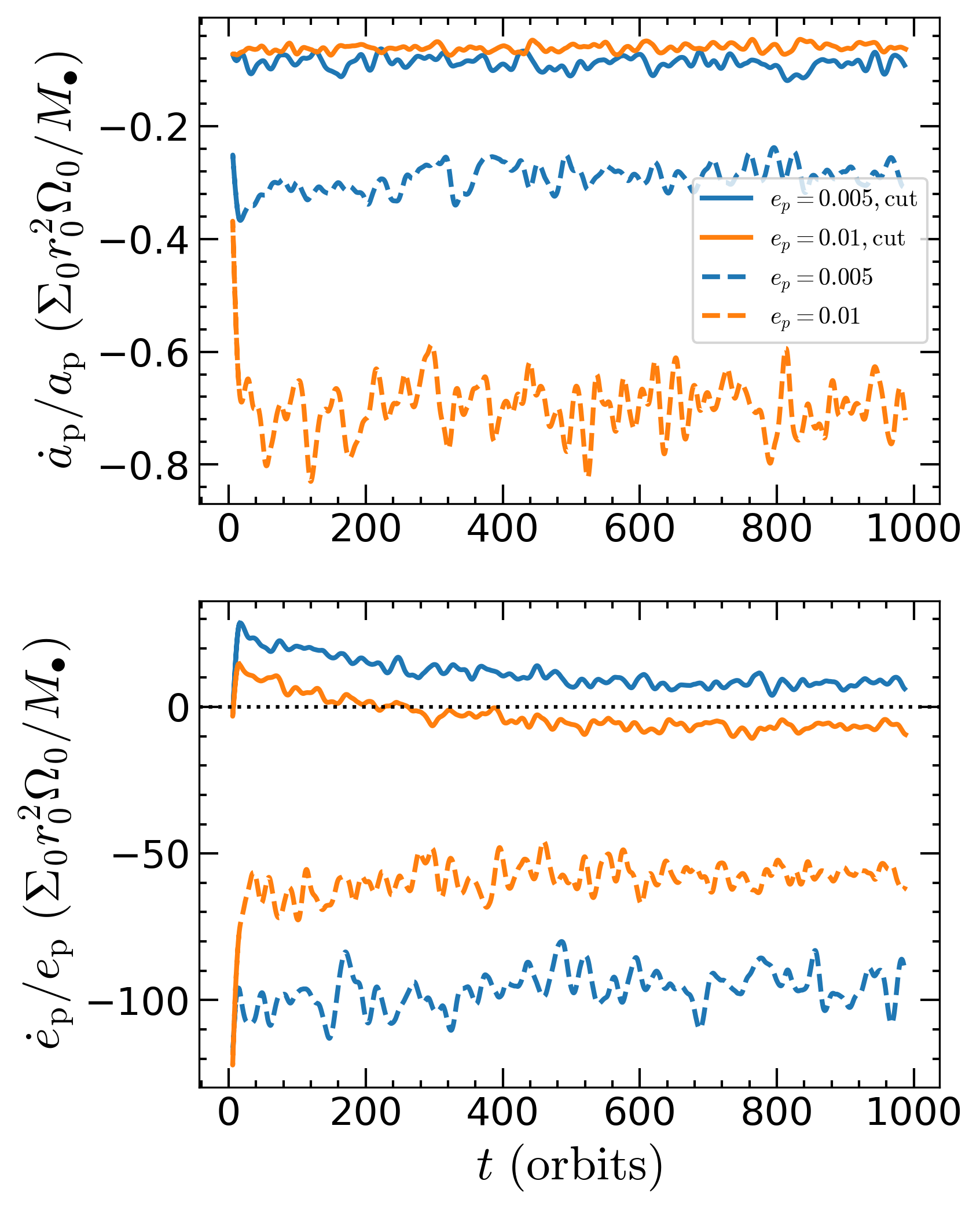}
\caption{$\dot{a}_{\rm p}/a_{\rm p}$ (upper panel) and $\dot{e}_{\rm p}/e_{\rm p}$ (lower panel) for our `fixed' orbit simulations with $h_{0}=0.01$. Orange (blue) line corresponds to a fixed orbital eccentricity of $e_{\rm p}=0.01$ ($e_{\rm p}=0.005$), and dashed (solid) line includes the full (cutoff with $r_{\rm c}=0.5\ R_{\rm H}$) disk force when calculating $\dot{a}_{\rm p}/a_{\rm p}$ and $\dot{e}_{\rm p}/e_{\rm p}$.}
 \label{fig:fixed_orbit}
\end{figure}

\begin{figure}
\centering
\includegraphics[width=0.45\textwidth,clip=true]{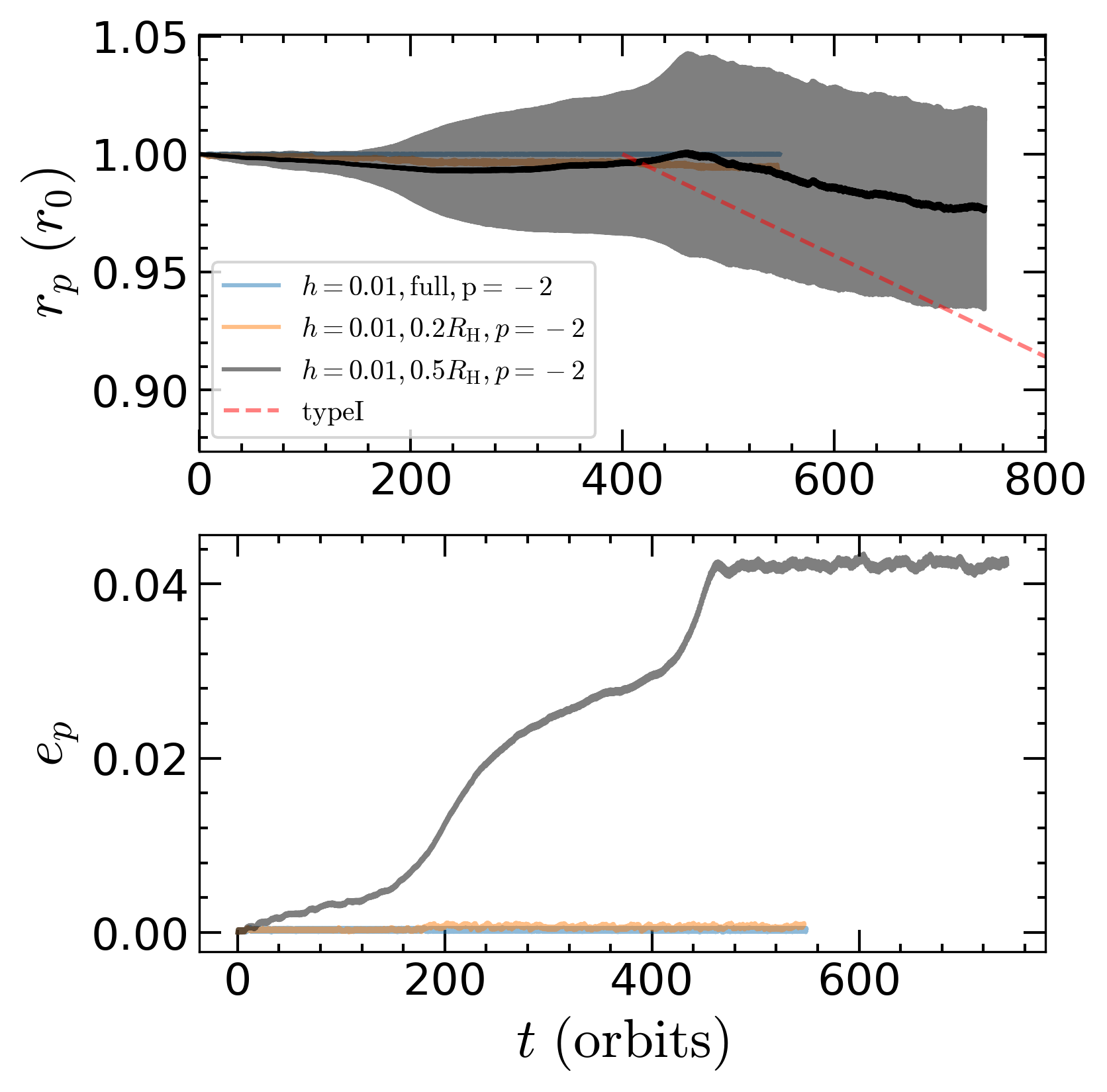}
\caption{Similar to Figure~\ref{fig:ecc_h01} but with a mid-plane density slope $p=-2.0$. The red dashed line shows the type I migration rate with an offset in the initial time point. 
}
 \label{fig:ecc_h01_p2}
\end{figure}

\subsubsection{The ``Fixed" Orbit Simulations}

We have further run two `fixed' orbit simulations with the orbital eccentricity of $e_{\rm p} =0.005$ and $e_{\rm p}=0.01$. For each one, we have monitored the disk force in two ways, one is the full disk force without any cutoff and another one is the disk force with a cutoff radius of $r_{\rm c}=0.5\ R_{\rm H}$. All other model parameters are the same as those of simulations in the planar Hill region.

We calculate the $\dot{a}_{\rm p}/a_{\rm p}$ and $\dot{e}_{\rm p}/e_{\rm p}$ according to Section~\ref{sec:eccdot}. The results for these four simulations are shown in Figure~\ref{fig:fixed_orbit}. 
As shown in the upper panel of Figure~\ref{fig:fixed_orbit}, the inward migration rate is significantly slow down when the torque cutoff is applied. 
Furthermore, we can clearly see that without torque cutoff (i.e., including the full disk force), both the cases with $e_{\rm p}=0.005$ and $e_{\rm p}=0.01$ tend to damp the orbital eccentricity efficiently ($\dot{e}_{\rm p}<0$). However, when the torque cutoff is implemented (i.e., exclude the disk force within $0.5\ R_{\rm H}$ around the sBH), the low eccentricity case tends to excite the sBH's eccentricity, while the high eccentricity case would damp its eccentricity instead. It clearly demonstrates that there is an equilibrium eccentricity around $e_{\rm eq}\sim0.01$ for the sBH when the torque cutoff is applied to the sBH. This is consistent with what we find in our ``live" sBH setup (see the black line in the lower panel of Figure~\ref{fig:ecc_h01}). 
All of these confirm the critical role of  the circumsingle disk on the dynamical evolution of the sBH in the planar Hill limit of $H<R_{\rm H}$.

\subsubsection{The Case with a Shallower Surface Density Gradient}

As the migration dynamics also depends on the disk density and temperature profiles \citep[e.g.,][]{Tanaka2002,Chen2020}, we further perform several simulations with a shallower disk density profile with $p=-2.0$ and keep the disk as globally isothermal (i.e. $\zeta=0.0$). The migration and eccentricity evolutions for different CSD cut-off are shown in Figure~\ref{fig:ecc_h01_p2}.
The migration rate with the full-disk torque is much slower than that of the steep density slope case, and the classical type I migration. This significant reduction in migration has not been reported in the past literature. With the materials within the CSD region removed, the migration rate changes modestly, but still quite far away from the type-I migration rate.
Strong eccentricity excitation is observed with a large CSD cut-off applied with $r_{\rm c}=0.5\ R_{\rm H}$, although the saturated eccentricity is a factor of a few times higher than the steep density slope case.
In this sense, the CSD appears to damp the eccentricity of the embedded stellar-mass object in a similar way for different disk models. Furthermore, with a diminished CSD, the density wave emission from the rest of disk materials can pump the sBH to a few percent level regardless of the disk density slope, although the exact value of the equilibrium orbital eccentricity is susceptible to detailed disk profiles. To fully understand the phenomena about the eccentricity evolution and the orbital migration observed here, a detailed analysis about density wave excitation away from the CSD and the dynamical role of circulating gas within the CSD is required.

\section{Discussion and Conclusion}\label{sec:con}

We have presented a three-dimensional global hydrodynamical numerical simulation of point masses embedded within high-viscosity thin accretion disks in a parameter range relevant for the migration of EMRIs in AGN disks. The feedback of sBHs within AGN likely involve super-Eddington accretion and strong outflows, which are not well understood and therefore not correspondingly modeled in these global simulations. As canonical type I and type II migration dynamics have been extensively studied in the past literature, in this work, we focus on a relatively new scenario that is less likely to present for protoplanetary disks (as it often requires high viscosity): no gap opening while the Hill radius is larger than the disk thickness. Such a scenario may be realized for sBHs and neutron stars orbiting around a massive black hole ($10^5-10^7 M_{\odot}$) at $\mathcal{O}(10^2)$ gravitational radii, right before the EMRI enters the gravitational-wave-dominated evolution stage.

The EMRI migration dynamics does show qualitatively distinct features in this new parameter regime. First, as the Hill radius is greater than the disk thickness, in the vicinity of the stellar-mass object where its own gravity dominates, the gas materials form a CSD disk, which is commonly seen in multi-dimensional hydrodynamical simulations if the Hill radius is greater than the disk thickness. Within our simulation setup, the gas trapped within the CSD disk interacts gravitationally with the stellar-mass object, which provides a damping mechanism for the orbital eccentricity. In simulations in which the CSD-point mass interaction is turned off, we also observe a new eccentricity-pump effect in operation (possibly associated with density-wave emissions), which brings the orbit eccentricity to $\mathcal{O}(0.01)$. This has been confirmed by both the ``live" and ``fixed" orbit simulations.
The underlying mechanisms for the eccentricity damping and pumping effects could be due to the external Lindblad resonances. 
Second, for different disk profiles with smaller radial density gradient $\rho \propto R^{-2}$, the difference in angular momentum carried by ingoing and outgoing density waves also decreases, and the corresponding migration is dramatically slower than the rate predicted by the type I formula. Notice that the existence of CSD and its circulating gas flow is a {\it non-perturbative} phenomenon, which is not included in a linear-theory analysis that assumes the gas flow only weakly deviates the mean global gas flow under the influence of the point mass. The size of CSD covers a good fraction of the Hill radius, which influences all the Lindblad resonances within its range. It is possible that the circulating gas re-balances the strength of ingoing and outgoing density waves within the range of CSD. However, the dependence on the density gradient of the disk is not understood, which warrants a more systematic study \citep[e.g.,][]{Chen2020}.

The critical role of CSD in both migration and eccentricity evolution highlights the importance of understanding the feedback of compact objects within the accretion disk \citep{Li2022b}, as the majority of the super-Eddington, infalling gas captured at the Bondi radius is expected to join the outflows compared to those accreted onto the compact objects (i.e., see the model used in \cite{Pan:2021xhv}). The outflow will definitely modify the structure of the CSDs seen in our simulations, possibly leading to very different effects on the migration rate and eccentricity. As initially discussed in \cite{Pan:2021ksp,Pan:2021oob}, eccentricity may be a key observable distinguishing EMRIs formed through wet and dry channels as in future observations from space-borne gravitational wave detectors. It is therefore important to quantify the eccentricity distribution in different formation channels for model selection analysis. Our study here is just a first step towards fully understanding eccentricity evolution of compact objects within AGN disks. Another possible disk effect on eccentricity excitation is through coupling with turbulent eddies in the disk. Given the high precision of LISA in measuring EMRI eccentricities ($\mathcal{O}(10^{-5})$), the gravitational wave measurement on EMRI eccentricities will likely be used to infer various properties of the AGN disk, if the underlying physics is well understood.

\vspace{0.2 cm}

\acknowledgments

We thank the referee for many constructive suggestions that improve the quality of the paper. We also thank  Douglas Lin, Shigeru Ida, Dong Lai, Richard Nelson for their beneficial discussions.
Y.P.L. is supported in part by the Natural Science Foundation of China (grants 12373070, and 12192223), the Natural Science Foundation of Shanghai (grant NO. 23ZR1473700). The calculations have made use of the High Performance Computing
Resource in the Core Facility for Advanced Research Computing
at Shanghai Astronomical Observatory.

\appendix

\section{Different Cut-off Schemes}

In our main text, we adopt a simply sharp CSD cut-off where only the disk force within $r_{\rm c}$ from the embedded object is excised. Here we test another smooth cut-off scheme, where the disk force  is  reduced by a factor of $\eta$ depending the distance from the embedded object $\delta r_{\rm p}$:

\begin{equation}
    \eta = 
    \begin{cases}
        0 & , \quad \delta r_{\rm p} < r_{\rm c} ; \\
         \sin^{2}\left(\frac{\pi}{2(R_{\rm H}-r_{\rm c})}\left(\delta r_{\rm p}-r_{\rm c}\right)\right) &, \quad r_{\rm c} \leq \delta r_{\rm p} \le R_{\rm H} ;  \\
        1 & , \quad \delta r_{\rm p} > R_{\rm H} . 
    \end{cases}
    \label{eq:cutoff_smooth}
\end{equation}

\begin{figure}
\centering
\includegraphics[width=0.45\textwidth,clip=true]{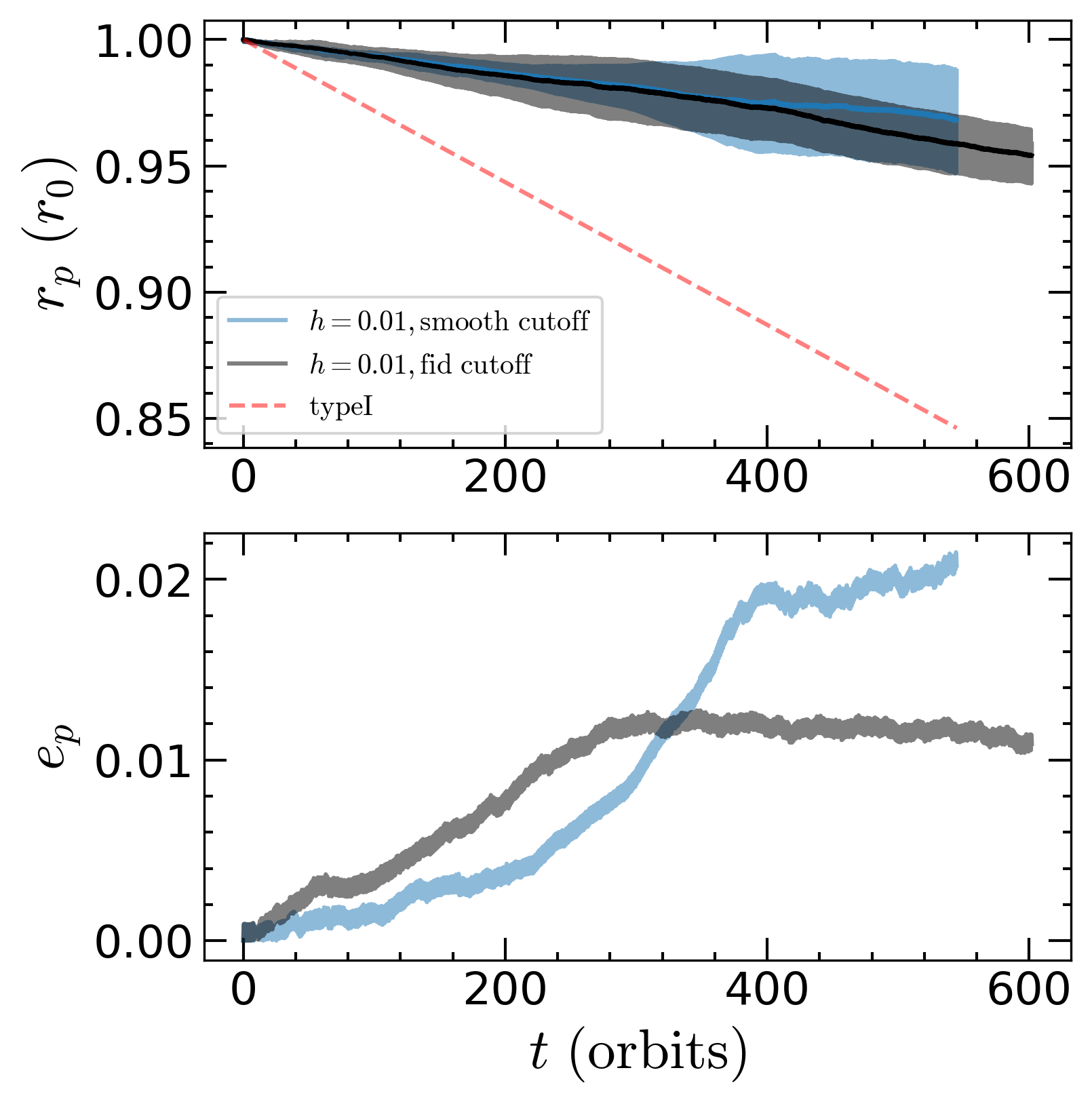}
\caption{Comparison between different cut-off schemes. The blue lines correspond to the smooth cut-off while the black lines show the fiducial sharp cutoff.}
 \label{fig:ecc_h01_smooth}
\end{figure}

The results of a smooth CSD cut-off with $r_{\rm c}=0.2\ R_{\rm H}$ are shown in Figure~\ref{fig:ecc_h01_smooth}. It can be seen that with a smaller $r_{\rm c}$ the embedded object can be excited into a distinguishable orbital eccentricity. This is simply because the disk force from the Hill sphere region outside $r_{\rm c}$ but within $R_{\rm H}$ is also modified by the smooth cut-off scheme.  It could be also expected that the smooth cut-off results in a different orbital eccentricity compared to our fiducial sharp cut-off method, although both of them show similar tendency. We have further tested with 2D simulations using \texttt{FARGO3D} \citep{Benitez-Llambay2016} in a similar parameter regime and a smooth cut-off method, and find a similar eccentricity evolution for the sBH.
All of these suggest that the CSD can indeed play a critical role in the orbital evolution of EMRI event in AGN disks.

\section{Calculating $\dot{e}_{\rm p}$ Based on Linear Theory} \label{sec:edot_ana}

We list the main formulae for the calculation of $\dot{e}_{\rm p}$. The complete equations are summarized in \citep{Duffell2015}. The angular momentum of the embedded point mass is $J_{\rm p}=M_{\rm p}\sqrt{GM_{\bullet}a_{\rm p}(1-e_{\rm p}^{2})}$. As listed in \citet{Goldreich2003}, we write down equations for eight  kinds of resonances except for the principal co-rotation resonance. The  eccentricity evolution for each resonance at azimuthal wavenumber $m$ is as follows.

For the first-order Lindblad resonances with $\ell = m + 1$, there are two Lindblad resonances, one is a “co-orbital” resonance located at the embedded object's semi-major axis with $\beta\equiv r/a_{\rm p}=1$, another is located at the external resonance with $\beta=\left(\frac{m-1}{m+1}\right)^{2/3}$. 
Each one will contribute to a change rate of eccentricity in the form of

\begin{equation}
\label{eq:edot_lp1}
\dot{e}_{\rm p}=-\frac{\Omega_{\rm p}}{m}\frac{J_{\rm p}^{2}}{(GM_{\bullet})^2}\frac{T_{m+1,m}^{\rm L}}{e_{\rm p}M_{\rm p}^{3}}\times\frac{1}{1+0.25(e_{\rm p}\mathcal{M})^{3}},
\end{equation}
where $\mathcal{M}=a_{\rm p}/H$ defines the Mach number of the disk. All the terms $T_{\ell,m}^{\rm L}$ (with explicit $\beta$ dependence) are listed in \citep{Duffell2015}.

For the first-order Lindblad resonances with $\ell = m - 1$, there are also two resonances associated with this potential. One is the “co-orbital” resonance located at the embedded object's semi-major axis ($\beta=1$), another is located at the external resonance with $\beta=\left(\frac{m+1}{m-1}\right)^{2/3}$. The eccentricity change rate for each resonance is

\begin{equation}
\label{eq:edot_lm1}
\dot{e}_{\rm p}=+\frac{\Omega_{\rm p}}{m}\frac{J_{\rm p}^{2}}{(GM_{\bullet})^2}\frac{T_{m-1,m}^{\rm L}}{e_{\rm p}M_{\rm p}^{3}}\times\frac{1}{1+0.25(e_{\rm p}\mathcal{M})^{3}}.
\end{equation}

Equations~\ref{eq:edot_lp1} and \ref{eq:edot_lm1} lead to two co-orbital resonances (summation of two terms with $\beta=1$) and two external resonances (summation of two terms with different $\beta$), with the former giving rise to a negative $\dot{e}_{\rm p}$ while the latter result in a positive $\dot{e}_{\rm p}$, shown as blue and green lines in Figure~\ref{fig:edot_ana}. These are usually the most important terms for the evolution of $e_{\rm p}$.

There are two first-order corotation resonances. One is for $\ell = m + 1$ which is located at $\beta=\left(\frac{m}{m+1}\right)^{2/3}$. The corresponding $\dot{e}_{\rm p}$ is

\begin{equation}\label{eq:edot_cp1}
\dot{e}_{\rm p}=-\frac{\Omega_{\rm p}}{m}\frac{J_{\rm p}^{2}}{(GM_{\bullet})^2}\frac{T_{m+1,m}^{\rm C}}{e_{\rm p}M_{\rm p}^{3}}\times F(p),
\end{equation}
where $T_{\ell,m}^{\rm C}$ and $F_{\rm p}$ are defined in \citep{Duffell2015}.
Another is for $\ell = m - 1$ located at $\beta=\left(\frac{m}{m-1}\right)^{2/3}$, and the $\dot{e}_{\rm p}$ is determined by

\begin{equation}\label{eq:edot_cm1}
\dot{e}_{\rm p}=+\frac{\Omega_{\rm p}}{m}\frac{J_{\rm p}^{2}}{(GM_{\bullet})^2}\frac{T_{m-1,m}^{\rm C}}{e_{\rm p}M_{\rm p}^{3}}\times F(p).
\end{equation}

The total first-order corotation resonances are obtained by the summation of Equations~\ref{eq:edot_cp1} and \ref{eq:edot_cm1} at different $m$. The $\dot{e}_{\rm p}$ from the first-order corotation resonances are usually negative, that is, damping the eccentricity but showing a dependence on the surface density profile. However, their contribution to $\dot{e}_{\rm p}$ is usually less dominated compared to the co-orbital and external resonances as shown in Figure~\ref{fig:edot_ana}.

Other two terms are associated with the principal inner and outer Lindblad resonances with $\ell = m$, which are located at $\beta=\left(\frac{m}{m+1}\right)^{2/3}$ and $\beta=\left(\frac{m}{m-1}\right)^{2/3}$. We can see that the principal inner and outer resonance locations are the same as the $\ell=m+1$ and $\ell=m+1$ corotation resonances, respectively. The $\dot{e}_{\rm p}$ for both the principal inner and outer resonance is given by

\begin{equation}\label{eq:edot_pp1}
\dot{e}_{\rm p}=\frac{e_{\rm p}\Omega_{\rm p}J_{\rm p}^{2}T_{m,m}^{\rm L}}{2(GM_{\bullet})^2M_{\rm p}^{3}}.
\end{equation}
Note that $T_{m,m}^{\rm L}$ are different for the principal inner and outer Lindblad resonances since the resonance location are different although both of them have the similar equation as above. The two principal Lindblad resonances also show much smaller magnitudes in our disk models as shown in Figure~\ref{fig:edot_ana}.

\bibliography{references}

\begin{thebibliography}{58}%
\makeatletter
\providecommand \@ifxundefined [1]{%
 \@ifx{#1\undefined}
}%
\providecommand \@ifnum [1]{%
 \ifnum #1\expandafter \@firstoftwo
 \else \expandafter \@secondoftwo
 \fi
}%
\providecommand \@ifx [1]{%
 \ifx #1\expandafter \@firstoftwo
 \else \expandafter \@secondoftwo
 \fi
}%
\providecommand \natexlab [1]{#1}%
\providecommand \enquote  [1]{``#1''}%
\providecommand \bibnamefont  [1]{#1}%
\providecommand \bibfnamefont [1]{#1}%
\providecommand \citenamefont [1]{#1}%
\providecommand \href@noop [0]{\@secondoftwo}%
\providecommand \href [0]{\begingroup \@sanitize@url \@href}%
\providecommand \@href[1]{\@@startlink{#1}\@@href}%
\providecommand \@@href[1]{\endgroup#1\@@endlink}%
\providecommand \@sanitize@url [0]{\catcode `\\12\catcode `\$12\catcode `\&12\catcode `\#12\catcode `\^12\catcode `\_12\catcode `\%12\relax}%
\providecommand \@@startlink[1]{}%
\providecommand \@@endlink[0]{}%
\providecommand \url  [0]{\begingroup\@sanitize@url \@url }%
\providecommand \@url [1]{\endgroup\@href {#1}{\urlprefix }}%
\providecommand \urlprefix  [0]{URL }%
\providecommand \Eprint [0]{\href }%
\providecommand \doibase [0]{http://dx.doi.org/}%
\providecommand \selectlanguage [0]{\@gobble}%
\providecommand \bibinfo  [0]{\@secondoftwo}%
\providecommand \bibfield  [0]{\@secondoftwo}%
\providecommand \translation [1]{[#1]}%
\providecommand \BibitemOpen [0]{}%
\providecommand \bibitemStop [0]{}%
\providecommand \bibitemNoStop [0]{.\EOS\space}%
\providecommand \EOS [0]{\spacefactor3000\relax}%
\providecommand \BibitemShut  [1]{\csname bibitem#1\endcsname}%
\let\auto@bib@innerbib\@empty
\bibitem [{\citenamefont {Evans}\ \emph {et~al.}(2021)\citenamefont {Evans} \emph {et~al.}}]{Evans:2021gyd}%
  \BibitemOpen
  \bibfield  {author} {\bibinfo {author} {\bibfnamefont {Matthew}\ \bibnamefont {Evans}} \emph {et~al.},\ }\bibfield  {title} {\enquote {\bibinfo {title} {{A Horizon Study for Cosmic Explorer: Science, Observatories, and Community}},}\ }\href@noop {} {\  (\bibinfo {year} {2021})},\ \Eprint {http://arxiv.org/abs/2109.09882} {arXiv:2109.09882 [astro-ph.IM]} \BibitemShut {NoStop}%
\bibitem [{\citenamefont {Punturo}\ \emph {et~al.}(2010)\citenamefont {Punturo} \emph {et~al.}}]{Punturo:2010zz}%
  \BibitemOpen
  \bibfield  {author} {\bibinfo {author} {\bibfnamefont {M.}~\bibnamefont {Punturo}} \emph {et~al.},\ }\bibfield  {title} {\enquote {\bibinfo {title} {{The Einstein Telescope: A third-generation gravitational wave observatory}},}\ }\href {\doibase 10.1088/0264-9381/27/19/194002} {\bibfield  {journal} {\bibinfo  {journal} {Class. Quant. Grav.}\ }\textbf {\bibinfo {volume} {27}},\ \bibinfo {pages} {194002} (\bibinfo {year} {2010})}\BibitemShut {NoStop}%
\bibitem [{\citenamefont {Baker}\ \emph {et~al.}(2019)\citenamefont {Baker} \emph {et~al.}}]{Baker:2019nia}%
  \BibitemOpen
  \bibfield  {author} {\bibinfo {author} {\bibfnamefont {John}\ \bibnamefont {Baker}} \emph {et~al.},\ }\bibfield  {title} {\enquote {\bibinfo {title} {{The Laser Interferometer Space Antenna: Unveiling the Millihertz Gravitational Wave Sky}},}\ }\href@noop {} {\  (\bibinfo {year} {2019})},\ \Eprint {http://arxiv.org/abs/1907.06482} {arXiv:1907.06482 [astro-ph.IM]} \BibitemShut {NoStop}%
\bibitem [{\citenamefont {Hu}\ and\ \citenamefont {Wu}(2017)}]{Hu:2017mde}%
  \BibitemOpen
  \bibfield  {author} {\bibinfo {author} {\bibfnamefont {Wen-Rui}\ \bibnamefont {Hu}}\ and\ \bibinfo {author} {\bibfnamefont {Yue-Liang}\ \bibnamefont {Wu}},\ }\bibfield  {title} {\enquote {\bibinfo {title} {{The Taiji Program in Space for gravitational wave physics and the nature of gravity}},}\ }\href {\doibase 10.1093/nsr/nwx116} {\bibfield  {journal} {\bibinfo  {journal} {Natl. Sci. Rev.}\ }\textbf {\bibinfo {volume} {4}},\ \bibinfo {pages} {685--686} (\bibinfo {year} {2017})}\BibitemShut {NoStop}%
\bibitem [{\citenamefont {Luo}\ \emph {et~al.}(2016)\citenamefont {Luo}, \citenamefont {Chen}, \citenamefont {Duan}, \citenamefont {Gong}, \citenamefont {Hu}, \citenamefont {Ji}, \citenamefont {Liu}, \citenamefont {Mei}, \citenamefont {Milyukov}, \citenamefont {Sazhin} \emph {et~al.}}]{luo2016tianqin}%
  \BibitemOpen
  \bibfield  {author} {\bibinfo {author} {\bibfnamefont {Jun}\ \bibnamefont {Luo}}, \bibinfo {author} {\bibfnamefont {Li-Sheng}\ \bibnamefont {Chen}}, \bibinfo {author} {\bibfnamefont {Hui-Zong}\ \bibnamefont {Duan}}, \bibinfo {author} {\bibfnamefont {Yun-Gui}\ \bibnamefont {Gong}}, \bibinfo {author} {\bibfnamefont {Shoucun}\ \bibnamefont {Hu}}, \bibinfo {author} {\bibfnamefont {Jianghui}\ \bibnamefont {Ji}}, \bibinfo {author} {\bibfnamefont {Qi}~\bibnamefont {Liu}}, \bibinfo {author} {\bibfnamefont {Jianwei}\ \bibnamefont {Mei}}, \bibinfo {author} {\bibfnamefont {Vadim}\ \bibnamefont {Milyukov}}, \bibinfo {author} {\bibfnamefont {Mikhail}\ \bibnamefont {Sazhin}},  \emph {et~al.},\ }\bibfield  {title} {\enquote {\bibinfo {title} {Tianqin: a space-borne gravitational wave detector},}\ }\href@noop {} {\bibfield  {journal} {\bibinfo  {journal} {Classical and Quantum Gravity}\ }\textbf {\bibinfo {volume} {33}},\ \bibinfo {pages} {035010} (\bibinfo {year} {2016})}\BibitemShut {NoStop}%
\bibitem [{\citenamefont {Aghanim}\ \emph {et~al.}(2020)\citenamefont {Aghanim} \emph {et~al.}}]{Planck:2018nkj}%
  \BibitemOpen
  \bibfield  {author} {\bibinfo {author} {\bibfnamefont {N.}~\bibnamefont {Aghanim}} \emph {et~al.} (\bibinfo {collaboration} {Planck}),\ }\bibfield  {title} {\enquote {\bibinfo {title} {{Planck 2018 results. I. Overview and the cosmological legacy of Planck}},}\ }\href {\doibase 10.1051/0004-6361/201833880} {\bibfield  {journal} {\bibinfo  {journal} {Astron. Astrophys.}\ }\textbf {\bibinfo {volume} {641}},\ \bibinfo {pages} {A1} (\bibinfo {year} {2020})},\ \Eprint {http://arxiv.org/abs/1807.06205} {arXiv:1807.06205 [astro-ph.CO]} \BibitemShut {NoStop}%
\bibitem [{\citenamefont {{Kanagawa}}\ \emph {et~al.}(2015)\citenamefont {{Kanagawa}}, \citenamefont {{Muto}}, \citenamefont {{Tanaka}}, \citenamefont {{Tanigawa}}, \citenamefont {{Takeuchi}}, \citenamefont {{Tsukagoshi}},\ and\ \citenamefont {{Momose}}}]{Kanagawa2015}%
  \BibitemOpen
  \bibfield  {author} {\bibinfo {author} {\bibfnamefont {Kazuhiro~D.}\ \bibnamefont {{Kanagawa}}}, \bibinfo {author} {\bibfnamefont {Takayuki}\ \bibnamefont {{Muto}}}, \bibinfo {author} {\bibfnamefont {Hidekazu}\ \bibnamefont {{Tanaka}}}, \bibinfo {author} {\bibfnamefont {Takayuki}\ \bibnamefont {{Tanigawa}}}, \bibinfo {author} {\bibfnamefont {Taku}\ \bibnamefont {{Takeuchi}}}, \bibinfo {author} {\bibfnamefont {Takashi}\ \bibnamefont {{Tsukagoshi}}}, \ and\ \bibinfo {author} {\bibfnamefont {Munetake}\ \bibnamefont {{Momose}}},\ }\bibfield  {title} {\enquote {\bibinfo {title} {{Mass Estimates of a Giant Planet in a Protoplanetary Disk from the Gap Structures}},}\ }\href {\doibase 10.1088/2041-8205/806/1/L15} {\bibfield  {journal} {\bibinfo  {journal} {\apjl}\ }\textbf {\bibinfo {volume} {806}},\ \bibinfo {eid} {L15} (\bibinfo {year} {2015})},\ \Eprint {http://arxiv.org/abs/1505.04482} {arXiv:1505.04482 [astro-ph.EP]} \BibitemShut {NoStop}%
\bibitem [{\citenamefont {{Lin}}\ and\ \citenamefont {{Papaloizou}}(1986)}]{Lin1986}%
  \BibitemOpen
  \bibfield  {author} {\bibinfo {author} {\bibfnamefont {D.~N.~C.}\ \bibnamefont {{Lin}}}\ and\ \bibinfo {author} {\bibfnamefont {John}\ \bibnamefont {{Papaloizou}}},\ }\bibfield  {title} {\enquote {\bibinfo {title} {{On the Tidal Interaction between Protoplanets and the Protoplanetary Disk. III. Orbital Migration of Protoplanets}},}\ }\href {\doibase 10.1086/164653} {\bibfield  {journal} {\bibinfo  {journal} {\apj}\ }\textbf {\bibinfo {volume} {309}},\ \bibinfo {pages} {846} (\bibinfo {year} {1986})}\BibitemShut {NoStop}%
\bibitem [{\citenamefont {{Duffell}}\ and\ \citenamefont {{MacFadyen}}(2013)}]{Duffell2013}%
  \BibitemOpen
  \bibfield  {author} {\bibinfo {author} {\bibfnamefont {Paul~C.}\ \bibnamefont {{Duffell}}}\ and\ \bibinfo {author} {\bibfnamefont {Andrew~I.}\ \bibnamefont {{MacFadyen}}},\ }\bibfield  {title} {\enquote {\bibinfo {title} {{Gap Opening by Extremely Low-mass Planets in a Viscous Disk}},}\ }\href {\doibase 10.1088/0004-637X/769/1/41} {\bibfield  {journal} {\bibinfo  {journal} {\apj}\ }\textbf {\bibinfo {volume} {769}},\ \bibinfo {eid} {41} (\bibinfo {year} {2013})},\ \Eprint {http://arxiv.org/abs/1302.1934} {arXiv:1302.1934 [astro-ph.EP]} \BibitemShut {NoStop}%
\bibitem [{\citenamefont {Afshordi}\ \emph {et~al.}(2023)\citenamefont {Afshordi} \emph {et~al.}}]{LISAConsortiumWaveformWorkingGroup:2023arg}%
  \BibitemOpen
  \bibfield  {author} {\bibinfo {author} {\bibfnamefont {Niayesh}\ \bibnamefont {Afshordi}} \emph {et~al.} (\bibinfo {collaboration} {LISA Consortium Waveform Working Group}),\ }\bibfield  {title} {\enquote {\bibinfo {title} {{Waveform Modelling for the Laser Interferometer Space Antenna}},}\ }\href@noop {} {\  (\bibinfo {year} {2023})},\ \Eprint {http://arxiv.org/abs/2311.01300} {arXiv:2311.01300 [gr-qc]} \BibitemShut {NoStop}%
\bibitem [{\citenamefont {Seoane}\ \emph {et~al.}(2023)\citenamefont {Seoane} \emph {et~al.}}]{LISA:2022yao}%
  \BibitemOpen
  \bibfield  {author} {\bibinfo {author} {\bibfnamefont {Pau~Amaro}\ \bibnamefont {Seoane}} \emph {et~al.} (\bibinfo {collaboration} {LISA}),\ }\bibfield  {title} {\enquote {\bibinfo {title} {{Astrophysics with the Laser Interferometer Space Antenna}},}\ }\href {\doibase 10.1007/s41114-022-00041-y} {\bibfield  {journal} {\bibinfo  {journal} {Living Rev. Rel.}\ }\textbf {\bibinfo {volume} {26}},\ \bibinfo {pages} {2} (\bibinfo {year} {2023})},\ \Eprint {http://arxiv.org/abs/2203.06016} {arXiv:2203.06016 [gr-qc]} \BibitemShut {NoStop}%
\bibitem [{\citenamefont {Barausse}\ \emph {et~al.}(2020)\citenamefont {Barausse} \emph {et~al.}}]{Barausse:2020rsu}%
  \BibitemOpen
  \bibfield  {author} {\bibinfo {author} {\bibfnamefont {Enrico}\ \bibnamefont {Barausse}} \emph {et~al.},\ }\bibfield  {title} {\enquote {\bibinfo {title} {{Prospects for Fundamental Physics with LISA}},}\ }\href {\doibase 10.1007/s10714-020-02691-1} {\bibfield  {journal} {\bibinfo  {journal} {Gen. Rel. Grav.}\ }\textbf {\bibinfo {volume} {52}},\ \bibinfo {pages} {81} (\bibinfo {year} {2020})},\ \Eprint {http://arxiv.org/abs/2001.09793} {arXiv:2001.09793 [gr-qc]} \BibitemShut {NoStop}%
\bibitem [{\citenamefont {Babak}\ \emph {et~al.}(2017)\citenamefont {Babak}, \citenamefont {Gair}, \citenamefont {Sesana}, \citenamefont {Barausse}, \citenamefont {Sopuerta}, \citenamefont {Berry}, \citenamefont {Berti}, \citenamefont {Amaro-Seoane}, \citenamefont {Petiteau},\ and\ \citenamefont {Klein}}]{Babak:2017tow}%
  \BibitemOpen
  \bibfield  {author} {\bibinfo {author} {\bibfnamefont {Stanislav}\ \bibnamefont {Babak}}, \bibinfo {author} {\bibfnamefont {Jonathan}\ \bibnamefont {Gair}}, \bibinfo {author} {\bibfnamefont {Alberto}\ \bibnamefont {Sesana}}, \bibinfo {author} {\bibfnamefont {Enrico}\ \bibnamefont {Barausse}}, \bibinfo {author} {\bibfnamefont {Carlos~F.}\ \bibnamefont {Sopuerta}}, \bibinfo {author} {\bibfnamefont {Christopher P.~L.}\ \bibnamefont {Berry}}, \bibinfo {author} {\bibfnamefont {Emanuele}\ \bibnamefont {Berti}}, \bibinfo {author} {\bibfnamefont {Pau}\ \bibnamefont {Amaro-Seoane}}, \bibinfo {author} {\bibfnamefont {Antoine}\ \bibnamefont {Petiteau}}, \ and\ \bibinfo {author} {\bibfnamefont {Antoine}\ \bibnamefont {Klein}},\ }\bibfield  {title} {\enquote {\bibinfo {title} {{Science with the space-based interferometer LISA. V: Extreme mass-ratio inspirals}},}\ }\href {\doibase 10.1103/PhysRevD.95.103012} {\bibfield  {journal} {\bibinfo  {journal} {Phys. Rev. D}\ }\textbf {\bibinfo {volume} {95}},\ \bibinfo {pages}
  {103012} (\bibinfo {year} {2017})},\ \Eprint {http://arxiv.org/abs/1703.09722} {arXiv:1703.09722 [gr-qc]} \BibitemShut {NoStop}%
\bibitem [{\citenamefont {Fan}\ \emph {et~al.}(2020)\citenamefont {Fan}, \citenamefont {Hu}, \citenamefont {Barausse}, \citenamefont {Sesana}, \citenamefont {Zhang}, \citenamefont {Zhang}, \citenamefont {Zi},\ and\ \citenamefont {Mei}}]{Fan:2020zhy}%
  \BibitemOpen
  \bibfield  {author} {\bibinfo {author} {\bibfnamefont {Hui-Min}\ \bibnamefont {Fan}}, \bibinfo {author} {\bibfnamefont {Yi-Ming}\ \bibnamefont {Hu}}, \bibinfo {author} {\bibfnamefont {Enrico}\ \bibnamefont {Barausse}}, \bibinfo {author} {\bibfnamefont {Alberto}\ \bibnamefont {Sesana}}, \bibinfo {author} {\bibfnamefont {Jian-dong}\ \bibnamefont {Zhang}}, \bibinfo {author} {\bibfnamefont {Xuefeng}\ \bibnamefont {Zhang}}, \bibinfo {author} {\bibfnamefont {Tie-Guang}\ \bibnamefont {Zi}}, \ and\ \bibinfo {author} {\bibfnamefont {Jianwei}\ \bibnamefont {Mei}},\ }\bibfield  {title} {\enquote {\bibinfo {title} {{Science with the TianQin observatory: Preliminary result on extreme-mass-ratio inspirals}},}\ }\href {\doibase 10.1103/PhysRevD.102.063016} {\bibfield  {journal} {\bibinfo  {journal} {Phys. Rev. D}\ }\textbf {\bibinfo {volume} {102}},\ \bibinfo {pages} {063016} (\bibinfo {year} {2020})},\ \Eprint {http://arxiv.org/abs/2005.08212} {arXiv:2005.08212 [astro-ph.HE]} \BibitemShut {NoStop}%
\bibitem [{\citenamefont {Levin}(2003)}]{Levin:2003ej}%
  \BibitemOpen
  \bibfield  {author} {\bibinfo {author} {\bibfnamefont {Yuri}\ \bibnamefont {Levin}},\ }\bibfield  {title} {\enquote {\bibinfo {title} {{Formation of massive stars and black holes in selfgravitating AGN discs, and gravitational waves in LISA band}},}\ }\href@noop {} {\  (\bibinfo {year} {2003})},\ \Eprint {http://arxiv.org/abs/astro-ph/0307084} {arXiv:astro-ph/0307084} \BibitemShut {NoStop}%
\bibitem [{\citenamefont {Pan}\ and\ \citenamefont {Yang}(2021{\natexlab{a}})}]{Pan:2021ksp}%
  \BibitemOpen
  \bibfield  {author} {\bibinfo {author} {\bibfnamefont {Zhen}\ \bibnamefont {Pan}}\ and\ \bibinfo {author} {\bibfnamefont {Huan}\ \bibnamefont {Yang}},\ }\bibfield  {title} {\enquote {\bibinfo {title} {{Formation Rate of Extreme Mass Ratio Inspirals in Active Galactic Nuclei}},}\ }\href {\doibase 10.1103/PhysRevD.103.103018} {\bibfield  {journal} {\bibinfo  {journal} {Phys. Rev. D}\ }\textbf {\bibinfo {volume} {103}},\ \bibinfo {pages} {103018} (\bibinfo {year} {2021}{\natexlab{a}})},\ \Eprint {http://arxiv.org/abs/2101.09146} {arXiv:2101.09146 [astro-ph.HE]} \BibitemShut {NoStop}%
\bibitem [{\citenamefont {Pan}\ \emph {et~al.}(2021)\citenamefont {Pan}, \citenamefont {Lyu},\ and\ \citenamefont {Yang}}]{Pan:2021oob}%
  \BibitemOpen
  \bibfield  {author} {\bibinfo {author} {\bibfnamefont {Zhen}\ \bibnamefont {Pan}}, \bibinfo {author} {\bibfnamefont {Zhenwei}\ \bibnamefont {Lyu}}, \ and\ \bibinfo {author} {\bibfnamefont {Huan}\ \bibnamefont {Yang}},\ }\bibfield  {title} {\enquote {\bibinfo {title} {{Wet extreme mass ratio inspirals may be more common for spaceborne gravitational wave detection}},}\ }\href {\doibase 10.1103/PhysRevD.104.063007} {\bibfield  {journal} {\bibinfo  {journal} {Phys. Rev. D}\ }\textbf {\bibinfo {volume} {104}},\ \bibinfo {pages} {063007} (\bibinfo {year} {2021})},\ \Eprint {http://arxiv.org/abs/2104.01208} {arXiv:2104.01208 [astro-ph.HE]} \BibitemShut {NoStop}%
\bibitem [{\citenamefont {Pan}\ \emph {et~al.}(2022)\citenamefont {Pan}, \citenamefont {Lyu},\ and\ \citenamefont {Yang}}]{Pan:2021lyw}%
  \BibitemOpen
  \bibfield  {author} {\bibinfo {author} {\bibfnamefont {Zhen}\ \bibnamefont {Pan}}, \bibinfo {author} {\bibfnamefont {Zhenwei}\ \bibnamefont {Lyu}}, \ and\ \bibinfo {author} {\bibfnamefont {Huan}\ \bibnamefont {Yang}},\ }\bibfield  {title} {\enquote {\bibinfo {title} {{Mass-gap extreme mass ratio inspirals}},}\ }\href {\doibase 10.1103/PhysRevD.105.083005} {\bibfield  {journal} {\bibinfo  {journal} {Phys. Rev. D}\ }\textbf {\bibinfo {volume} {105}},\ \bibinfo {pages} {083005} (\bibinfo {year} {2022})},\ \Eprint {http://arxiv.org/abs/2112.10237} {arXiv:2112.10237 [astro-ph.HE]} \BibitemShut {NoStop}%
\bibitem [{\citenamefont {{Wang}}\ \emph {et~al.}(2023)\citenamefont {{Wang}}, \citenamefont {{Ma}},\ and\ \citenamefont {{Wu}}}]{Wang2023}%
  \BibitemOpen
  \bibfield  {author} {\bibinfo {author} {\bibfnamefont {Mengye}\ \bibnamefont {{Wang}}}, \bibinfo {author} {\bibfnamefont {Yiqiu}\ \bibnamefont {{Ma}}}, \ and\ \bibinfo {author} {\bibfnamefont {Qingwen}\ \bibnamefont {{Wu}}},\ }\bibfield  {title} {\enquote {\bibinfo {title} {{Accretion-modified stellar-mass black hole distribution and milli-Hz gravitational wave backgrounds from galaxy centre}},}\ }\href {\doibase 10.1093/mnras/stad422} {\bibfield  {journal} {\bibinfo  {journal} {\mnras}\ }\textbf {\bibinfo {volume} {520}},\ \bibinfo {pages} {4502--4516} (\bibinfo {year} {2023})},\ \Eprint {http://arxiv.org/abs/2212.05724} {arXiv:2212.05724 [astro-ph.HE]} \BibitemShut {NoStop}%
\bibitem [{\citenamefont {Zhou}\ \emph {et~al.}(2024{\natexlab{a}})\citenamefont {Zhou}, \citenamefont {Huang}, \citenamefont {Guo}, \citenamefont {Li},\ and\ \citenamefont {Pan}}]{Zhou:2024bjt}%
  \BibitemOpen
  \bibfield  {author} {\bibinfo {author} {\bibfnamefont {Cong}\ \bibnamefont {Zhou}}, \bibinfo {author} {\bibfnamefont {Lei}\ \bibnamefont {Huang}}, \bibinfo {author} {\bibfnamefont {Kangrou}\ \bibnamefont {Guo}}, \bibinfo {author} {\bibfnamefont {Ya-Ping}\ \bibnamefont {Li}}, \ and\ \bibinfo {author} {\bibfnamefont {Zhen}\ \bibnamefont {Pan}},\ }\bibfield  {title} {\enquote {\bibinfo {title} {{Probing orbits of stellar mass objects deep in galactic nuclei with quasiperiodic eruptions}},}\ }\href {\doibase 10.1103/PhysRevD.109.103031} {\bibfield  {journal} {\bibinfo  {journal} {Phys. Rev. D}\ }\textbf {\bibinfo {volume} {109}},\ \bibinfo {pages} {103031} (\bibinfo {year} {2024}{\natexlab{a}})},\ \Eprint {http://arxiv.org/abs/2401.11190} {arXiv:2401.11190 [astro-ph.HE]} \BibitemShut {NoStop}%
\bibitem [{\citenamefont {Zhou}\ \emph {et~al.}(2024{\natexlab{b}})\citenamefont {Zhou}, \citenamefont {Zhong}, \citenamefont {Zeng}, \citenamefont {Huang},\ and\ \citenamefont {Pan}}]{Zhou:2024nuu}%
  \BibitemOpen
  \bibfield  {author} {\bibinfo {author} {\bibfnamefont {Cong}\ \bibnamefont {Zhou}}, \bibinfo {author} {\bibfnamefont {Binyu}\ \bibnamefont {Zhong}}, \bibinfo {author} {\bibfnamefont {Yuhe}\ \bibnamefont {Zeng}}, \bibinfo {author} {\bibfnamefont {Lei}\ \bibnamefont {Huang}}, \ and\ \bibinfo {author} {\bibfnamefont {Zhen}\ \bibnamefont {Pan}},\ }\bibfield  {title} {\enquote {\bibinfo {title} {{Probing orbits of stellar mass objects deep in galactic nuclei with quasiperiodic eruptions. II. Population analysis}},}\ }\href {\doibase 10.1103/PhysRevD.110.083019} {\bibfield  {journal} {\bibinfo  {journal} {Phys. Rev. D}\ }\textbf {\bibinfo {volume} {110}},\ \bibinfo {pages} {083019} (\bibinfo {year} {2024}{\natexlab{b}})},\ \Eprint {http://arxiv.org/abs/2405.06429} {arXiv:2405.06429 [astro-ph.HE]} \BibitemShut {NoStop}%
\bibitem [{\citenamefont {Zhou}\ \emph {et~al.}(2024{\natexlab{c}})\citenamefont {Zhou}, \citenamefont {Zeng},\ and\ \citenamefont {Pan}}]{Zhou:2024vwj}%
  \BibitemOpen
  \bibfield  {author} {\bibinfo {author} {\bibfnamefont {Cong}\ \bibnamefont {Zhou}}, \bibinfo {author} {\bibfnamefont {Yuhe}\ \bibnamefont {Zeng}}, \ and\ \bibinfo {author} {\bibfnamefont {Zhen}\ \bibnamefont {Pan}},\ }\bibfield  {title} {\enquote {\bibinfo {title} {{Probing orbits of stellar mass objects deep in galactic nuclei with quasi-periodic eruptions -- III: Long term evolution}},}\ }\href@noop {} {\  (\bibinfo {year} {2024}{\natexlab{c}})},\ \Eprint {http://arxiv.org/abs/2411.18046} {arXiv:2411.18046 [astro-ph.HE]} \BibitemShut {NoStop}%
\bibitem [{\citenamefont {{Goldreich}}\ and\ \citenamefont {{Tremaine}}(1980)}]{GT1980}%
  \BibitemOpen
  \bibfield  {author} {\bibinfo {author} {\bibfnamefont {P.}~\bibnamefont {{Goldreich}}}\ and\ \bibinfo {author} {\bibfnamefont {S.}~\bibnamefont {{Tremaine}}},\ }\bibfield  {title} {\enquote {\bibinfo {title} {{Disk-satellite interactions.}}}\ }\href {\doibase 10.1086/158356} {\bibfield  {journal} {\bibinfo  {journal} {\apj}\ }\textbf {\bibinfo {volume} {241}},\ \bibinfo {pages} {425--441} (\bibinfo {year} {1980})}\BibitemShut {NoStop}%
\bibitem [{\citenamefont {{Kley}}\ and\ \citenamefont {{Nelson}}(2012)}]{KleyNelson2012}%
  \BibitemOpen
  \bibfield  {author} {\bibinfo {author} {\bibfnamefont {W.}~\bibnamefont {{Kley}}}\ and\ \bibinfo {author} {\bibfnamefont {R.~P.}\ \bibnamefont {{Nelson}}},\ }\bibfield  {title} {\enquote {\bibinfo {title} {{Planet-Disk Interaction and Orbital Evolution}},}\ }\href {\doibase 10.1146/annurev-astro-081811-125523} {\bibfield  {journal} {\bibinfo  {journal} {\araa}\ }\textbf {\bibinfo {volume} {50}},\ \bibinfo {pages} {211--249} (\bibinfo {year} {2012})},\ \Eprint {http://arxiv.org/abs/1203.1184} {arXiv:1203.1184 [astro-ph.EP]} \BibitemShut {NoStop}%
\bibitem [{\citenamefont {{Chen}}\ \emph {et~al.}(2020)\citenamefont {{Chen}}, \citenamefont {{Zhang}}, \citenamefont {{Li}}, \citenamefont {{Li}},\ and\ \citenamefont {{Lin}}}]{Chen2020}%
  \BibitemOpen
  \bibfield  {author} {\bibinfo {author} {\bibfnamefont {Yi-Xian}\ \bibnamefont {{Chen}}}, \bibinfo {author} {\bibfnamefont {Xiaojia}\ \bibnamefont {{Zhang}}}, \bibinfo {author} {\bibfnamefont {Ya-Ping}\ \bibnamefont {{Li}}}, \bibinfo {author} {\bibfnamefont {Hui}\ \bibnamefont {{Li}}}, \ and\ \bibinfo {author} {\bibfnamefont {Douglas N.~C.}\ \bibnamefont {{Lin}}},\ }\bibfield  {title} {\enquote {\bibinfo {title} {{Retention of Long-period Gas Giant Planets: Type II Migration Revisited}},}\ }\href {\doibase 10.3847/1538-4357/abaab6} {\bibfield  {journal} {\bibinfo  {journal} {\apj}\ }\textbf {\bibinfo {volume} {900}},\ \bibinfo {eid} {44} (\bibinfo {year} {2020})},\ \Eprint {http://arxiv.org/abs/2007.14905} {arXiv:2007.14905 [astro-ph.EP]} \BibitemShut {NoStop}%
\bibitem [{\citenamefont {{Paardekooper}}\ \emph {et~al.}(2023)\citenamefont {{Paardekooper}}, \citenamefont {{Dong}}, \citenamefont {{Duffell}}, \citenamefont {{Fung}}, \citenamefont {{Masset}}, \citenamefont {{Ogilvie}},\ and\ \citenamefont {{Tanaka}}}]{Paardekooper2023}%
  \BibitemOpen
  \bibfield  {author} {\bibinfo {author} {\bibfnamefont {S.}~\bibnamefont {{Paardekooper}}}, \bibinfo {author} {\bibfnamefont {R.}~\bibnamefont {{Dong}}}, \bibinfo {author} {\bibfnamefont {P.}~\bibnamefont {{Duffell}}}, \bibinfo {author} {\bibfnamefont {J.}~\bibnamefont {{Fung}}}, \bibinfo {author} {\bibfnamefont {F.~S.}\ \bibnamefont {{Masset}}}, \bibinfo {author} {\bibfnamefont {G.}~\bibnamefont {{Ogilvie}}}, \ and\ \bibinfo {author} {\bibfnamefont {H.}~\bibnamefont {{Tanaka}}},\ }\bibfield  {title} {\enquote {\bibinfo {title} {{Planet-Disk Interactions and Orbital Evolution}},}\ }in\ \href {\doibase 10.48550/arXiv.2203.09595} {\emph {\bibinfo {booktitle} {Protostars and Planets VII}}},\ \bibinfo {series} {Astronomical Society of the Pacific Conference Series}, Vol.\ \bibinfo {volume} {534},\ \bibinfo {editor} {edited by\ \bibinfo {editor} {\bibfnamefont {S.}~\bibnamefont {{Inutsuka}}}, \bibinfo {editor} {\bibfnamefont {Y.}~\bibnamefont {{Aikawa}}}, \bibinfo {editor} {\bibfnamefont {T.}~\bibnamefont
  {{Muto}}}, \bibinfo {editor} {\bibfnamefont {K.}~\bibnamefont {{Tomida}}}, \ and\ \bibinfo {editor} {\bibfnamefont {M.}~\bibnamefont {{Tamura}}}}\ (\bibinfo {year} {2023})\ p.\ \bibinfo {pages} {685},\ \Eprint {http://arxiv.org/abs/2203.09595} {arXiv:2203.09595 [astro-ph.EP]} \BibitemShut {NoStop}%
\bibitem [{\citenamefont {{Li}}\ \emph {et~al.}(2024)\citenamefont {{Li}}, \citenamefont {{Chen}},\ and\ \citenamefont {{Lin}}}]{Li2024}%
  \BibitemOpen
  \bibfield  {author} {\bibinfo {author} {\bibfnamefont {Ya-Ping}\ \bibnamefont {{Li}}}, \bibinfo {author} {\bibfnamefont {Yi-Xian}\ \bibnamefont {{Chen}}}, \ and\ \bibinfo {author} {\bibfnamefont {Douglas N.~C.}\ \bibnamefont {{Lin}}},\ }\bibfield  {title} {\enquote {\bibinfo {title} {{Concurrent Accretion and Migration of Giant Planets in Their Natal Disks with Consistent Accretion Torque}},}\ }\href {\doibase 10.3847/1538-4357/ad5a06} {\bibfield  {journal} {\bibinfo  {journal} {\apj}\ }\textbf {\bibinfo {volume} {971}},\ \bibinfo {eid} {130} (\bibinfo {year} {2024})},\ \Eprint {http://arxiv.org/abs/2406.12716} {arXiv:2406.12716 [astro-ph.EP]} \BibitemShut {NoStop}%
\bibitem [{\citenamefont {{Pinte}}\ \emph {et~al.}(2016)\citenamefont {{Pinte}}, \citenamefont {{Dent}}, \citenamefont {{M{\'e}nard}}, \citenamefont {{Hales}}, \citenamefont {{Hill}}, \citenamefont {{Cortes}},\ and\ \citenamefont {{de Gregorio-Monsalvo}}}]{Pinte2016}%
  \BibitemOpen
  \bibfield  {author} {\bibinfo {author} {\bibfnamefont {C.}~\bibnamefont {{Pinte}}}, \bibinfo {author} {\bibfnamefont {W.~R.~F.}\ \bibnamefont {{Dent}}}, \bibinfo {author} {\bibfnamefont {F.}~\bibnamefont {{M{\'e}nard}}}, \bibinfo {author} {\bibfnamefont {A.}~\bibnamefont {{Hales}}}, \bibinfo {author} {\bibfnamefont {T.}~\bibnamefont {{Hill}}}, \bibinfo {author} {\bibfnamefont {P.}~\bibnamefont {{Cortes}}}, \ and\ \bibinfo {author} {\bibfnamefont {I.}~\bibnamefont {{de Gregorio-Monsalvo}}},\ }\bibfield  {title} {\enquote {\bibinfo {title} {{Dust and Gas in the Disk of HL Tauri: Surface Density, Dust Settling, and Dust-to-gas Ratio}},}\ }\href {\doibase 10.3847/0004-637X/816/1/25} {\bibfield  {journal} {\bibinfo  {journal} {\apj}\ }\textbf {\bibinfo {volume} {816}},\ \bibinfo {eid} {25} (\bibinfo {year} {2016})},\ \Eprint {http://arxiv.org/abs/1508.00584} {arXiv:1508.00584 [astro-ph.SR]} \BibitemShut {NoStop}%
\bibitem [{\citenamefont {{Flaherty}}\ \emph {et~al.}(2017)\citenamefont {{Flaherty}}, \citenamefont {{Hughes}}, \citenamefont {{Rose}}, \citenamefont {{Simon}}, \citenamefont {{Qi}}, \citenamefont {{Andrews}}, \citenamefont {{K{\'o}sp{\'a}l}}, \citenamefont {{Wilner}}, \citenamefont {{Chiang}}, \citenamefont {{Armitage}},\ and\ \citenamefont {{Bai}}}]{Flaherty2017}%
  \BibitemOpen
  \bibfield  {author} {\bibinfo {author} {\bibfnamefont {Kevin~M.}\ \bibnamefont {{Flaherty}}}, \bibinfo {author} {\bibfnamefont {A.~Meredith}\ \bibnamefont {{Hughes}}}, \bibinfo {author} {\bibfnamefont {Sanaea~C.}\ \bibnamefont {{Rose}}}, \bibinfo {author} {\bibfnamefont {Jacob~B.}\ \bibnamefont {{Simon}}}, \bibinfo {author} {\bibfnamefont {Chunhua}\ \bibnamefont {{Qi}}}, \bibinfo {author} {\bibfnamefont {Sean~M.}\ \bibnamefont {{Andrews}}}, \bibinfo {author} {\bibfnamefont {{\'A}gnes}\ \bibnamefont {{K{\'o}sp{\'a}l}}}, \bibinfo {author} {\bibfnamefont {David~J.}\ \bibnamefont {{Wilner}}}, \bibinfo {author} {\bibfnamefont {Eugene}\ \bibnamefont {{Chiang}}}, \bibinfo {author} {\bibfnamefont {Philip~J.}\ \bibnamefont {{Armitage}}}, \ and\ \bibinfo {author} {\bibfnamefont {Xue-ning}\ \bibnamefont {{Bai}}},\ }\bibfield  {title} {\enquote {\bibinfo {title} {{A Three-dimensional View of Turbulence: Constraints on Turbulent Motions in the HD 163296 Protoplanetary Disk Using DCO$^{+}$}},}\ }\href {\doibase
  10.3847/1538-4357/aa79f9} {\bibfield  {journal} {\bibinfo  {journal} {\apj}\ }\textbf {\bibinfo {volume} {843}},\ \bibinfo {eid} {150} (\bibinfo {year} {2017})},\ \Eprint {http://arxiv.org/abs/1706.04504} {arXiv:1706.04504 [astro-ph.EP]} \BibitemShut {NoStop}%
\bibitem [{\citenamefont {Martin}\ \emph {et~al.}(2019)\citenamefont {Martin}, \citenamefont {Nixon}, \citenamefont {Pringle},\ and\ \citenamefont {Livio}}]{Martin:2019qfg}%
  \BibitemOpen
  \bibfield  {author} {\bibinfo {author} {\bibfnamefont {R.~G.}\ \bibnamefont {Martin}}, \bibinfo {author} {\bibfnamefont {C.~J.}\ \bibnamefont {Nixon}}, \bibinfo {author} {\bibfnamefont {J.~E.}\ \bibnamefont {Pringle}}, \ and\ \bibinfo {author} {\bibfnamefont {M.}~\bibnamefont {Livio}},\ }\bibfield  {title} {\enquote {\bibinfo {title} {{On the physical nature of accretion disc viscosity}},}\ }\href {\doibase 10.1016/j.newast.2019.01.001} {\bibfield  {journal} {\bibinfo  {journal} {New Astron.}\ }\textbf {\bibinfo {volume} {70}},\ \bibinfo {pages} {7--11} (\bibinfo {year} {2019})},\ \Eprint {http://arxiv.org/abs/1901.01580} {arXiv:1901.01580 [astro-ph.HE]} \BibitemShut {NoStop}%
\bibitem [{\citenamefont {{Chiang}}\ and\ \citenamefont {{Goldreich}}(1997)}]{Chiang1997}%
  \BibitemOpen
  \bibfield  {author} {\bibinfo {author} {\bibfnamefont {E.~I.}\ \bibnamefont {{Chiang}}}\ and\ \bibinfo {author} {\bibfnamefont {P.}~\bibnamefont {{Goldreich}}},\ }\bibfield  {title} {\enquote {\bibinfo {title} {{Spectral Energy Distributions of T Tauri Stars with Passive Circumstellar Disks}},}\ }\href {\doibase 10.1086/304869} {\bibfield  {journal} {\bibinfo  {journal} {\apj}\ }\textbf {\bibinfo {volume} {490}},\ \bibinfo {pages} {368--376} (\bibinfo {year} {1997})},\ \Eprint {http://arxiv.org/abs/astro-ph/9706042} {arXiv:astro-ph/9706042 [astro-ph]} \BibitemShut {NoStop}%
\bibitem [{\citenamefont {{Abramowicz}}\ \emph {et~al.}(1988)\citenamefont {{Abramowicz}}, \citenamefont {{Czerny}}, \citenamefont {{Lasota}},\ and\ \citenamefont {{Szuszkiewicz}}}]{Abramowicz1988}%
  \BibitemOpen
  \bibfield  {author} {\bibinfo {author} {\bibfnamefont {M.~A.}\ \bibnamefont {{Abramowicz}}}, \bibinfo {author} {\bibfnamefont {B.}~\bibnamefont {{Czerny}}}, \bibinfo {author} {\bibfnamefont {J.~P.}\ \bibnamefont {{Lasota}}}, \ and\ \bibinfo {author} {\bibfnamefont {E.}~\bibnamefont {{Szuszkiewicz}}},\ }\bibfield  {title} {\enquote {\bibinfo {title} {{Slim Accretion Disks}},}\ }\href {\doibase 10.1086/166683} {\bibfield  {journal} {\bibinfo  {journal} {\apj}\ }\textbf {\bibinfo {volume} {332}},\ \bibinfo {pages} {646} (\bibinfo {year} {1988})}\BibitemShut {NoStop}%
\bibitem [{\citenamefont {{Narayan}}\ and\ \citenamefont {{Yi}}(1994)}]{Narayan1994}%
  \BibitemOpen
  \bibfield  {author} {\bibinfo {author} {\bibfnamefont {Ramesh}\ \bibnamefont {{Narayan}}}\ and\ \bibinfo {author} {\bibfnamefont {Insu}\ \bibnamefont {{Yi}}},\ }\bibfield  {title} {\enquote {\bibinfo {title} {{Advection-dominated Accretion: A Self-similar Solution}},}\ }\href {\doibase 10.1086/187381} {\bibfield  {journal} {\bibinfo  {journal} {\apjl}\ }\textbf {\bibinfo {volume} {428}},\ \bibinfo {pages} {L13} (\bibinfo {year} {1994})},\ \Eprint {http://arxiv.org/abs/astro-ph/9403052} {arXiv:astro-ph/9403052 [astro-ph]} \BibitemShut {NoStop}%
\bibitem [{\citenamefont {{Yuan}}\ and\ \citenamefont {{Narayan}}(2014)}]{Yuan2014}%
  \BibitemOpen
  \bibfield  {author} {\bibinfo {author} {\bibfnamefont {Feng}\ \bibnamefont {{Yuan}}}\ and\ \bibinfo {author} {\bibfnamefont {Ramesh}\ \bibnamefont {{Narayan}}},\ }\bibfield  {title} {\enquote {\bibinfo {title} {{Hot Accretion Flows Around Black Holes}},}\ }\href {\doibase 10.1146/annurev-astro-082812-141003} {\bibfield  {journal} {\bibinfo  {journal} {\araa}\ }\textbf {\bibinfo {volume} {52}},\ \bibinfo {pages} {529--588} (\bibinfo {year} {2014})},\ \Eprint {http://arxiv.org/abs/1401.0586} {arXiv:1401.0586 [astro-ph.HE]} \BibitemShut {NoStop}%
\bibitem [{\citenamefont {{Guo}}\ \emph {et~al.}(2024)\citenamefont {{Guo}}, \citenamefont {{Stone}}, \citenamefont {{Quataert}},\ and\ \citenamefont {{Kim}}}]{Guo2024}%
  \BibitemOpen
  \bibfield  {author} {\bibinfo {author} {\bibfnamefont {Minghao}\ \bibnamefont {{Guo}}}, \bibinfo {author} {\bibfnamefont {James~M.}\ \bibnamefont {{Stone}}}, \bibinfo {author} {\bibfnamefont {Eliot}\ \bibnamefont {{Quataert}}}, \ and\ \bibinfo {author} {\bibfnamefont {Chang-Goo}\ \bibnamefont {{Kim}}},\ }\bibfield  {title} {\enquote {\bibinfo {title} {{Magnetized Accretion onto and Feedback from Supermassive Black Holes in Elliptical Galaxies}},}\ }\href {\doibase 10.3847/1538-4357/ad5fe7} {\bibfield  {journal} {\bibinfo  {journal} {\apj}\ }\textbf {\bibinfo {volume} {973}},\ \bibinfo {eid} {141} (\bibinfo {year} {2024})},\ \Eprint {http://arxiv.org/abs/2405.11711} {arXiv:2405.11711 [astro-ph.HE]} \BibitemShut {NoStop}%
\bibitem [{\citenamefont {{Hopkins}}\ \emph {et~al.}(2024)\citenamefont {{Hopkins}}, \citenamefont {{Squire}}, \citenamefont {{Su}}, \citenamefont {{Steinwandel}}, \citenamefont {{Kremer}}, \citenamefont {{Shi}}, \citenamefont {{Grudic}}, \citenamefont {{Wellons}}, \citenamefont {{Faucher-Giguere}}, \citenamefont {{Angles-Alcazar}}, \citenamefont {{Murray}},\ and\ \citenamefont {{Quataert}}}]{Hopkins2024}%
  \BibitemOpen
  \bibfield  {author} {\bibinfo {author} {\bibfnamefont {Philip~F.}\ \bibnamefont {{Hopkins}}}, \bibinfo {author} {\bibfnamefont {Jonathan}\ \bibnamefont {{Squire}}}, \bibinfo {author} {\bibfnamefont {Kung-Yi}\ \bibnamefont {{Su}}}, \bibinfo {author} {\bibfnamefont {Ulrich~P.}\ \bibnamefont {{Steinwandel}}}, \bibinfo {author} {\bibfnamefont {Kyle}\ \bibnamefont {{Kremer}}}, \bibinfo {author} {\bibfnamefont {Yanlong}\ \bibnamefont {{Shi}}}, \bibinfo {author} {\bibfnamefont {Michael~Y.}\ \bibnamefont {{Grudic}}}, \bibinfo {author} {\bibfnamefont {Sarah}\ \bibnamefont {{Wellons}}}, \bibinfo {author} {\bibfnamefont {Claude-Andre}\ \bibnamefont {{Faucher-Giguere}}}, \bibinfo {author} {\bibfnamefont {Daniel}\ \bibnamefont {{Angles-Alcazar}}}, \bibinfo {author} {\bibfnamefont {Norman}\ \bibnamefont {{Murray}}}, \ and\ \bibinfo {author} {\bibfnamefont {Eliot}\ \bibnamefont {{Quataert}}},\ }\bibfield  {title} {\enquote {\bibinfo {title} {{FORGE'd in FIRE II: The Formation of Magnetically-Dominated Quasar Accretion
  Disks from Cosmological Initial Conditions}},}\ }\href {\doibase 10.21105/astro.2310.04506} {\bibfield  {journal} {\bibinfo  {journal} {The Open Journal of Astrophysics}\ }\textbf {\bibinfo {volume} {7}},\ \bibinfo {eid} {19} (\bibinfo {year} {2024})},\ \Eprint {http://arxiv.org/abs/2310.04506} {arXiv:2310.04506 [astro-ph.HE]} \BibitemShut {NoStop}%
\bibitem [{\citenamefont {Pan}\ and\ \citenamefont {Yang}(2021{\natexlab{b}})}]{Pan:2021xhv}%
  \BibitemOpen
  \bibfield  {author} {\bibinfo {author} {\bibfnamefont {Zhen}\ \bibnamefont {Pan}}\ and\ \bibinfo {author} {\bibfnamefont {Huan}\ \bibnamefont {Yang}},\ }\bibfield  {title} {\enquote {\bibinfo {title} {{Supercritical Accretion of Stellar-mass Compact Objects in Active Galactic Nuclei}},}\ }\href {\doibase 10.3847/1538-4357/ac249c} {\bibfield  {journal} {\bibinfo  {journal} {Astrophys. J.}\ }\textbf {\bibinfo {volume} {923}},\ \bibinfo {pages} {173} (\bibinfo {year} {2021}{\natexlab{b}})},\ \Eprint {http://arxiv.org/abs/2108.00267} {arXiv:2108.00267 [astro-ph.HE]} \BibitemShut {NoStop}%
\bibitem [{\citenamefont {Sirko}\ and\ \citenamefont {Goodman}(2003)}]{Sirko:2002ex}%
  \BibitemOpen
  \bibfield  {author} {\bibinfo {author} {\bibfnamefont {Edwin}\ \bibnamefont {Sirko}}\ and\ \bibinfo {author} {\bibfnamefont {Jeremy}\ \bibnamefont {Goodman}},\ }\bibfield  {title} {\enquote {\bibinfo {title} {{Spectral energy distributions of selfgravitating QSO discs}},}\ }\href {\doibase 10.1046/j.1365-8711.2003.06431.x} {\bibfield  {journal} {\bibinfo  {journal} {Mon. Not. Roy. Astron. Soc.}\ }\textbf {\bibinfo {volume} {341}},\ \bibinfo {pages} {501} (\bibinfo {year} {2003})},\ \Eprint {http://arxiv.org/abs/astro-ph/0209469} {arXiv:astro-ph/0209469} \BibitemShut {NoStop}%
\bibitem [{\citenamefont {{Papaloizou}}\ and\ \citenamefont {{Larwood}}(2000)}]{Papaloizou2000}%
  \BibitemOpen
  \bibfield  {author} {\bibinfo {author} {\bibfnamefont {J.~C.~B.}\ \bibnamefont {{Papaloizou}}}\ and\ \bibinfo {author} {\bibfnamefont {J.~D.}\ \bibnamefont {{Larwood}}},\ }\bibfield  {title} {\enquote {\bibinfo {title} {{On the orbital evolution and growth of protoplanets embedded in a gaseous disc}},}\ }\href {\doibase 10.1046/j.1365-8711.2000.03466.x} {\bibfield  {journal} {\bibinfo  {journal} {\mnras}\ }\textbf {\bibinfo {volume} {315}},\ \bibinfo {pages} {823--833} (\bibinfo {year} {2000})},\ \Eprint {http://arxiv.org/abs/astro-ph/9911431} {arXiv:astro-ph/9911431 [astro-ph]} \BibitemShut {NoStop}%
\bibitem [{\citenamefont {{Goldreich}}\ and\ \citenamefont {{Sari}}(2003)}]{Goldreich2003}%
  \BibitemOpen
  \bibfield  {author} {\bibinfo {author} {\bibfnamefont {Peter}\ \bibnamefont {{Goldreich}}}\ and\ \bibinfo {author} {\bibfnamefont {Re'em}\ \bibnamefont {{Sari}}},\ }\bibfield  {title} {\enquote {\bibinfo {title} {{Eccentricity Evolution for Planets in Gaseous Disks}},}\ }\href {\doibase 10.1086/346202} {\bibfield  {journal} {\bibinfo  {journal} {\apj}\ }\textbf {\bibinfo {volume} {585}},\ \bibinfo {pages} {1024--1037} (\bibinfo {year} {2003})},\ \Eprint {http://arxiv.org/abs/astro-ph/0202462} {arXiv:astro-ph/0202462 [astro-ph]} \BibitemShut {NoStop}%
\bibitem [{\citenamefont {{Li}}\ \emph {et~al.}(2019)\citenamefont {{Li}}, \citenamefont {{Li}}, \citenamefont {{Li}},\ and\ \citenamefont {{Lin}}}]{Li2019}%
  \BibitemOpen
  \bibfield  {author} {\bibinfo {author} {\bibfnamefont {Ya-Ping}\ \bibnamefont {{Li}}}, \bibinfo {author} {\bibfnamefont {Hui}\ \bibnamefont {{Li}}}, \bibinfo {author} {\bibfnamefont {Shengtai}\ \bibnamefont {{Li}}}, \ and\ \bibinfo {author} {\bibfnamefont {Douglas N.~C.}\ \bibnamefont {{Lin}}},\ }\bibfield  {title} {\enquote {\bibinfo {title} {{On the Dust Signatures Induced by Eccentric Super-Earths in Protoplanetary Disks}},}\ }\href {\doibase 10.3847/1538-4357/ab4bc8} {\bibfield  {journal} {\bibinfo  {journal} {\apj}\ }\textbf {\bibinfo {volume} {886}},\ \bibinfo {eid} {62} (\bibinfo {year} {2019})},\ \Eprint {http://arxiv.org/abs/1910.03130} {arXiv:1910.03130 [astro-ph.EP]} \BibitemShut {NoStop}%
\bibitem [{\citenamefont {{Ida}}\ \emph {et~al.}(2020)\citenamefont {{Ida}}, \citenamefont {{Muto}}, \citenamefont {{Matsumura}},\ and\ \citenamefont {{Brasser}}}]{Ida2020}%
  \BibitemOpen
  \bibfield  {author} {\bibinfo {author} {\bibfnamefont {Shigeru}\ \bibnamefont {{Ida}}}, \bibinfo {author} {\bibfnamefont {Takayuki}\ \bibnamefont {{Muto}}}, \bibinfo {author} {\bibfnamefont {Soko}\ \bibnamefont {{Matsumura}}}, \ and\ \bibinfo {author} {\bibfnamefont {Ramon}\ \bibnamefont {{Brasser}}},\ }\bibfield  {title} {\enquote {\bibinfo {title} {{A new and simple prescription for planet orbital migration and eccentricity damping by planet-disc interactions based on dynamical friction}},}\ }\href {\doibase 10.1093/mnras/staa1073} {\bibfield  {journal} {\bibinfo  {journal} {\mnras}\ }\textbf {\bibinfo {volume} {494}},\ \bibinfo {pages} {5666--5674} (\bibinfo {year} {2020})},\ \Eprint {http://arxiv.org/abs/2004.07481} {arXiv:2004.07481 [astro-ph.EP]} \BibitemShut {NoStop}%
\bibitem [{\citenamefont {{Papaloizou}}\ \emph {et~al.}(2001)\citenamefont {{Papaloizou}}, \citenamefont {{Nelson}},\ and\ \citenamefont {{Masset}}}]{Papaloizou2001}%
  \BibitemOpen
  \bibfield  {author} {\bibinfo {author} {\bibfnamefont {J.~C.~B.}\ \bibnamefont {{Papaloizou}}}, \bibinfo {author} {\bibfnamefont {R.~P.}\ \bibnamefont {{Nelson}}}, \ and\ \bibinfo {author} {\bibfnamefont {F.}~\bibnamefont {{Masset}}},\ }\bibfield  {title} {\enquote {\bibinfo {title} {{Orbital eccentricity growth through disc-companion tidal interaction}},}\ }\href {\doibase 10.1051/0004-6361:20000011} {\bibfield  {journal} {\bibinfo  {journal} {\aap}\ }\textbf {\bibinfo {volume} {366}},\ \bibinfo {pages} {263--275} (\bibinfo {year} {2001})}\BibitemShut {NoStop}%
\bibitem [{\citenamefont {{D'Angelo}}\ \emph {et~al.}(2006)\citenamefont {{D'Angelo}}, \citenamefont {{Lubow}},\ and\ \citenamefont {{Bate}}}]{DAngelo2006}%
  \BibitemOpen
  \bibfield  {author} {\bibinfo {author} {\bibfnamefont {Gennaro}\ \bibnamefont {{D'Angelo}}}, \bibinfo {author} {\bibfnamefont {Stephen~H.}\ \bibnamefont {{Lubow}}}, \ and\ \bibinfo {author} {\bibfnamefont {Matthew~R.}\ \bibnamefont {{Bate}}},\ }\bibfield  {title} {\enquote {\bibinfo {title} {{Evolution of Giant Planets in Eccentric Disks}},}\ }\href {\doibase 10.1086/508451} {\bibfield  {journal} {\bibinfo  {journal} {\apj}\ }\textbf {\bibinfo {volume} {652}},\ \bibinfo {pages} {1698--1714} (\bibinfo {year} {2006})},\ \Eprint {http://arxiv.org/abs/astro-ph/0608355} {arXiv:astro-ph/0608355 [astro-ph]} \BibitemShut {NoStop}%
\bibitem [{\citenamefont {{Duffell}}\ and\ \citenamefont {{Chiang}}(2015)}]{Duffell2015}%
  \BibitemOpen
  \bibfield  {author} {\bibinfo {author} {\bibfnamefont {Paul~C.}\ \bibnamefont {{Duffell}}}\ and\ \bibinfo {author} {\bibfnamefont {Eugene}\ \bibnamefont {{Chiang}}},\ }\bibfield  {title} {\enquote {\bibinfo {title} {{Eccentric Jupiters via Disk-Planet Interactions}},}\ }\href {\doibase 10.1088/0004-637X/812/2/94} {\bibfield  {journal} {\bibinfo  {journal} {\apj}\ }\textbf {\bibinfo {volume} {812}},\ \bibinfo {eid} {94} (\bibinfo {year} {2015})},\ \Eprint {http://arxiv.org/abs/1507.08667} {arXiv:1507.08667 [astro-ph.EP]} \BibitemShut {NoStop}%
\bibitem [{\citenamefont {{Stone}}\ \emph {et~al.}(2020)\citenamefont {{Stone}}, \citenamefont {{Tomida}}, \citenamefont {{White}},\ and\ \citenamefont {{Felker}}}]{Stone2020}%
  \BibitemOpen
  \bibfield  {author} {\bibinfo {author} {\bibfnamefont {James~M.}\ \bibnamefont {{Stone}}}, \bibinfo {author} {\bibfnamefont {Kengo}\ \bibnamefont {{Tomida}}}, \bibinfo {author} {\bibfnamefont {Christopher~J.}\ \bibnamefont {{White}}}, \ and\ \bibinfo {author} {\bibfnamefont {Kyle~G.}\ \bibnamefont {{Felker}}},\ }\bibfield  {title} {\enquote {\bibinfo {title} {{The Athena++ Adaptive Mesh Refinement Framework: Design and Magnetohydrodynamic Solvers}},}\ }\href {\doibase 10.3847/1538-4365/ab929b} {\bibfield  {journal} {\bibinfo  {journal} {\apjs}\ }\textbf {\bibinfo {volume} {249}},\ \bibinfo {eid} {4} (\bibinfo {year} {2020})},\ \Eprint {http://arxiv.org/abs/2005.06651} {arXiv:2005.06651 [astro-ph.IM]} \BibitemShut {NoStop}%
\bibitem [{\citenamefont {{Nelson}}\ \emph {et~al.}(2013)\citenamefont {{Nelson}}, \citenamefont {{Gressel}},\ and\ \citenamefont {{Umurhan}}}]{Nelson2013}%
  \BibitemOpen
  \bibfield  {author} {\bibinfo {author} {\bibfnamefont {Richard~P.}\ \bibnamefont {{Nelson}}}, \bibinfo {author} {\bibfnamefont {Oliver}\ \bibnamefont {{Gressel}}}, \ and\ \bibinfo {author} {\bibfnamefont {Orkan~M.}\ \bibnamefont {{Umurhan}}},\ }\bibfield  {title} {\enquote {\bibinfo {title} {{Linear and non-linear evolution of the vertical shear instability in accretion discs}},}\ }\href {\doibase 10.1093/mnras/stt1475} {\bibfield  {journal} {\bibinfo  {journal} {\mnras}\ }\textbf {\bibinfo {volume} {435}},\ \bibinfo {pages} {2610--2632} (\bibinfo {year} {2013})},\ \Eprint {http://arxiv.org/abs/1209.2753} {arXiv:1209.2753 [astro-ph.EP]} \BibitemShut {NoStop}%
\bibitem [{\citenamefont {{Shakura}}\ and\ \citenamefont {{Sunyaev}}(1973)}]{Shakura1973}%
  \BibitemOpen
  \bibfield  {author} {\bibinfo {author} {\bibfnamefont {N.~I.}\ \bibnamefont {{Shakura}}}\ and\ \bibinfo {author} {\bibfnamefont {R.~A.}\ \bibnamefont {{Sunyaev}}},\ }\bibfield  {title} {\enquote {\bibinfo {title} {{Black holes in binary systems. Observational appearance.}}}\ }\href@noop {} {\bibfield  {journal} {\bibinfo  {journal} {\aap}\ }\textbf {\bibinfo {volume} {24}},\ \bibinfo {pages} {337--355} (\bibinfo {year} {1973})}\BibitemShut {NoStop}%
\bibitem [{\citenamefont {{Li}}\ \emph {et~al.}(2009)\citenamefont {{Li}}, \citenamefont {{Lubow}}, \citenamefont {{Li}},\ and\ \citenamefont {{Lin}}}]{Li2009}%
  \BibitemOpen
  \bibfield  {author} {\bibinfo {author} {\bibfnamefont {H.}~\bibnamefont {{Li}}}, \bibinfo {author} {\bibfnamefont {S.~H.}\ \bibnamefont {{Lubow}}}, \bibinfo {author} {\bibfnamefont {S.}~\bibnamefont {{Li}}}, \ and\ \bibinfo {author} {\bibfnamefont {D.~N.~C.}\ \bibnamefont {{Lin}}},\ }\bibfield  {title} {\enquote {\bibinfo {title} {{Type I Planet Migration in Nearly Laminar Disks}},}\ }\href {\doibase 10.1088/0004-637X/690/1/L52} {\bibfield  {journal} {\bibinfo  {journal} {\apjl}\ }\textbf {\bibinfo {volume} {690}},\ \bibinfo {pages} {L52--L55} (\bibinfo {year} {2009})},\ \Eprint {http://arxiv.org/abs/0812.0586} {arXiv:0812.0586 [astro-ph]} \BibitemShut {NoStop}%
\bibitem [{\citenamefont {{Ben{\'\i}tez-Llambay}}\ and\ \citenamefont {{Masset}}(2016)}]{Benitez-Llambay2016}%
  \BibitemOpen
  \bibfield  {author} {\bibinfo {author} {\bibfnamefont {Pablo}\ \bibnamefont {{Ben{\'\i}tez-Llambay}}}\ and\ \bibinfo {author} {\bibfnamefont {Fr{\'e}d{\'e}ric~S.}\ \bibnamefont {{Masset}}},\ }\bibfield  {title} {\enquote {\bibinfo {title} {{FARGO3D: A New GPU-oriented MHD Code}},}\ }\href {\doibase 10.3847/0067-0049/223/1/11} {\bibfield  {journal} {\bibinfo  {journal} {\apjs}\ }\textbf {\bibinfo {volume} {223}},\ \bibinfo {eid} {11} (\bibinfo {year} {2016})},\ \Eprint {http://arxiv.org/abs/1602.02359} {arXiv:1602.02359 [astro-ph.IM]} \BibitemShut {NoStop}%
\bibitem [{\citenamefont {{Machida}}\ \emph {et~al.}(2008)\citenamefont {{Machida}}, \citenamefont {{Kokubo}}, \citenamefont {{Inutsuka}},\ and\ \citenamefont {{Matsumoto}}}]{Machida2008}%
  \BibitemOpen
  \bibfield  {author} {\bibinfo {author} {\bibfnamefont {Masahiro~N.}\ \bibnamefont {{Machida}}}, \bibinfo {author} {\bibfnamefont {Eiichiro}\ \bibnamefont {{Kokubo}}}, \bibinfo {author} {\bibfnamefont {Shu-ichiro}\ \bibnamefont {{Inutsuka}}}, \ and\ \bibinfo {author} {\bibfnamefont {Tomoaki}\ \bibnamefont {{Matsumoto}}},\ }\bibfield  {title} {\enquote {\bibinfo {title} {{Angular Momentum Accretion onto a Gas Giant Planet}},}\ }\href {\doibase 10.1086/590421} {\bibfield  {journal} {\bibinfo  {journal} {\apj}\ }\textbf {\bibinfo {volume} {685}},\ \bibinfo {pages} {1220--1236} (\bibinfo {year} {2008})},\ \Eprint {http://arxiv.org/abs/0801.3305} {arXiv:0801.3305 [astro-ph]} \BibitemShut {NoStop}%
\bibitem [{\citenamefont {{Fung}}\ \emph {et~al.}(2019)\citenamefont {{Fung}}, \citenamefont {{Zhu}},\ and\ \citenamefont {{Chiang}}}]{Fung2019}%
  \BibitemOpen
  \bibfield  {author} {\bibinfo {author} {\bibfnamefont {Jeffrey}\ \bibnamefont {{Fung}}}, \bibinfo {author} {\bibfnamefont {Zhaohuan}\ \bibnamefont {{Zhu}}}, \ and\ \bibinfo {author} {\bibfnamefont {Eugene}\ \bibnamefont {{Chiang}}},\ }\bibfield  {title} {\enquote {\bibinfo {title} {{Circumplanetary Disk Dynamics in the Isothermal and Adiabatic Limits}},}\ }\href {\doibase 10.3847/1538-4357/ab53da} {\bibfield  {journal} {\bibinfo  {journal} {\apj}\ }\textbf {\bibinfo {volume} {887}},\ \bibinfo {eid} {152} (\bibinfo {year} {2019})},\ \Eprint {http://arxiv.org/abs/1909.09655} {arXiv:1909.09655 [astro-ph.EP]} \BibitemShut {NoStop}%
\bibitem [{\citenamefont {{Li}}\ \emph {et~al.}(2023)\citenamefont {{Li}}, \citenamefont {{Chen}},\ and\ \citenamefont {{Lin}}}]{Li2023}%
  \BibitemOpen
  \bibfield  {author} {\bibinfo {author} {\bibfnamefont {Ya-Ping}\ \bibnamefont {{Li}}}, \bibinfo {author} {\bibfnamefont {Yi-Xian}\ \bibnamefont {{Chen}}}, \ and\ \bibinfo {author} {\bibfnamefont {Douglas N.~C.}\ \bibnamefont {{Lin}}},\ }\bibfield  {title} {\enquote {\bibinfo {title} {{3D global simulations of accretion onto gap-opening planets: implications for circumplanetary disc structures and accretion rates}},}\ }\href {\doibase 10.1093/mnras/stad3049} {\bibfield  {journal} {\bibinfo  {journal} {\mnras}\ }\textbf {\bibinfo {volume} {526}},\ \bibinfo {pages} {5346--5364} (\bibinfo {year} {2023})},\ \Eprint {http://arxiv.org/abs/2310.02822} {arXiv:2310.02822 [astro-ph.EP]} \BibitemShut {NoStop}%
\bibitem [{\citenamefont {{Li}}\ \emph {et~al.}(2022{\natexlab{a}})\citenamefont {{Li}}, \citenamefont {{Dempsey}}, \citenamefont {{Li}}, \citenamefont {{Li}},\ and\ \citenamefont {{Li}}}]{Li2022b}%
  \BibitemOpen
  \bibfield  {author} {\bibinfo {author} {\bibfnamefont {Ya-Ping}\ \bibnamefont {{Li}}}, \bibinfo {author} {\bibfnamefont {Adam~M.}\ \bibnamefont {{Dempsey}}}, \bibinfo {author} {\bibfnamefont {Hui}\ \bibnamefont {{Li}}}, \bibinfo {author} {\bibfnamefont {Shengtai}\ \bibnamefont {{Li}}}, \ and\ \bibinfo {author} {\bibfnamefont {Jiaru}\ \bibnamefont {{Li}}},\ }\bibfield  {title} {\enquote {\bibinfo {title} {{Hot Circumsingle Disks Drive Binary Black Hole Mergers in Active Galactic Nucleus Disks}},}\ }\href {\doibase 10.3847/2041-8213/ac60fd} {\bibfield  {journal} {\bibinfo  {journal} {\apjl}\ }\textbf {\bibinfo {volume} {928}},\ \bibinfo {eid} {L19} (\bibinfo {year} {2022}{\natexlab{a}})},\ \Eprint {http://arxiv.org/abs/2112.11057} {arXiv:2112.11057 [astro-ph.HE]} \BibitemShut {NoStop}%
\bibitem [{\citenamefont {{de Val-Borro}}\ \emph {et~al.}(2006)\citenamefont {{de Val-Borro}}, \citenamefont {{Edgar}}, \citenamefont {{Artymowicz}}, \citenamefont {{Ciecielag}}, \citenamefont {{Cresswell}}, \citenamefont {{D'Angelo}}, \citenamefont {{Delgado-Donate}}, \citenamefont {{Dirksen}}, \citenamefont {{Fromang}}, \citenamefont {{Gawryszczak}}, \citenamefont {{Klahr}}, \citenamefont {{Kley}}, \citenamefont {{Lyra}}, \citenamefont {{Masset}}, \citenamefont {{Mellema}}, \citenamefont {{Nelson}}, \citenamefont {{Paardekooper}}, \citenamefont {{Peplinski}}, \citenamefont {{Pierens}}, \citenamefont {{Plewa}}, \citenamefont {{Rice}}, \citenamefont {{Sch{\"a}fer}},\ and\ \citenamefont {{Speith}}}]{deValborro2006}%
  \BibitemOpen
  \bibfield  {author} {\bibinfo {author} {\bibfnamefont {M.}~\bibnamefont {{de Val-Borro}}}, \bibinfo {author} {\bibfnamefont {R.~G.}\ \bibnamefont {{Edgar}}}, \bibinfo {author} {\bibfnamefont {P.}~\bibnamefont {{Artymowicz}}}, \bibinfo {author} {\bibfnamefont {P.}~\bibnamefont {{Ciecielag}}}, \bibinfo {author} {\bibfnamefont {P.}~\bibnamefont {{Cresswell}}}, \bibinfo {author} {\bibfnamefont {G.}~\bibnamefont {{D'Angelo}}}, \bibinfo {author} {\bibfnamefont {E.~J.}\ \bibnamefont {{Delgado-Donate}}}, \bibinfo {author} {\bibfnamefont {G.}~\bibnamefont {{Dirksen}}}, \bibinfo {author} {\bibfnamefont {S.}~\bibnamefont {{Fromang}}}, \bibinfo {author} {\bibfnamefont {A.}~\bibnamefont {{Gawryszczak}}}, \bibinfo {author} {\bibfnamefont {H.}~\bibnamefont {{Klahr}}}, \bibinfo {author} {\bibfnamefont {W.}~\bibnamefont {{Kley}}}, \bibinfo {author} {\bibfnamefont {W.}~\bibnamefont {{Lyra}}}, \bibinfo {author} {\bibfnamefont {F.}~\bibnamefont {{Masset}}}, \bibinfo {author} {\bibfnamefont {G.}~\bibnamefont {{Mellema}}},
  \bibinfo {author} {\bibfnamefont {R.~P.}\ \bibnamefont {{Nelson}}}, \bibinfo {author} {\bibfnamefont {S.~J.}\ \bibnamefont {{Paardekooper}}}, \bibinfo {author} {\bibfnamefont {A.}~\bibnamefont {{Peplinski}}}, \bibinfo {author} {\bibfnamefont {A.}~\bibnamefont {{Pierens}}}, \bibinfo {author} {\bibfnamefont {T.}~\bibnamefont {{Plewa}}}, \bibinfo {author} {\bibfnamefont {K.}~\bibnamefont {{Rice}}}, \bibinfo {author} {\bibfnamefont {C.}~\bibnamefont {{Sch{\"a}fer}}}, \ and\ \bibinfo {author} {\bibfnamefont {R.}~\bibnamefont {{Speith}}},\ }\bibfield  {title} {\enquote {\bibinfo {title} {{A comparative study of disc-planet interaction}},}\ }\href {\doibase 10.1111/j.1365-2966.2006.10488.x} {\bibfield  {journal} {\bibinfo  {journal} {\mnras}\ }\textbf {\bibinfo {volume} {370}},\ \bibinfo {pages} {529--558} (\bibinfo {year} {2006})},\ \Eprint {http://arxiv.org/abs/astro-ph/0605237} {arXiv:astro-ph/0605237 [astro-ph]} \BibitemShut {NoStop}%
\bibitem [{\citenamefont {{Li}}\ \emph {et~al.}(2022{\natexlab{b}})\citenamefont {{Li}}, \citenamefont {{Chen}}, \citenamefont {{Lin}},\ and\ \citenamefont {{Wang}}}]{Li2022a}%
  \BibitemOpen
  \bibfield  {author} {\bibinfo {author} {\bibfnamefont {Ya-Ping}\ \bibnamefont {{Li}}}, \bibinfo {author} {\bibfnamefont {Yi-Xian}\ \bibnamefont {{Chen}}}, \bibinfo {author} {\bibfnamefont {Douglas N.~C.}\ \bibnamefont {{Lin}}}, \ and\ \bibinfo {author} {\bibfnamefont {Zhuoxiao}\ \bibnamefont {{Wang}}},\ }\bibfield  {title} {\enquote {\bibinfo {title} {{Spin Evolution of Stellar-mass Black Holes Embedded in AGN Disks: Orbital Eccentricity Produces Retrograde Circumstellar Flows}},}\ }\href {\doibase 10.3847/2041-8213/ac5b61} {\bibfield  {journal} {\bibinfo  {journal} {\apjl}\ }\textbf {\bibinfo {volume} {928}},\ \bibinfo {eid} {L1} (\bibinfo {year} {2022}{\natexlab{b}})},\ \Eprint {http://arxiv.org/abs/2203.05539} {arXiv:2203.05539 [astro-ph.HE]} \BibitemShut {NoStop}%
\bibitem [{\citenamefont {{Tanaka}}\ \emph {et~al.}(2002)\citenamefont {{Tanaka}}, \citenamefont {{Takeuchi}},\ and\ \citenamefont {{Ward}}}]{Tanaka2002}%
  \BibitemOpen
  \bibfield  {author} {\bibinfo {author} {\bibfnamefont {Hidekazu}\ \bibnamefont {{Tanaka}}}, \bibinfo {author} {\bibfnamefont {Taku}\ \bibnamefont {{Takeuchi}}}, \ and\ \bibinfo {author} {\bibfnamefont {William~R.}\ \bibnamefont {{Ward}}},\ }\bibfield  {title} {\enquote {\bibinfo {title} {{Three-Dimensional Interaction between a Planet and an Isothermal Gaseous Disk. I. Corotation and Lindblad Torques and Planet Migration}},}\ }\href {\doibase 10.1086/324713} {\bibfield  {journal} {\bibinfo  {journal} {\apj}\ }\textbf {\bibinfo {volume} {565}},\ \bibinfo {pages} {1257--1274} (\bibinfo {year} {2002})}\BibitemShut {NoStop}%
\bibitem [{\citenamefont {{Szul{\'a}gyi}}\ \emph {et~al.}(2016)\citenamefont {{Szul{\'a}gyi}}, \citenamefont {{Masset}}, \citenamefont {{Lega}}, \citenamefont {{Crida}}, \citenamefont {{Morbidelli}},\ and\ \citenamefont {{Guillot}}}]{Szulagyi2016}%
  \BibitemOpen
  \bibfield  {author} {\bibinfo {author} {\bibfnamefont {J.}~\bibnamefont {{Szul{\'a}gyi}}}, \bibinfo {author} {\bibfnamefont {F.}~\bibnamefont {{Masset}}}, \bibinfo {author} {\bibfnamefont {E.}~\bibnamefont {{Lega}}}, \bibinfo {author} {\bibfnamefont {A.}~\bibnamefont {{Crida}}}, \bibinfo {author} {\bibfnamefont {A.}~\bibnamefont {{Morbidelli}}}, \ and\ \bibinfo {author} {\bibfnamefont {T.}~\bibnamefont {{Guillot}}},\ }\bibfield  {title} {\enquote {\bibinfo {title} {{Circumplanetary disc or circumplanetary envelope?}}}\ }\href {\doibase 10.1093/mnras/stw1160} {\bibfield  {journal} {\bibinfo  {journal} {\mnras}\ }\textbf {\bibinfo {volume} {460}},\ \bibinfo {pages} {2853--2861} (\bibinfo {year} {2016})},\ \Eprint {http://arxiv.org/abs/1605.04586} {arXiv:1605.04586 [astro-ph.EP]} \BibitemShut {NoStop}%
\end{thebibliography}%

\end{document}